\documentclass[sigconf]{acmart} %manuscript,screen,review

\AtBeginDocument{%
  \providecommand\BibTeX{{%
    \normalfont B\kern-0.5em{\scshape i\kern-0.25em b}\kern-0.8em\TeX}}}

\copyrightyear{2025} 
\acmYear{2025} 
\setcopyright{cc}
\setcctype{by}
\acmConference[CHI '25]{CHI Conference on Human Factors in Computing Systems}{April 26-May 1, 2025}{Yokohama, Japan}
\acmBooktitle{CHI Conference on Human Factors in Computing Systems (CHI '25), April 26-May 1, 2025, Yokohama, Japan}\acmDOI{10.1145/3706598.3714077}
\acmISBN{979-8-4007-1394-1/25/04}

\usepackage{enumitem}
\setlist[itemize]{left=0pt, label=--, topsep=3pt}

\usepackage{subcaption} 
\usepackage{url}
\usepackage{verbatim}
\makeatletter
\g@addto@macro{\UrlBreaks}{\UrlOrds}
\makeatother
\usepackage{multirow} %tabular cells spanning multiple rows
\usepackage{color} %colour management
\usepackage{xspace}
\acmSubmissionID{4377}

\newcommand{\m}{\textit{M=}}

\newcommand{\md}{\textit{Mdn=}}
\newcommand{\sd}{\textit{SD=}}

\newcommand{\N}{\textit{N=}}

\newcommand{\F}[3]{$F({#1},{#2})={#3}$}
\newcommand{\p}{\textit{p=}}

\newcommand{\pminor}{\textit{p$<$}}

\newcommand{\condition}{interaction method\xspace}
\newcommand{\movementCondition}{movement\xspace}

\newcommand{\physicalDemand}{\textit{physical demand}\xspace}

\begin{document}

\title{Bumpy Ride? Understanding the Effects of External Forces on Spatial Interactions in Moving Vehicles}

\author{Markus Sasalovici}
\orcid{0000-0001-9883-2398}
\affiliation{
  \institution{Mercedes-Benz Tech Motion GmbH}
  \city{Böblingen}
  \country{Germany}
}
\affiliation{%
  \institution{Institute of Media Informatics, Ulm University}
  \city{Ulm}
  \country{Germany}
}
\email{markus.sasalovici@mercedes-benz.com}

\author{Albin Zeqiri}
\orcid{0000-0001-6516-3810}
\affiliation{
  \institution{Institute of Media Informatics, Ulm University}
  \city{Ulm}
  \country{Germany}
}
\email{albin.zeqiri@uni-ulm.de}

\author{Robin Connor Schramm}
\orcid{0000-0002-4775-4219}
\affiliation{
  \institution{Mercedes-Benz Tech Motion GmbH}
  \city{Böblingen}
  \country{Germany}
}
\affiliation{
  \institution{RheinMain University of Applied Sciences}
  \city{Wiesbaden}
  \country{Germany}
}
\email{robin.schramm@mercedes-benz.com}

\author{Oscar Javier Ariza Nuñez}
\orcid{0000-0001-8130-840X}
\affiliation{
  \institution{Mercedes-Benz Tech Motion GmbH}
  \city{Böblingen}
  \country{Germany}
}
\email{oscar.ariza@mercedes-benz.com}

\author{Pascal Jansen}
\orcid{0000-0002-9335-5462}
\affiliation{
 \institution{Institute of Media Informatics, Ulm University}
 \city{Ulm}
 \country{Germany}
}
\email{pascal.jansen@uni-ulm.de}

\author{Jann Philipp Freiwald}
\orcid{0000-0002-1977-5186}
\affiliation{
  \institution{Mercedes-Benz Tech Motion GmbH}
  \city{Böblingen}
  \country{Germany}
}
\email{jann_philipp.freiwald@mercedes-benz.com}

\author{Mark Colley}
\orcid{0000-0001-5207-5029}
\email{m.colley@ucl.ac.uk}
\affiliation{
  \institution{Institute of Media Informatics, Ulm University}
  \city{Ulm}
  \country{Germany}
}

\affiliation{
  \institution{UCL Interaction Centre}
  \city{London}
  \country{United Kingdom}
}

\author{Christian Winkler}
\orcid{0009-0007-5406-9846}
\affiliation{
  \institution{Mercedes-Benz Tech Motion GmbH}
  \city{Böblingen}
  \country{Germany}
}
\email{christian.w.winkler@mercedes-benz.com}

\author{Enrico Rukzio}
\orcid{0000-0002-4213-2226}
\affiliation{
  \institution{Institute of Media Informatics, Ulm University}
  \city{Ulm}
  \country{Germany}
}
\email{enrico.rukzio@uni-ulm.de}

\renewcommand{\shortauthors}{Sasalovici et al.}

\begin{abstract}
As the use of Head-Mounted Displays in moving vehicles increases, passengers can immerse themselves in visual experiences independent of their physical environment. However, interaction methods are susceptible to physical motion, leading to input errors and reduced task performance. This work investigates the impact of G-forces, vibrations, and unpredictable maneuvers on 3D interaction methods. We conducted a field study with 24 participants in both stationary and moving vehicles to examine the effects of vehicle motion on four interaction methods: (1) Gaze\&Pinch, (2) DirectTouch, (3) Handray, and (4) HeadGaze. Participants performed selections in a Fitts' Law task. Our findings reveal a significant effect of vehicle motion on interaction accuracy and duration across the tested combinations of Interaction Method $\times$ Road Type $\times$ Curve Type. We found a significant impact of movement on throughput, error rate, and perceived workload. Finally, we propose future research considerations and recommendations on interaction methods during vehicle movement.
\end{abstract}




\begin{teaserfigure}
    \includegraphics[width=\textwidth]{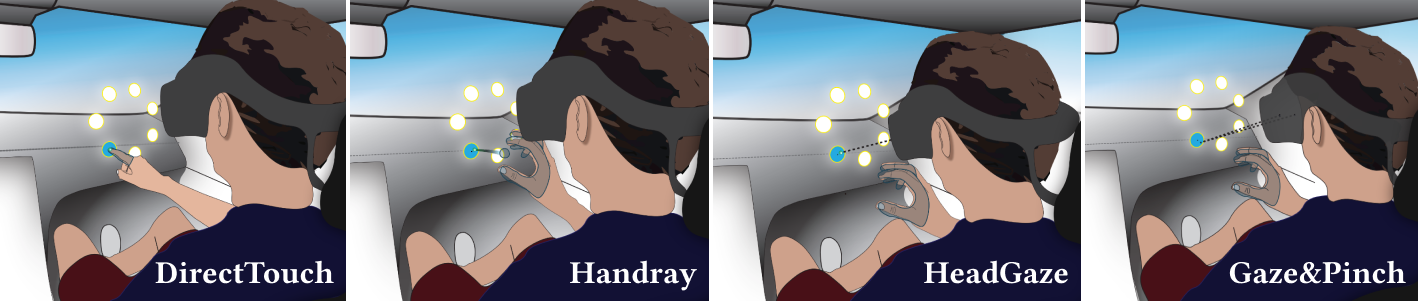}
    \caption{We investigate the effects of external forces in a moving vehicle on spatial interactions in the context of in-car augmented reality using a Fitts' law task. The methods are, from left to right, DirectTouch, Handray, HeadGaze, and Gaze\&Pinch.}
    \label{fig:teaser}
    \Description{The figure shows a person sitting on the passenger seat inside a car wearing a Mixed Reality headset. The person is interacting with virtual objects projected in front of him. The virtual objects represent buttons of the Fitts' Law Task. Seven of these buttons are arranged in a circular layout. One is highlighted in blue, prompting a selection from the person. A total of four interaction methods are illustrated: DirectTouch, Handray, HeadGaze, Gaze\&Pinch.}
\end{teaserfigure}

\maketitle

\section{Introduction}

In 2022, workers in the US commuted for $\approx$230 hours to and from work~\cite{uscensus_commutetime_2022}. Commuting time negatively affects well-being~\cite{colley2023much, raffelhuschen2018deutsche} due to the limited possibility of engaging in meaningful activities. However, with automated vehicles (AVs), commute times could be used to perform non-driving related tasks (NDRT). This will enable former drivers to use the freed-up time to work~\cite{Mathis.2021}, watch movies, or even sleep~\cite{colley2021resync}. In a survey by \citet{Mathis.2021}, 58.8\% of the participants indicated that they could imagine using the newly gained time to perform work-related tasks. Due to the current commuting behavior, 60.7\% state that they do not usually use this time for work-related tasks. Current challenges hindering work-related tasks include manual driving, lack of infrastructure for connectivity, and privacy in public transport~\cite{Mathis.2021}.

Using augmented reality (AR) and virtual reality (VR) Head-Mounted Displays (HMD) during commutes enables users to engage in working tasks, freely customize their environment, and keep content private. Layouts can also be adapted to task-specific requirements and respect user-based ergonomics to ensure an optimal working environment~\cite{Mcgill.2020b}. However, the display of digital content at head level can disrupt the visibility of the surrounding environment. This can lead to users experiencing increased symptoms of motion sickness (MS), as visual motion perception is reduced~\cite{Diels.2016, sasalovici2023InCarOfficeCan}. 

HMDs are currently still in their development but are starting to mature with notable launches of the Apple Vision Pro~\cite{AppleInc.2023} and the Meta Quest 3~\cite{Meta.2023}. The experiences available could inspire users to try using such devices during commutes, such as watching movies on large virtual screens. Meta aims to bring the Quest 3 into vehicles\footnote{Meta and BMW: Taking AR and VR Experiences on the Road. \url{https://about.fb.com/news/2023/05/meta-bmw-ar-vr-experiences/}; Accessed 22.08.2024} and multiple car vendors have conducted research on using HMDs in their vehicles~\cite{Bach.2022,Haeling.2018,AUDIAG.2022,Ghiurau.2020,BMWMGmbH.2022}. To engage in NDRTs, interactions within the vehicle are necessary. These occur while standing, for example, at a traffic light, or during movement~\cite{ahmadTouchscreenUsabilityInput2015, mayerEffectRoadBumps2018, colleySwiVRCarSeatExploringVehicle2021}. Here, multiple driving-related factors such as G-Forces, vibrations, dynamic lighting, unpredictable maneuvers, and the constrained space within the vehicle present challenges for passengers to efficiently use HMDs during commutes. While previous work has already shown effects of this motion in simulators with HMDs~\cite{colleySwiVRCarSeatExploringVehicle2021} and for touch-screen interaction~\cite{mayerEffectRoadBumps2018}, an evaluation of HMD-relevant interaction methods in a real-world setting is missing. 

In a within-subject study with \N{24} participants, we evaluated interaction methods that do not require proprietary hardware other than the headset.
We evaluated the interaction methods \textit{Gaze\&Pinch, DirectTouch, Handray}, and \textit{HeadGaze} using a Fitts' Law task. The impact of vehicular motion on interactions was analyzed regarding task performance, perceived workload, system usability, motion sickness, perceived safety, and trust using standardized questionnaires. Furthermore, we identified patterns between vehicle movement and interaction errors.
In this work, we answer the following research questions (RQs):

\begin{itemize}[itemsep=5pt]
\item \textbf{RQ1} \label{rq1} {\itshape What impact does vehicle motion have on perceived workload, usability and task performance for the interaction methods regarding selection tasks?}
\item \textbf{RQ2} \label{rq2} {\itshape Which type of vehicle motion (i.e., standstill, bumpy road, long-curve, short-curve) impacts the evaluated interaction methods significantly?}
\item \textbf{RQ3} \label{rq3} {\itshape How should the interaction methods Gaze\&Pinch, DirectTouch, Handray and HeadGaze be adapted, considering their usage within a moving vehicle?}
\end{itemize}

\textit{Contribution Statement:} This study (1) investigated the interactions {\itshape Gaze\&Pinch, DirectTouch, Handray} and {\itshape HeadGaze} employable for AR and VR HMD applications in moving and standing contexts.
Expanding on this, (2) the collection of sensor data and subsequent labeling into three road and curve types, allowed for the assessment of selection precision and time for each combination of Interaction Method $\times$ Road Type $\times$ Curve Type. This enabled a detailed analysis of how the interaction methods performed during each combination.
Additionally, (3) we performed semi-structured interviews to gather qualitative data on using these interaction methods in moving vehicles. 
Finally, (4) we propose a set of guidelines with recommendations on interaction methods for selection tasks during vehicle movement, and considerations for future research.

The results help define which interaction method should best be used during movement, with the assessment of the impact of road and curve types during vehicle movement on eye-tracking-based interactions being an additional novelty.

\section{Related Work}
In the following, we introduce general input modalities in the automotive context, provide an in-depth description of AR and VR interaction, and discuss motion effects on interaction. 

\subsection{Input Modalities in Vehicles}
\label{sec:input-modalities-rw}

Vehicle interfaces commonly utilize input modalities such as touch, gaze, and gestures, particularly in the front area, where interaction is most frequent~\cite{jansenDesignSpaceHuman2022}. 
% tactile - touch
%There are various approaches to embed visual, auditory, and tactile/haptic input modalities into vehicles.
%Regarding tactile/haptic input, 
Previous studies on AVs explored the use of touch panels for drivers to initiate maneuvers at the automation limit~\cite{walch_towards_2016, walch_touch_2017} or to select specific AV maneuvers, such as lane changes~\cite{kauer_how_2010}. These touch panels were typically placed either on the steering wheel~\cite{koyama2014multi, pfeiffer2010multi, doring2011gestural} within the center/middle console~\cite{rumelin_free-hand_2013, ng2016investigating, ahmadTouchscreenUsabilityInput2015, walch2019cooperation, colley2021orias}, or on a separate tablet~\cite{colley2022systematic100m}.
% tactile - gesture
Hand gestures were also used for maneuver-based intervention~\cite{detjen_user-defined_2019, colley2022systematic100m} and lateral and longitudinal motion~\cite{manawadu_hand_2016}. 
Similarly, \citet{rumelin_free-hand_2013} and \citet{colley2022systematic100m} utilized free-hand pointing gestures for input, while \citet{fujimura_driver_2013} employed hand-constrained pointing gestures.
% visual - gaze
Eye-gaze as a standalone input was utilized by \citet{poitschke_gaze-based_2011} for referencing or selecting objects~\cite{roider_effects_2017, neselrath_combining_2016}. Additionally, multimodal input was employed to address the challenges associated with unimodal interaction. For instance, gaze was used to localize the target, while hand gestures were used to coordinate pointing~\cite{kim_cascaded_2020, roiderSeeYourPoint2018, colley2023effectsurgency}.
% auditory - speech
Speech input has been implemented to facilitate driver-vehicle cooperation and to select vehicle maneuvers~\cite{ataya_how_2021}. For instance, \citet{roider_effects_2017}, \citet{neselrath_combining_2016}, and \citet{sezgin_multimodal_2009} examined the use of speech commands for selecting objects within the vehicle. However, voice input may be less effective in noisy environments (e.g., during group conversations), and drivers may have limited trust in speech recognition systems or may become confused about the appropriate commands needed to initiate the desired actions~\cite{bengler_hmi_2020, detjen_user-defined_2019}.

Most studies were conducted using low-fidelity driving simulators without motion feedback (e.g.,\cite{gomaa_studying_2020, roider_effects_2017, rumelin_free-hand_2013, roiderSeeYourPoint2018, riegler2020gaze}). However, the vehicle motions induced by road and driving conditions likely impact the results significantly. They may alter the considerations for real in-vehicle interaction proposed in these studies. This is particularly important for studies that measure interaction precision~\cite{gomaa_studying_2020} or completion time~\cite{ng2016investigating}.

\subsection{Interactions in Augmented and Virtual Reality in Vehicles}

Performing mid-air interactions in moving vehicles introduces a distinct set of challenges, stemming primarily from unpredictable vehicular motion. Prior research investigated the usage of touchscreens in moving vehicles~\cite{ahmadInteractiveDisplaysVehicles2014, ahmadTouchscreenUsabilityInput2015, mayerEffectRoadBumps2018, aslanLeapTouchProximity2015, pampelFittsGoesAutobahn2019}, with studies such as those performed by \citet{mayerEffectRoadBumps2018} and Ahmad et al.~\cite{ahmadInteractiveDisplaysVehicles2014, ahmadTouchscreenUsabilityInput2015} having specifically investigated the impact such movements have on touchscreen interactions. \citet{ahmadInteractiveDisplaysVehicles2014} state that road perturbations and vehicle motion can increase erroneous selections. Furthermore, they state that this behavior requires drivers to dedicate more time to performing selection tasks, potentially diverting attention from driving and raising safety concerns. \citet{mayerEffectRoadBumps2018} further explored this topic by using a motion simulator, investigating the effects of road bumps on touchscreen interactions under varying vehicle speeds. They identified a significant reduction in selection accuracy, with vehicle speed not influencing task performance. Furthermore, previous research has considered various aspects like input prediction~\cite{ahmadInteractiveDisplaysVehicles2014, mayerEffectRoadBumps2018} and multimodal input~\cite{roiderSeeYourPoint2018} to improve the usability of such interactions.

However, only limited research was conducted regarding the investigation of interactions performed within AR or VR in vehicle contexts~\cite{schramm2023AssessingAugmentedReality, colleySwiVRCarSeatExploringVehicle2021, kariHandyCastPhonebasedBimanual2023, tsengFingerMapperMappingFinger2023}. Studies by \citet{tsengFingerMapperMappingFinger2023}, and \citet{kariHandyCastPhonebasedBimanual2023} investigated interaction methods that improve interactions within constrained spaces persisting within cars. 
\citet{colleySwiVRCarSeatExploringVehicle2021} used a 1-Degree of Freedom (DoF) motion platform to investigate common interaction methods such as touch, speech, gesture and eye-gaze input in VR regarding task performance. They found that movement negatively affected task performance for eye-gaze and gesture, with touch and speech remaining largely unaffected. Furthermore, \citet{schramm2023AssessingAugmentedReality} investigated multiple interaction methods performed within AR regarding workload, usability, and task performance in a moving vehicle. They found Eye-Gaze with a hardware button as the selection method to be the fastest interaction methods, providing the lowest workload. HeadGaze techniques featured low error rates and a comparably low workload. Hand-pointing with a gesture as confirmation was described as highly frustrating for participants and featured high physical demand.

\subsection{Effects of Vehicle Motion on Interaction}

According to Hock, Colley et al.~\cite{hock2022introducing}, motion and visual fidelity dimensions are essential to classify approaches on the Simulator Continuum. Motion fidelity can range from no motion over motion cues to using a real vehicle. Visual fidelity can range from a 2D screen to the real world. While studies in the real world naturally exhibit the highest external validity, reproducibility or specific situations might only be possible in simulators. Therefore, \citet{colleySwiVRCarSeatExploringVehicle2021} introduced the SwiVR-Car-Seat, representing longitudinal and lateral vehicle dynamics using a 1-DoF rotation. In the SwiVR-Car-Seat, vehicle dynamics in curves are also matched to the chair’s rotation; however, it cannot provide simultaneous motion feedback for both longitudinal and lateral dynamics. Thus, the VAMPIRE by Hock, Colley et al.~\cite{hock2022introducing} introduces a 2-DoF approach where a wheelchair drives in circles to simulate motion forces.

Additionally, on-road driving simulation~\cite{goedicke_vr-oom_2018, bu2024portobello, mcgill20222passengxr, goedicke2022xroom} and the use of a Wizard-of-Oz (WoZ) driver to control the vehicle~\cite{baltodano_rrads_2015, detjen2020wizard} have been proposed. However, these works have not yet used these setups to evaluate the effects of motion on interaction but for visualization purposes only.

Regarding interaction effects of motion, \citet{ng2016investigating} compared pressure input and haptic feedback for in-car touchscreens between a low-fidelity driving simulator and a real vehicle, finding that while accuracy was similar, selection time was worse in the real vehicle. Similarly, \citet{ahmadTouchscreenUsabilityInput2015} showed that vehicle motion increases the effort required for selection. Similar findings were also found by \citet{goode_impact_nodate, salmon_effects_2011, kim_evaluation_2014}.

\section{User Study: Understanding External Forces}
To answer the RQs, a within-subject user study with 24 participants was conducted. We varied the \condition (Gaze\&Pinch, DirectTouch, Handray, and HeadGaze) and the \movementCondition (standstill, movement).

\subsection{Interaction Methods} \label{c:interactionMethods}
This section describes the evaluated interaction methods, focusing on approaches for selecting near objects in virtual environments.
According to \citet{hertel2021TaxonomyInteractionTechniques}, selecting near objects usually involves the collision of a handheld controller or the hand itself with the virtual object.
For more distant objects, techniques based on ray-casts are used, originating from the user's head, eyes, or hand. They may require a secondary confirmation step.
Building on this and previous studies focusing on AR/VR interaction research in vehicles \cite{schramm2023AssessingAugmentedReality,colleySwiVRCarSeatExploringVehicle2021}, we investigate (1) Gaze\&Pinch, (2) DirectTouch, (3) Handray, and (4) HeadGaze as interaction methods (see \autoref{fig:teaser}).
Each method offers a different approach to user interaction, allowing us to assess various aspects of usability and responsiveness. For the implementation, we used Unity 2022.3.34f1 along with \href{https://docs.unity3d.com/Packages/com.unity.xr.interaction.toolkit@2.5/manual/index.html}{XR Interaction Toolkit (XRI) 2.5.4}, \href{https://docs.unity3d.com/Packages/com.unity.xr.hands@1.4/manual/index.html}{XR Hands 1.4.1}, and the \href{https://github.com/ultraleap/UnityPlugin/releases/tag/com.ultraleap.tracking\%2F6.15.1}{Unity Ultraleap Tracking Package 6.15.1} with the tracking service 6.0.0 (Hyperion). For immediate feedback, selected objects in the environment responded visually by changing colors from white to red (see \autoref{c:FittsTask}). We validated the usability of all interaction methods by performing internal tests with three participants from our institution, including trials with prescription glasses to ensure the functionality of Gaze\&Pinch.

\subsubsection{Gaze\&Pinch}
The user moves their eyes to focus on an object or interface element they wish to select and then confirms the selection by performing a pinch gesture with the dominant hand~\cite{pfeuffer2017GazePinchInteraction, mutasim2021PinchClickDwell}. Gaze\&Pinch allows for quick and natural targeting, as humans can swiftly shift their attention by moving their eyes, far quicker than moving a cursor with a traditional input device \cite{pfeuffer2017GazePinchInteraction}. We continuously monitor the user’s gaze, identifying the point in the environment the user is looking at and highlighting potential interactive objects.

We did not implement a visual representation of the cursor for Gaze\&Pinch, as \citet{sidenmark2019EyeHeadSynergetic} suggests it to be distracting, especially when it follows every eye movement. Visual feedback was limited to the colored highlighting of the targets.

\subsubsection{DirectTouch}
For the implementation of DirectTouch, we utilized the \href{https://docs.unity3d.com/Packages/com.unity.xr.interaction.toolkit@2.5/manual/xr-poke-interactor.html}{XR Poke Interactor} of the XRI. When the user's dominant index finger pokes a virtual object, it starts following the finger tip regarding depth, is visually highlighted, and becomes selected when the user releases the virtual object thereafter. In line with \citet{kim2022PseudohapticButtonImproving, speicher2019PseudohapticControlsMidair}, we used protrusion of virtual objects to improve the sense of reality and spatiotemporal perception during interactions. This behavior mimics the push of a physical button, making it natural and easy to learn, potentially increasing the feeling of presence and immersion.

\subsubsection{Handray}
Handray uses a ray projection from the user's dominant hand for object selection and was implemented using the \href{https://docs.unity3d.com/Packages/com.unity.xr.interaction.toolkit@2.5/manual/xr-ray-interactor.html}{XR Ray Interactor}. To address precision issues of ray projection and user hand movements, we applied smoothing by using the \href{https://docs.unity3d.com/Packages/com.unity.xr.interaction.toolkit@2.5/manual/xr-transform-stabilizer.html}{XR Transform Stabilizer} to compensate for the Heisenberg effect in spatial interaction~\cite{wolf2020understanding}. 
The ray extends from the center between the thumb and index finger until it reaches the target object or a predefined distance (10m), while a pinch gesture confirms the selection. This method combines intuitive pointing with a simple, natural gesture for confirmation, providing a seamless and user-friendly interaction experience.

\subsubsection{HeadGaze}

HeadGaze was implemented with the \href{https://docs.unity3d.com/Packages/com.unity.xr.interaction.toolkit@2.5/manual/xr-gaze-interactor.html}{XR Gaze Interactor}. The system tracks the user's head movements and emits a raycast in the direction of gaze. The cursor is placed in the center of the view, which has to be aligned with the target object for selection. A pinch gesture confirms the selection.

\subsection{Apparatus \& Test Environment}

\begin{figure}[ht]
	\centering
  	\includegraphics[width=\linewidth]{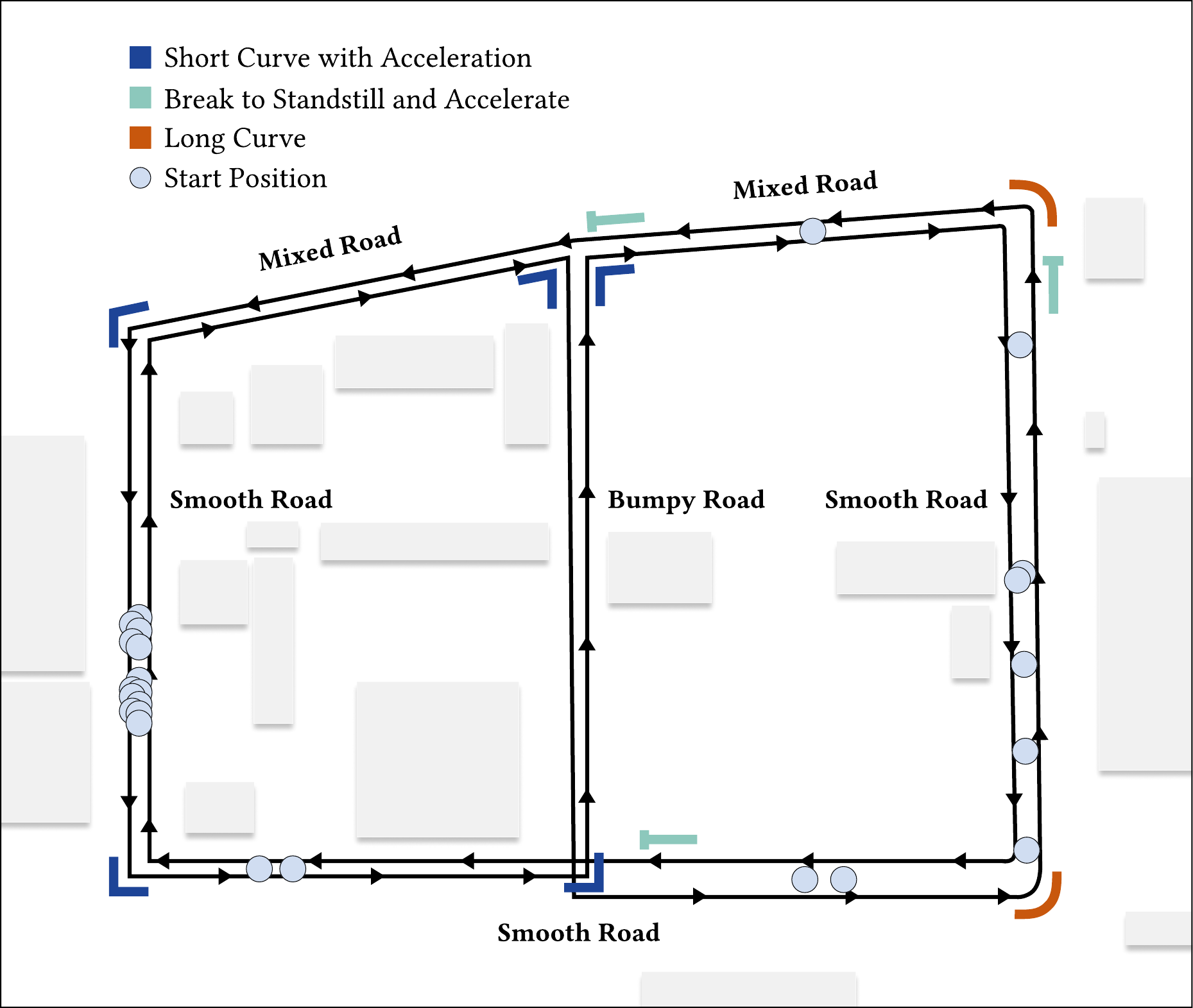} 
  	\caption{Route, instructions, and starting positions for all 24 participants. 30km/h is the standard speed employed on the course.}
  	\label{fig:study-course}
  	\Description{The figure displays a map of a driving course with several road types and curve types being highlighted. Driving instructions are given with the position along the course where they should be applied. The direction to be driven is indicated by black arrows, which guide the driver through the following sections: Short-Curves with Acceleration (highlighted in blue), areas where the driver must Break to Standstill and Accelerate (highlighted in pink), Long-Curves (highlighted in orange). Furthermore, start positions are depicted as a light-blue circle with black outline.}
\end{figure}

The study took place in a midsize-estate, with participants seated in the front passenger seat. The vehicle was equipped with a Varjo XR-3 HMD and a 6-DoF tracking system implemented using middleware by LP-Research~\cite{Petersen.2019b}. An additional IMU was fixed to the vehicle's dashboard in front of the co-driver and used for data recording. Hand tracking was implemented using an Ultraleap Leap Motion 2 attached to the front of the HMD with a 15° downward-facing angle. As we performed a field study, outdoor lighting conditions could vary across participants, from high sunlight exposure to cloudy and rainy days. We, therefore, utilized Ultraleap's {\itshape Hinting API} to ensure stable tracking of users' hands across weather conditions. The parameters can be obtained from \autoref{sec:hyperion_hinting}.

We investigated two driving scenarios, namely standstill and movement, with each interaction method performed during each scenario. Therefore, our study design includes the independent variable interaction method with four levels and movement condition with two levels, resulting in a  4 $\times$ 2 within-subject study. 

\begin{figure*}[ht!]
\centering
\small
    \begin{subfigure}[c]{0.245\linewidth}
        \includegraphics[width=\linewidth, trim=0 27 0 20, clip]{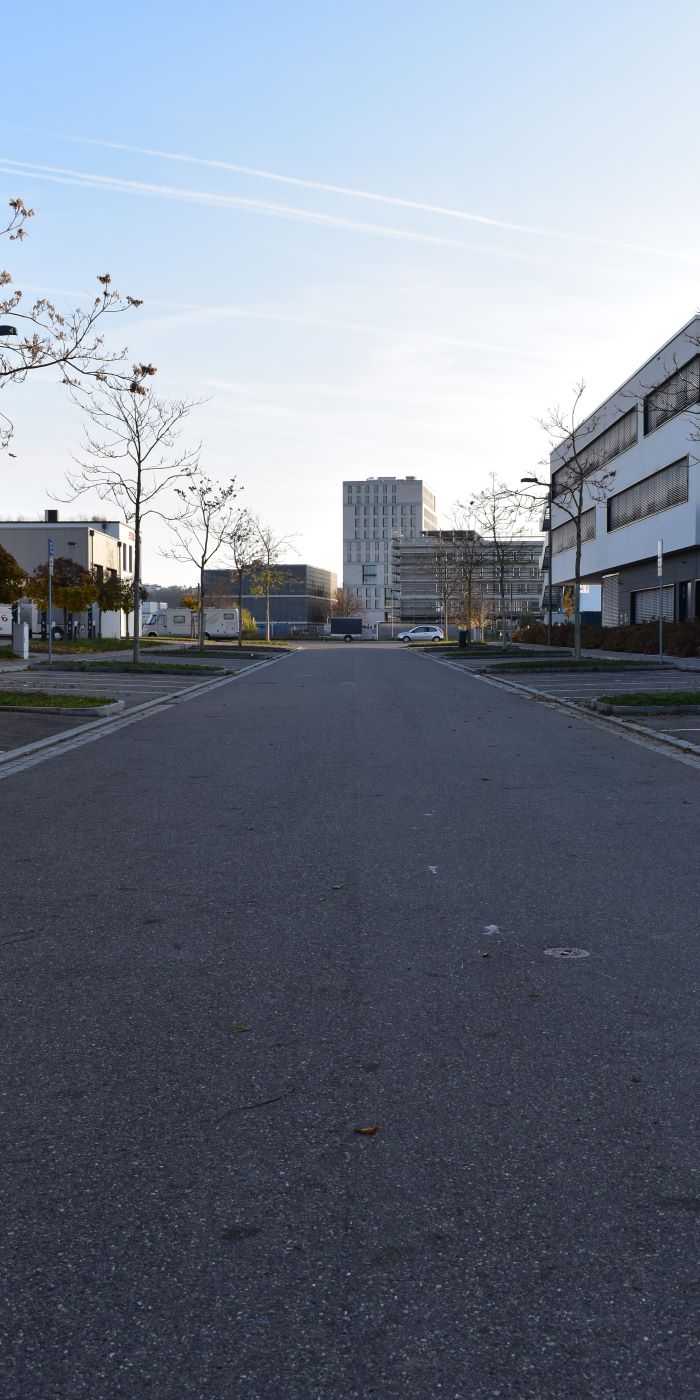}
             \caption{SmoothRoad}~\label{fig:smooth_road}
        \Description{The picture shows a straight road with a well-maintained, smooth surface. The road contains close to no potholes or cracks.}
    \end{subfigure}
    \begin{subfigure}[c]{0.245\linewidth}
        \includegraphics[width=\linewidth, trim=0 20 0 27, clip]{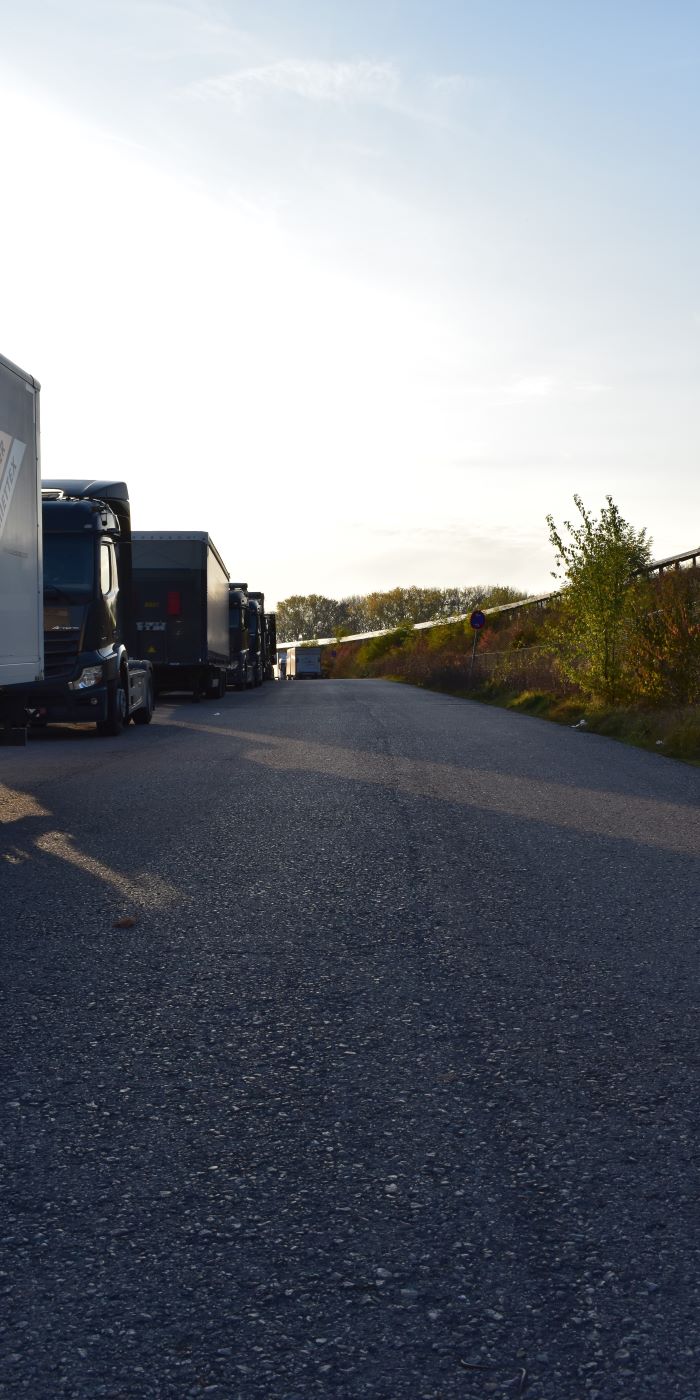}
        \caption{MixedRoad}~\label{fig:mixed_road_2}
        \Description{The picture shows a straight road with a rough surface.}
    \end{subfigure}
    \begin{subfigure}[c]{0.245\linewidth}
        \includegraphics[width=\linewidth, trim=0 20 0 27, clip]{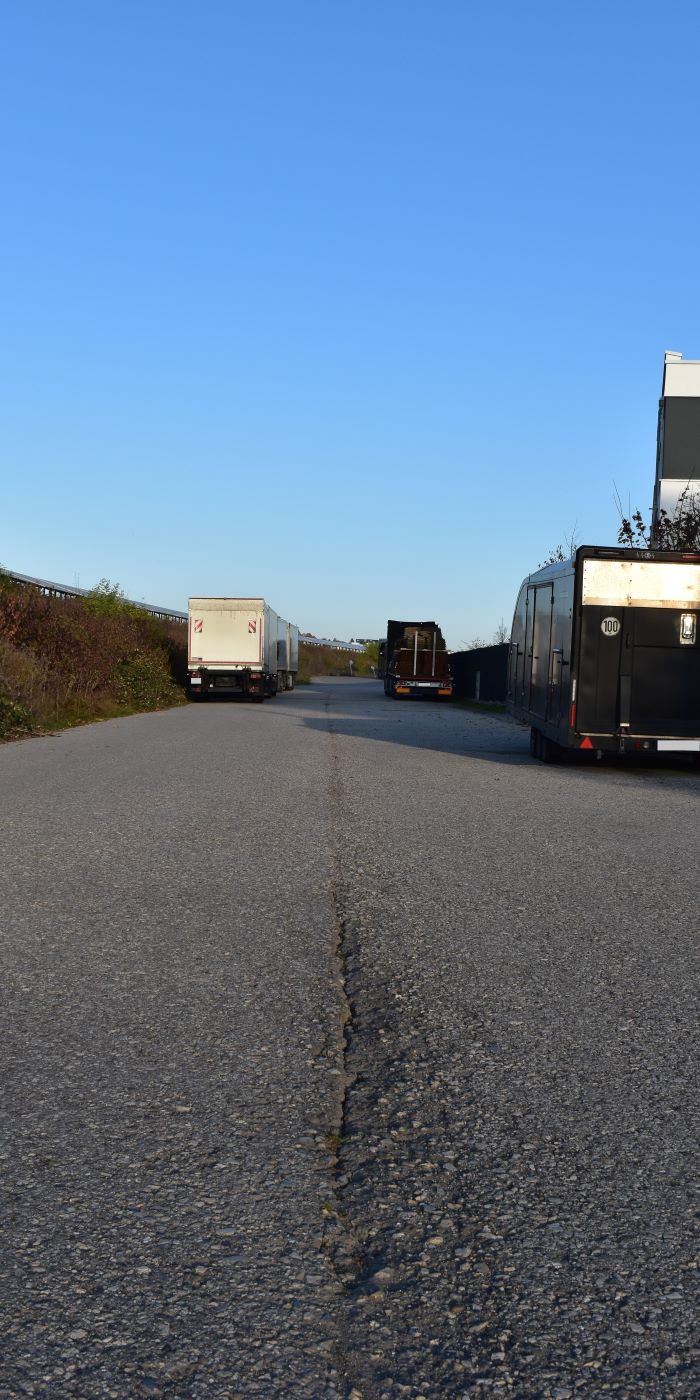}
        \caption{MixedRoad}~\label{fig:mixed_road_1}
        \Description{The picture shows a straight road with a rough surface. The road contains occasional cracks along the center.}
    \end{subfigure}
    \begin{subfigure}[c]{0.245\linewidth}
        \includegraphics[width=\linewidth, trim=0 23 0 24, clip]{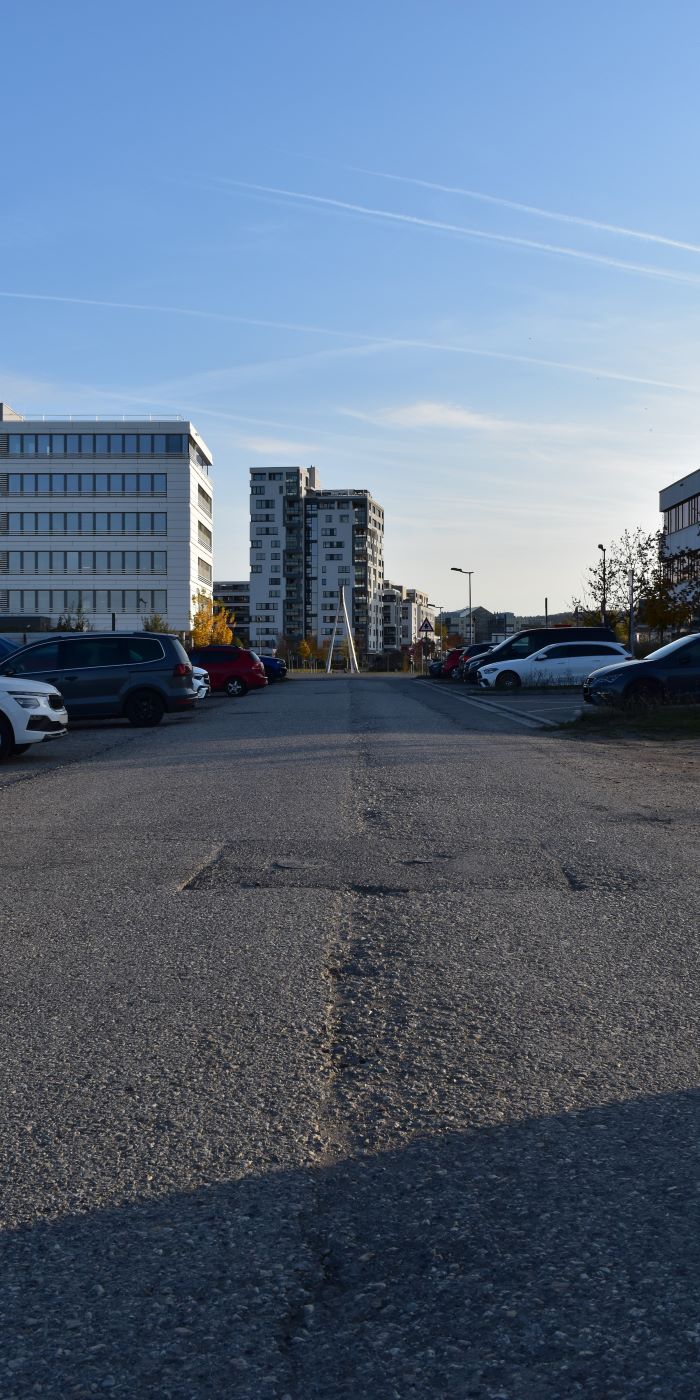}
        \caption{BumpyRoad}~\label{fig:bumpy_road}
        \Description{The picture shows a straight road with a rough surface. The road contains multiple cracks, potholes, and sections that have been patched.}
    \end{subfigure}
   \caption{Overview of road types investigated in our study, depicting variations in surface quality.}~\label{fig:road_conditions}
   \Description{Overview of road types investigated in our study, depicting variations in surface quality. This includes SmoothRoad, MixedRoad, and BumpyRoad.}
\end{figure*}

The course was set in a traffic-calmed environment. We employed a reproducible driving style across all conditions to ensure internal validity~\cite{Muhlbacher.2020}. All participants were driven by the same driver who performed preliminary training of the course to ensure a uniform driving style~\cite{Jones.2018}. We used a speed limiter to ensure a uniform maximum speed of 30km/h. The course featured an equal number of directory turns and straight parts to ensure equal driving style variation (see \autoref{fig:study-course}). In total, one round along the course featured three sharp curves and two long-curves in each direction, two straight areas in which the vehicle was brought to a standstill (braking) and then accelerated to 30km/h again. One of the long-curves was always driven with a steady speed of 30km/h, while the second one contained parallel acceleration from 0km/h to 30km/h. This acceleration behavior was also applied to short-curves along the route.
Furthermore, road conditions (see \autoref{fig:road_conditions}) differed across the route, containing a paved road with close to no bumps (\textit{SmoothRoad}), a section with a bumpy road containing multiple potholes (\textit{BumpyRoad}), and a third section which could be described as a mixture of the previous two (\textit{MixedRoad}). 

\begin{figure*}[p]
\centering
\small
    \begin{subfigure}[t]{0.8\linewidth}
        \includegraphics[width=\linewidth]{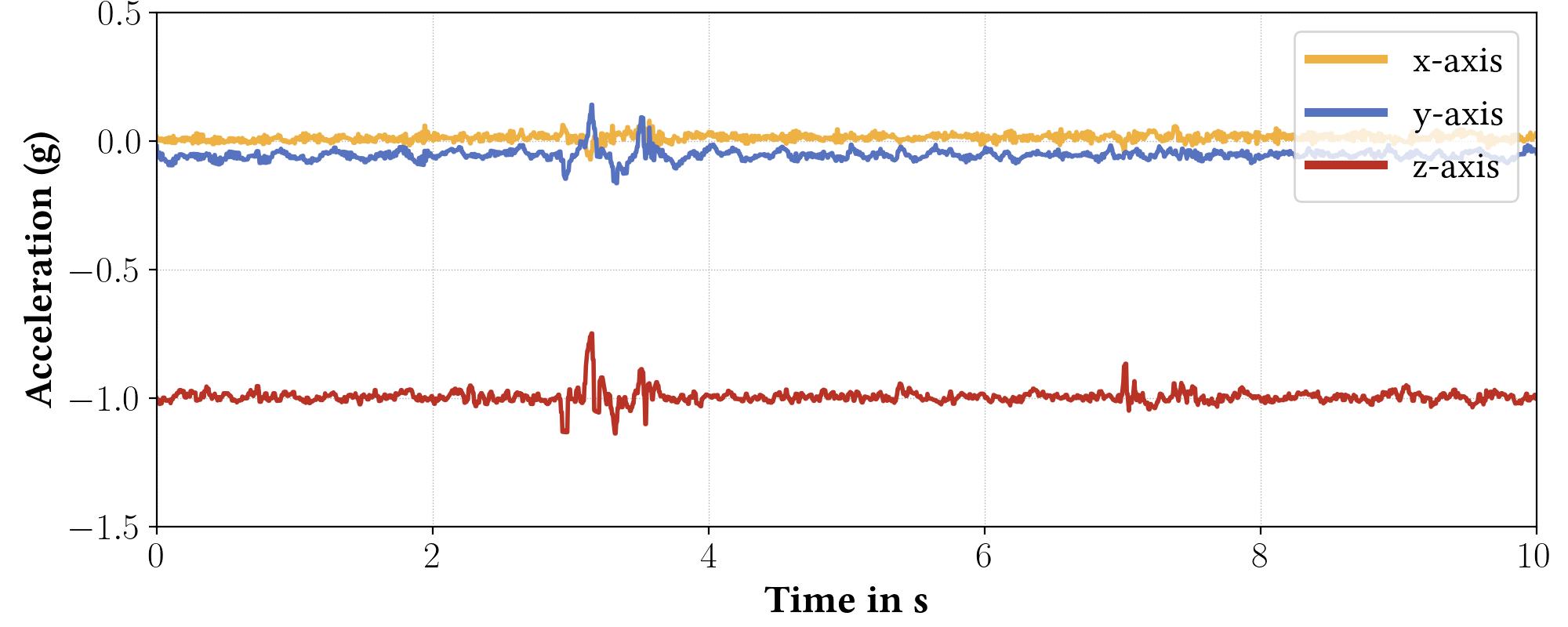}
             \caption{SmoothRoad}~\label{fig:acc_smooth_road}
        \Description{The figure depicts a time series graph over a ten-second interval, visualizing the acceleration (g) experienced by participants while driving on SmoothRoad. All three depicted axis (x,y,z) display changes in acceleration with a very low amplitude, indicating that the road contains close to no potholes or cracks.}
    \end{subfigure}
    \vspace{1em}
    \begin{subfigure}[t]{0.8\linewidth}
        \includegraphics[width=\linewidth]{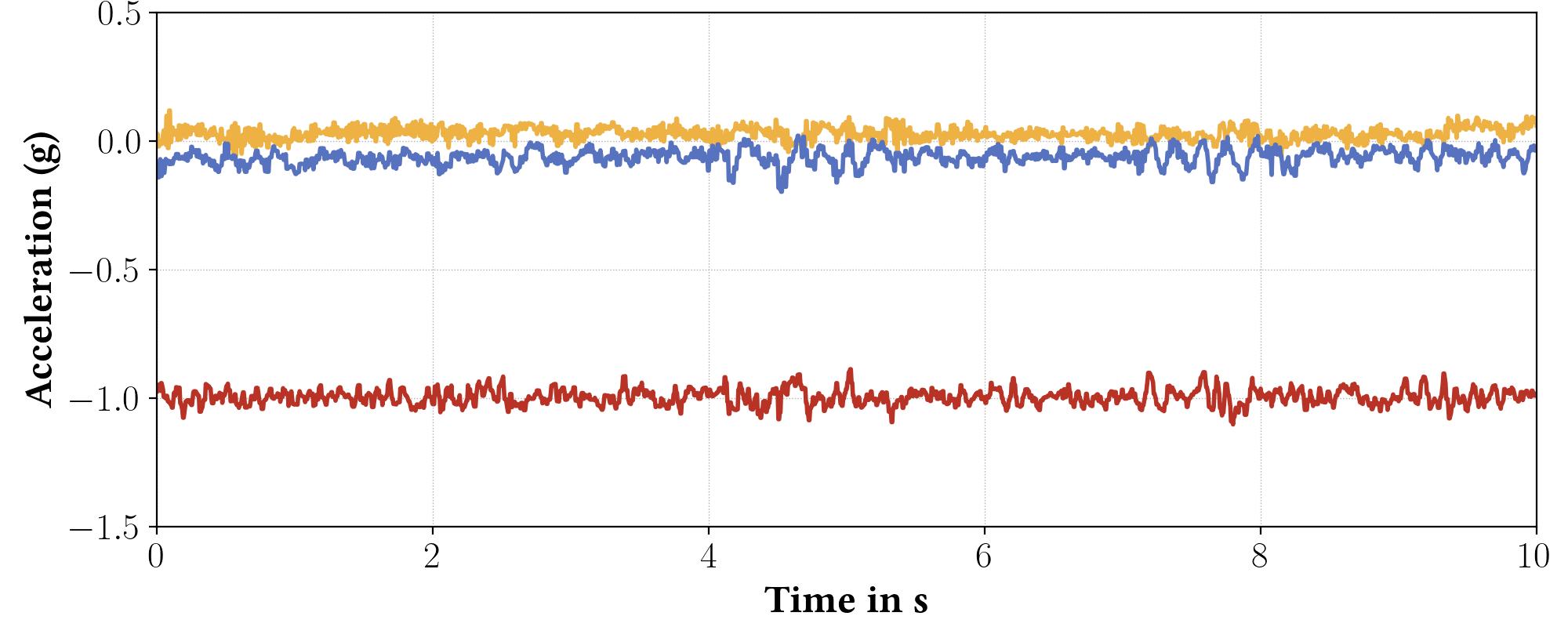}
        \caption{MixedRoad}~\label{fig:acc_mixed_road}
        \Description{The figure depicts a time series graph over a ten-second interval, visualizing the acceleration (g) experienced by participants while driving on MixedRoad. All three depicted axis (x,y,z) display changes in acceleration with a higher amplitude compared to SmoothRoad, indicating that the road contains a rougher surface with occasional cracks along the center.}
    \end{subfigure}
    \vspace{1em}
    \begin{subfigure}[t]{0.8\linewidth}
        \includegraphics[width=\linewidth]{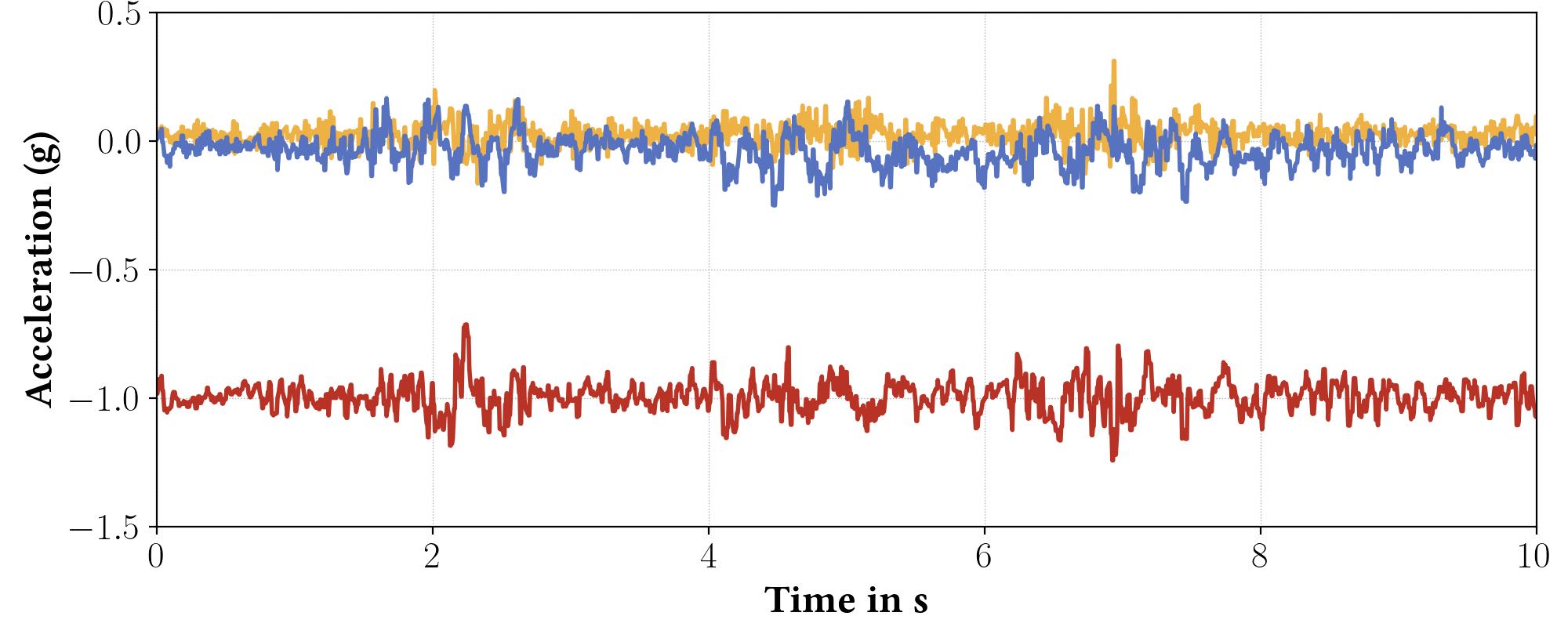}
        \caption{BumpyRoad}~\label{fig:acc_bumpy_road}
        \Description{The figure depicts a time series graph over a ten-second interval, visualizing the acceleration (g) experienced by participants while driving on Bumpy Road. All three depicted axis (x,y,z) display changes in acceleration with a higher amplitude compared to MixedRoad, indicating that the road contains an even rougher surface with multiple cracks, potholes, and sections that have been patched.}
    \end{subfigure}
   \caption{Vehicle acceleration on the three road conditions: SmoothRoad, MixedRoad, and BumpyRoad. Recordings from P08 during Gaze\&Pinch. The x-axis describes longitudinal, the y-axis lateral, and the z-axis vertical acceleration.}~\label{fig:acc_road_conditions}
   \Description{We display three sub-figures all depicting a time series graph over a ten-second interval, visualizing the acceleration (g) experienced by participants while driving on the three road conditions.}
\end{figure*}

\autoref{fig:acc_road_conditions} visualizes the motion profile (vibrations, accelerations) of the road conditions recorded using the IMU specified in \autoref{objectiveMeasures}. This IMU was mounted on the dashboard in front of the co-driver, so that the x-axis was parallel to the vehicle's forward direction, the y-axis was orthogonal to the vehicle's forward vector, and the z-axis represented vertical acceleration (e.g., road bumps). The start position was randomized per participant, as far as possible considering the road and traffic conditions. This was performed to vary the timing and sequence of the road condition occurrences, thus alleviating, for example, fatigue symptoms occurring at similar points.

\begin{figure}[H]
	\centering
  	\includegraphics[width=\linewidth]{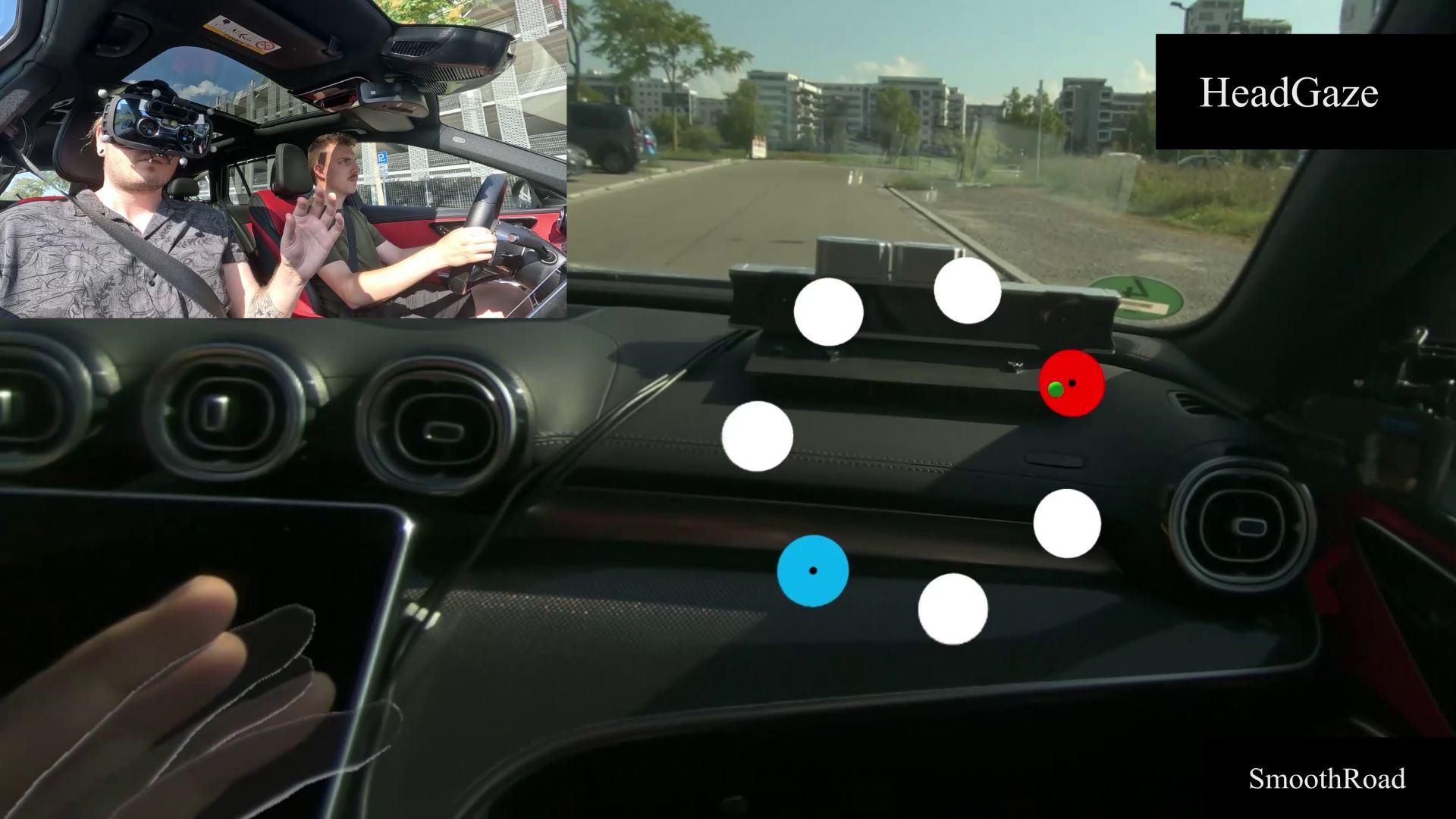}
  	\caption{Fitts' Law Task as observed through the Varjo XR-3 with passthrough enabled. The cursor of each interaction method is shown in green and highlights targets in red when hovering, indicating the ability to select the target. One of the seven targets is always highlighted in blue to indicate it should be selected next, until a successful selection is performed.}
  	\label{fig:fittsTaskVis}
  	\Description{The figure shows the Fitts' Law Task as visible for participants during the study. Seven circular targets are visible, of which one is highlighted in blue (bottom-left) and one in red (top-right). A small, round green sphere is visible in front of the red target, representing the cursor position. The targets are arranged in a circular layout.}
\end{figure}

\subsection{Procedure}
First, participants signed a privacy consent form and were informed about the study procedure. Furthermore, they were informed that they could abort the study at any point in time, for example, due to MS symptoms. They then entered demographic data, including age, gender, handedness, vision impairments, and frequency of AR/VR usage on a five-point likert scale (\textit{never} to \textit{always}). Afterwards, they were introduced to the Varjo XR-3 and the Fitts' Law Task and could practice each interaction method until they felt comfortable using it~\cite{bergstrom2021HowEvaluateObject}.
Subsequently, the first measurement of the Misery Scale (MISC) \cite{Bos.2005} was collected, followed by participants putting on the HMD and performing the One-Dot Eye-Tracking Calibration as provided by Varjo Base. 
To ensure high accuracy and precision for Gaze\&Pinch, participants were instructed to fixate the HMD comfortably but tightly to prevent the headset from moving and having a negative effect on eye-tracking. Subsequently the five-dot calibration of the Varjo XR-3 was performed for increased tracking precision when performing eye-tracking interactions. The calibration quality was then validated before the trial could start. We ensured high calibration quality for Gaze\&Pinch by checking the calibration result obtained by VarjoAPI, as well as accuracy and precision measures obtained by using GazeMetrics~\cite{b.adhanom2020GazeMetricsOpenSourceTool}. We ensured that the accuracy would not exceed half the diameter of the Fitts' Law task targets. 
Participants performed each interaction method in a moving vehicle and in a standstill environment.

Then, participants started with the Fitts' Law Tasks. Every time participants finished the tasks, they filled out the questionnaires outlined in Section~\ref{sec:subjective}, followed by a semi-structured interview. The driving session was aborted if a MISC $\geq$ 6 was reported~\cite{Dam.2021,Kuiper.2018}. This process was then repeated for each interaction method. After participants completed each interaction method and movement combination, they concluded their participation by performing a final semi-structured interview. We ensured counterbalancing by applying a balanced latin square. The study took about 2.5 hours. Participation was voluntary.

\subsection{Fitts' Law Task} \label{c:FittsTask}

We employed a Fitts' Law Task to investigate the implications of standstill vs vehicle movement on point-and-select tasks with different interaction methods. To analyze user performance, we calculated throughput based on \citet{batmaz2022EffectiveThroughputAnalysis}. Here, the following formulation of effective throughput is used, with movement time representing the task completion time:

$$ID_e = \log_2 \left( \frac{A_e}{W_e} + 1 \right)$$
$$\text{Throughput} = \left( \frac{ID_e}{\text{MovementTime}} \right)$$

$A_e$ represents the actual traveled euclidean distance in three-dimensional space between the last and current selection. $W_e$ represents the effective target width and is calculated as $W_e$ = $4.133 \cdot SDx$. Similar to \citet{batmaz2022EffectiveThroughputAnalysis}, we projected the standard deviation (SD) of the distance between the current selection point and the target center onto the task axis. Effective throughput was calculated for each repetition of seven targets.

Based on Mixed Reality design guidelines by \citet{Microsoft.2021}, we placed the targets on a two-dimensional plane centered at 40cm in front of participants' heads, with a downwards offset of 20°. This ensured that the center of the interaction plane was positioned within the range of the resting gaze angle, providing a reduced risk of eye strain~\cite{pfeuffer2024DesignPrinciplesChallenges} while also providing an ergonomic area for head movements. The targets were arranged in a circular layout with an amplitude of 20° (i.e., the distance between the targets, 14.106cm at 40cm depth), requiring small head movements for selections performed with Gaze\&Pinch~\cite{sidenmark2019EyeHeadSynergetic}. The interaction plane on which the targets were positioned always faced the participant. Based on an internal evaluation (see \autoref{c:interactionMethods}) of the eye-tracker, we set the target size for the subsequent Fitts' Law Task to an angular size of 4.25° (2.9684cm width at 40cm depth), ensuring that tracking inaccuracy, which could occur to differing degrees based on the calibration quality and target position was accounted for. A green cursor of 1° size (0.69815cm at 40cm depth) was visible across all conditions except for Gaze\&Pinch. We used these values across all input modalities to compare task-related metrics. Targets were displayed as white by default, with them being highlighted in red on hover. The next target to be selected was highlighted in blue, containing a smaller black circle in the middle to aid precision during Gaze\&Pinch trials \cite{wagnerFittsLawStudy2023} (see \autoref{fig:fittsTaskVis}).

The first target at the start of the block was highlighted in purple. All seven targets remained visible to the participants throughout the study. To provide additional feedback, correct and incorrect selections were accompanied by an individual acoustic signal. Selections that occurred in the area outside of targets also behaved this way. 
Participants were instructed to perform selections as fast and as precisely as possible. The task contained 44 repetitions, each involving the selection of seven targets, leading to 308 correct selection to be performed in total. For the Fitts' Law implementation, we used the open source implementation \textit{3DFitts} by jlcouto\footnote{jlcouto: 3DFitts. \url{https://github.com/jlcouto/3DFitts}, commit 7eef967; accessed 20.06.2024)}.
We instructed participants not to rest their hands while using interaction methods which required pre-selection by hand movement. This was done due to the challenge of adjusting hand rests to accommodate for differences in body height and arm length in the vehicle.
Because of the Gorilla-Arm effect~\cite{hincapie-ramos2014ConsumedEnduranceMetric}, we introduced breaks across all interaction methods where participants could rest. Breaks took place after every eight repetitions and lasted for eight seconds. To ensure a correct analysis of the Fitts' Law Task, the first repetition after each break was excluded from the statistical analysis, with 266 correct selections remaining in the dataset. This ensured comparable starting positions for the cursor in context of the first target selection in each repetition.

\subsection{Subjective Measures}\label{sec:subjective}

\subsubsection{Trust and Perceived Safety}
We assessed the implications of the interaction methods on the participants' perceived safety and trust towards the AV. For trust, we employed the two sub-scales "Understanding/Predictability" and "Trust in Automation" of the Trust in Automation (TiA) questionnaire by \citet{korber2019TheoreticalConsiderationsDevelopment}. The subscale "Understanding/Predictability" assesses participants' ability to understand the reason behind performed maneuvers.
Additionally, we measured subjective perceived safety using semantic differentials (-3 to +3) by \citet{faas2020LongitudinalVideoStudy}. 

\subsubsection{Usability and Workload}
We used the System Usability Scale (SUS)~\cite{brookeSUSQuickDirty1995} to assess the subjective usability metrics. Furthermore, the NASA-TLX~\cite{Hart.2006} was employed to assess the subjective workload exhibited by each method. Related scores were calculated based on the raw-TLX~\cite{Hart.2006} (NASA-rTLX). For the total score, sub-scales were summed and divided by their count.

\subsubsection{Motion Sickness}
Due to the varying susceptibility of participants to MS symptoms, we continuously assessed MS during the study by administering the MISC questionnaire~\cite{Bos.2005}. We terminated the study session if a value $\geq$ 6 was reported~\cite{Dam.2021,Kuiper.2018}.

\subsubsection{Post-Condition and Post-Study Questionnaire}
Post-Condition and Post-Study semi-structured interviews were performed regarding user preferences, comparisons across conditions, challenges of use, possible improvements, and future usage of HMDs concerning potential benefits of their usage in vehicles (see \autoref{c:appendixInterview}). We evaluated the gathered data by performing a thematic analysis.
Quotes obtained in a language other than English were translated using \href{https://www.deepl.com/de/translator}{DeepL}.

\subsubsection{Fitts' Law Metrics}
For the entire Fitts' Law Task duration, we logged cursor movement and the position and rotation of participants' heads, hand palms, and index tips of the chosen handedness. Task-related metrics we assessed include correct and incorrect selections, the distance of the selection endpoint to the target center, movement time, throughput, effective width, and effective amplitude.

\begin{table*}[h!]
\centering
\caption{Classification of Road Types, Curve Types, and Maneuvers}
\label{tab:typesRoadCurveManeuver}
\begin{tabular}{l|l|l}
\toprule
\textbf{Road Type} & \textbf{Curve Type} & \textbf{Maneuver} \\ \midrule
SmoothRoad & Short [Left/Right] Curve with Acceleration & Braking \\ 
MixedRoad & Long [Left/Right] Curve with Acceleration & Accelerating \\
BumpyRoad & Long [Left/Right] Curve with Steady Acceleration & \\ 
\bottomrule
\end{tabular}
\end{table*}

\subsection{Objective Measures} \label{objectiveMeasures}
To analyze Fitts' Law-related data regarding the effects of vehicle motions, we collected vehicle acceleration, vibrations, and angular motion during driving sessions using a car-mounted LPMS-IG1P IMU, with a sampling rate of 250Hz and the x-axis values representing longitudinal accelerations\footnote{\url{https://www.lp-research.com/9-axis-imu-with-gps-receiver-series/}; Accessed 26.08.2024}.
Additionally, vehicle speed in km/h, movement of the acceleration and brake pedals, and the exact position along the course were recorded, with first preliminary labels being applied in real-time according to the current road conditions (e.g., BumpyRoad, SmoothRoad, MixedRoad) and curve categories (e.g., Short-Left Curve, Long-Left Curve with Steady Speed).

To assess the impact of vehicular motion on the body parts used for or related to performing interactions, we recorded the movement of the hands in space by utilizing the Ultraleap hand tracking, as well as the position and rotation of the participant's head. Furthermore, Eye-Tracking features (e.g., Focus Point, Pupillary Index) were recorded with a frequency of 100Hz using the standard Varjo filter during the whole study. 
For Gaze\&Pinch this includes a post-calibration validation step based on GazeMetrics, measuring accuracy and precision later used to assess the related interactions. Reporting of these metrics is based on the RMS and accuracy formulas presented by \citet{b.adhanom2020GazeMetricsOpenSourceTool}. 

While \citet{b.adhanom2020GazeMetricsOpenSourceTool} states that a target arrangement similar to the system's native calibration procedure tends to result in better accuracy, we decided to use a circular layout consisting of nine targets, while using a radius of 20° at a depth of 0.4m to resemble the target visualization of the utilized Fitts' Law Task. Furthermore, based on~\cite{b.adhanom2020GazeMetricsOpenSourceTool}, samples collected during the first 800ms are excluded with targets being visible for two seconds each.

\subsection{Data Preparation} \label{c:dataPrep}

The raw data obtained within this study contained the parameters specified in \autoref{objectiveMeasures} for \N{24} participants, consisting of four interaction methods each performed during vehicle movement and standstill. 
As each data stream was recorded into a separate file due to differing sampling rates, raw data was first resampled to 200Hz, interpolated, and then synchronized based on UnixTime in milliseconds. This resulted in one file for each combination of factors.

Due to technical issues with the hand-tracking sensor, and based on the reports of participants, we recognized that unintentional selections (pinch gestures) were picked up by the hardware. By running an internal test to reproduce this issue, we found that selections with a duration of less than or equal to 70ms and a time interval from each other under 10ms to be erroneous and filtered them accordingly. Additionally, the technical issue also resulted in two selections being logged simultaneously. In such cases, the first recorded selection attempt was kept with the subsequent ones being removed. An exception to this filtering approach were correct selections - as they always led to a continuation of the Fitts' Law Task. This led to an average of 23.22 selections being filtered across factor combinations. A detailed analysis can be obtained in \autoref{c:filteredSelectionCount}.

Afterwards, the data was automatically labeled. Vehicle movement was subdivided into the following labels, which extend over three categories taking place in parallel: Road Type, Curve Type, and Maneuver. See \autoref{tab:typesRoadCurveManeuver} for an overview regarding the assigned sub-labels.
Road Types were assigned using the vehicle position along the pre-defined course, mapping each segment accordingly. Curves were categorized based on a feature combination of the vehicle speed in km/h, the z-axis values of the employed gyroscope, and the position along the course. The label Breaking was assigned based on values obtained from the vehicle's breaking pedal, while accelerations were categorized by using the vehicle speed.

\section{User Study: Results}

For the statistical analysis, we used RStudio (2024.09.1) and R (Version 4.4.2). All packages used were up-to-date as of December 2024. Analysis was performed by utilizing \textit{rCode} by \citet{colley2024rcode}. Data was checked for normal distribution and homogeneity of variance for every statistical test. For non-normally distributed data, and if not stated differently, Aligned Rank Transform (ART) using the ARTool package~\cite{wobbrock2011AlignedRankTransform}
was applied. Post-hoc analysis was performed using Dunn's test with Holm correction~\cite{maxwell2017DesigningExperimentsAnalyzing}.

\subsection{Participants} 
24 participants aged between 23 and 60 years (\m{33.0}, \sd{8.37}, \md{31}) participated in the study (3 female, 21 male, 0 non-binary). 5 participants were left- and 19 participants were right-handed. All participants except one were employees of an automotive company.
Participants responses regarding the usage frequency of AR and VR devices ranged from \textit{always (\N{2})}, \textit{often (\N{6})}, \textit{sometimes (\N{4})}, \textit{rarely (\N{8})} to \textit{never (\N{4})}. None of the participants had to abort the study due to motion sickness.

Out of 24 participants, 20.8\% (\N{5}) stated that they were nearsighted, while 25\% (\N{6}) were nearsighted while also having astigmatism or were nearsighted and partially sighted on their left eye (4.2\%, \N{1}). One participant stated to be farsighted (4.2\%, \N{1}), and one was farsighted while having astigmatism (4.2\%, \N{1}). 41.7\% (\N{10}) had no vision problems. 
54.2\% (\N{13}) of participants reported regularly using prescription glasses, while 4.2\% (\N{1}) reported using both, prescription glasses and contact lenses. 41.1\% (\N{10}) do not use either. None of the participants stated to have a glaucoma or cataract.

\subsection{Eye-Tracking Validation} 
Before using Gaze\&Pinch, the accuracy and precision of the performed Eye-Tracking Calibration were validated using GazeMetrics~\cite{b.adhanom2020GazeMetricsOpenSourceTool}.
The data was sampled at 100Hz, with Varjo Base filtering set to standard. Nine targets were displayed for 2s each, and samples within the initial 800ms were excluded. Targets were arranged in a circular layout, resembling the Fitts' Law Task. They were positioned at 40cm depth, featured an amplitude of 20° (14.106cm), and had an angular size of 5.0102° (3.5cm width).
We obtained an average accuracy of 1.32° (\sd{0.90}, \md{1.10}) and an RMS precision of 0.0848° (\sd{0.180}, \md{0.0339}). For participant 26 (P26), One-Dot Calibration was performed in Gaze\&Pinch Movement, as it yielded better calibration quality according to Varjo Base. Participant 12 did not use vision corrections while performing Gaze\&Pinch during Standstill as opposed to the movement condition. For this participant, data is missing due to technical logging issues. Furthermore, we can differentiate between participants using prescription glasses. \autoref{tab:gazemetrics_visionaids} contains the measurements per vision correction group.

\subsection{Perceived Workload} \label{c:perceivedWorkloadArt} 
\begin{figure}[hb!]
    \centering
    \includegraphics[width=0.5\textwidth]{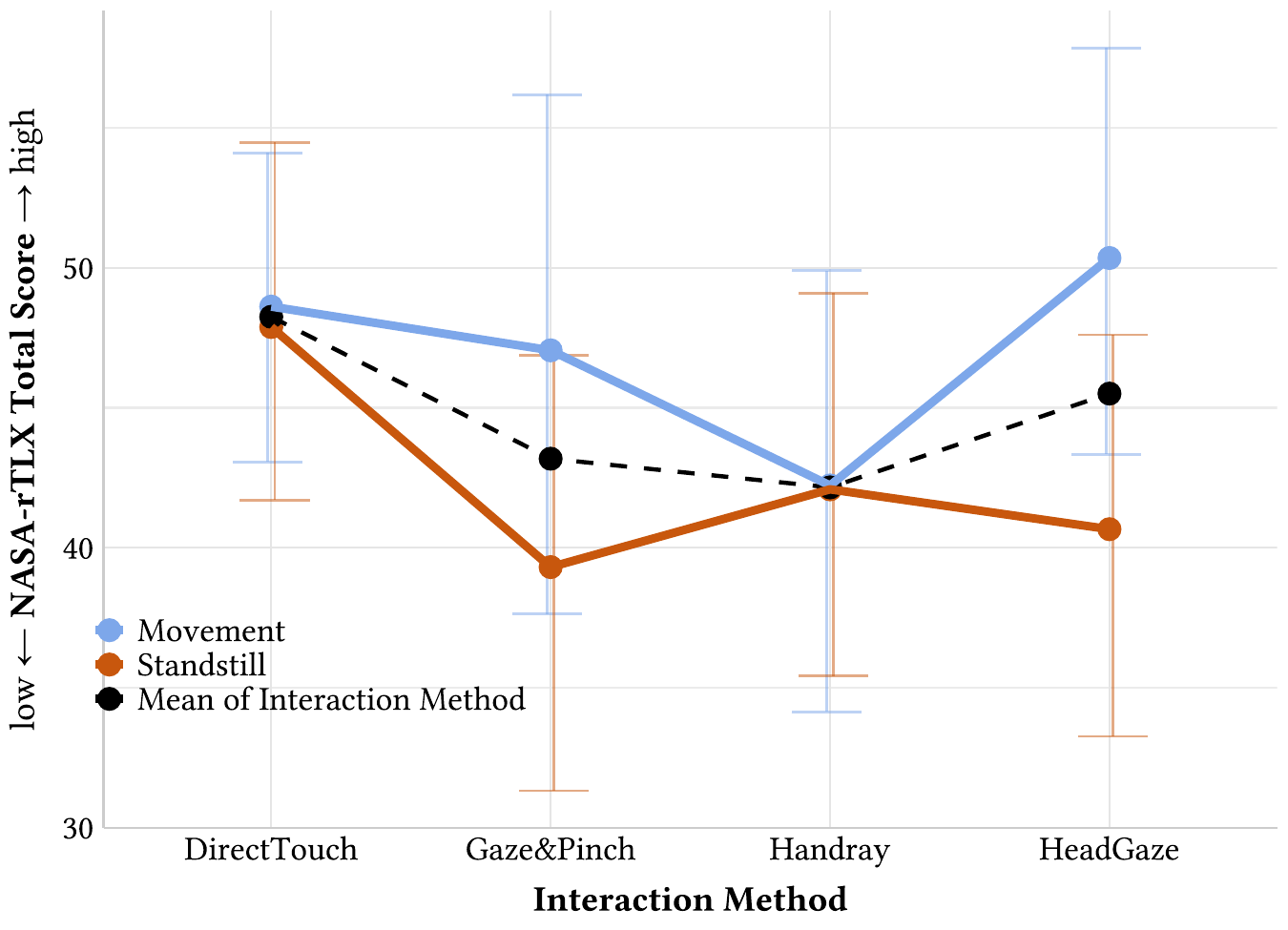}
    \caption{Interaction effect on NASA-rTLX: Total Score}
    \label{fig:ie_nasa_totalScore}
    \Description{The figure depicts a line graph visualizing the interaction effect on NASA-rTLX Total Score. The Y-Axis represents the Total Score, the X-Axis contains the interaction methods. Across all interaction methods, values are higher for movement than for standstill. The highest workload during movement is exhibited for HeadGaze, while Handray features the lowest value. During standstill, DirectTouch shows the highest workload, while Gaze\&Pinch features the lowest value.}
\end{figure}

%NASA-rtlx results
In this section, we present results obtained from the NASA-rTLX questionnaire.
The statistical analysis via ART found no significant effects on temporal demand or effort. However, the ART found a significant main effect of \movementCondition on the NASA-rTLX Total Score (\F{1}{23}{9.52}, \p{0.005}). Total Score was significantly higher during movement (\m{47.06}, \sd{18.82}) than in standstill (\m{42.48}, \sd{18.01}). The ART found a significant interaction effect of \condition $\times$ \movementCondition on NASA-rTLX Total Score (\F{3}{69}{3.04}, \p{0.035}; see \autoref{fig:ie_nasa_totalScore}). 
The Total Score of DirectTouch and Handray remain largely unaffected by movement, resulting in similar scores compared to standstill. For Gaze\&Pinch and HeadGaze, the Total Score is higher during Movement than during standstill. The highest score is reached for HeadGaze during movement (see \autoref{tab:nasartlxTotalScore}).
The ART found a significant main effect of \movementCondition on mental demand (\F{1}{23}{4.77}, \p{0.039}). Mental demand was significantly higher during movement (\m{41.72}, \sd{27.80}) than during standstill (\m{36.88}, \sd{26.45}).

The ART found a significant main effect of \condition on physical demand (\F{3}{69}{9.10}, \pminor{0.001}). \autoref{tab:posthoc-condition-physical} contains the results of the post-hoc test. \autoref{tab:MeanSdPhysicalDemand} displays descriptive statistics.

\begin{table}[ht]
    \centering
    \caption{NASA-rTLX Total Score (Mean and Standard Deviation for different interaction methods)}
    \label{tab:nasartlxTotalScore}
    \begin{tabular}{lcc}
        \toprule
        Interaction Method & Mean & Standard Deviation \\
        \midrule
        DirectTouch Movement          & 48.6             & 13.8             \\
        DirectTouch Standstill        & 47.9             & 16.6             \\ 
        \midrule
        Gaze\&Pinch Movement            & 47.0             & 22.6             \\
        Gaze\&Pinch Standstill          & 39.3             & 19.9             \\ 
        \midrule
        Handray Movement              & 42.2             & 20.4             \\
        Handray Standstill            & 42.1             & 17.6             \\ 
        \midrule
        HeadGaze Movement             & 50.3             & 17.6             \\
        HeadGaze Standstill           & 40.7             & 17.7             \\ 
        \bottomrule
    \end{tabular}
\end{table}

The ART found a significant main effect of \movementCondition on performance (\F{1}{23}{31.13}, \pminor{0.001}). Performance was significantly lower with movement (\m{43.39}, \sd{19.58}) than during standstill (\m{32.66}, \sd{18.11}). As a high rating on this scale represents low self-perceived performance, participants reported performing better during standstill than during movement.

The ART found a significant main effect of \movementCondition on frustration (\F{1}{23}{5.84}, \p{0.024}). Frustration was significantly higher with movement (\m{43.65}, \sd{27.17}) than during standstill (\m{38.54}, \sd{25.76}).

\subsection{System Usability} \label{c:usabilityArt}
%SUS

\begin{figure}[hb]
    \centering
    \includegraphics[width=0.5\textwidth]{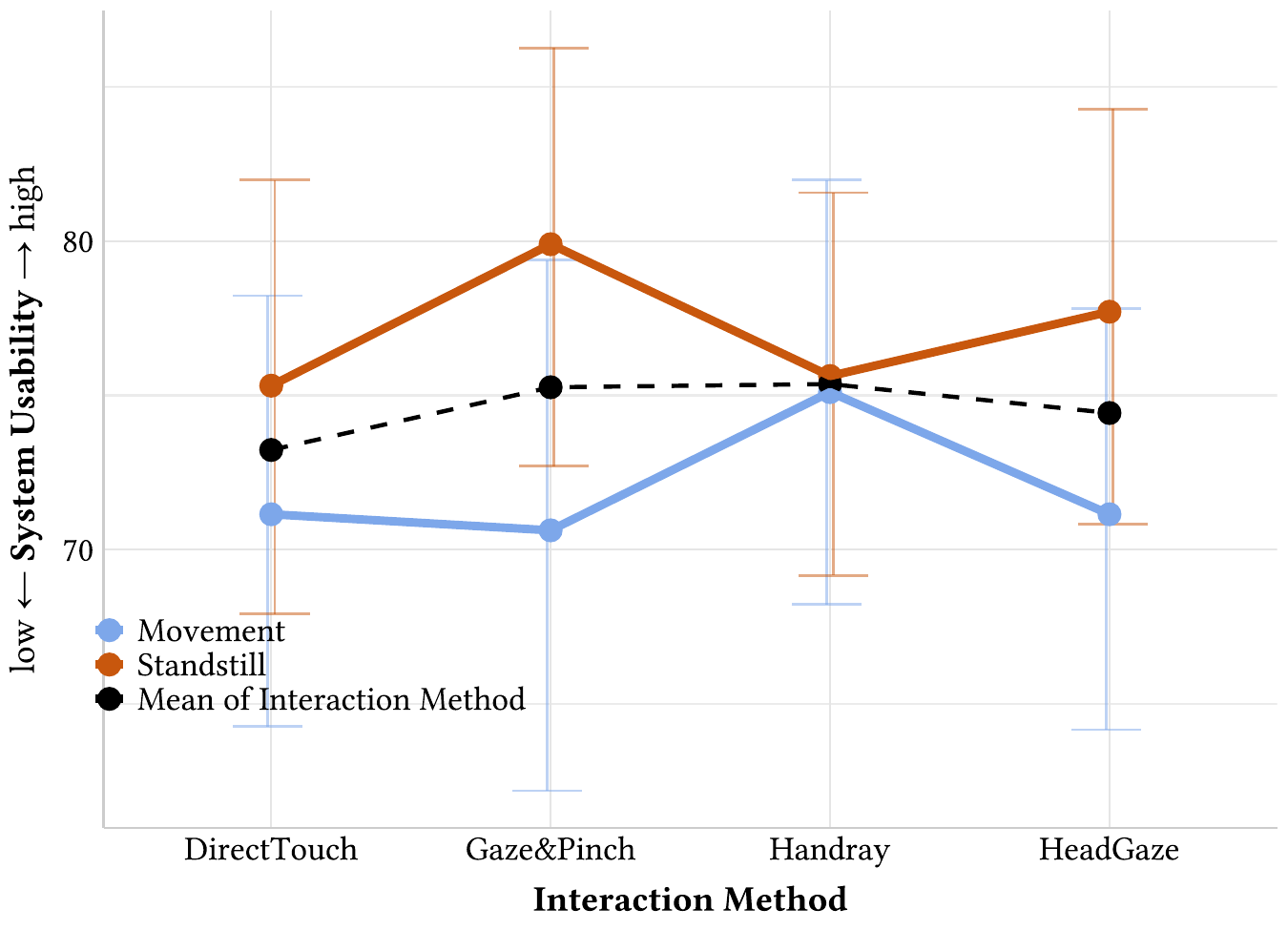}
    \caption{Results of the System Usability Scale}
    \label{fig:sus_plot}
    \Description{The figure depicts a line graph visualizing the values obtained from the System Usability Scale (SUS). The Y-Axis represents the SUS Total Score, the X-Axis contains the interaction methods. Across all interaction methods, values are higher for standstill than for movement. The highest usability score during movement is exhibited for Handray, while Gaze\&Pinch features the lowest value. During standstill, Gaze\&Pinch shows the highest usability score, while DirectTouch features the lowest value.}
\end{figure}

The ART found a significant main effect of \movementCondition on SUS Score (\F{1}{23}{14.18}, \p{0.001}). Usability was significantly lower with movement (\m{72.01}, \sd{18.19}) than during standstill (\m{77.14}, \sd{16.83}), see \autoref{fig:sus_plot}.

\begin{table}[ht]
\centering
\caption{Post-hoc comparisons for independent variable \condition and dependent variable \physicalDemand. Positive Z-values mean that the first-named level is significantly higher than the second-named. For negative Z-values, the opposite is true.}
\label{tab:posthoc-condition-physical}
\begin{tabular}{lrr}
    \toprule
    Comparison & Z & p-adjusted \\ 
    \midrule
    DirectTouch - Gaze\&Pinch & 5.3139 & $<$0.001 \\ 
    DirectTouch - Handray & 2.3585 & 0.0275 \\ 
    DirectTouch - HeadGaze & 2.3244 & 0.0201 \\ 
    Gaze\&Pinch - Handray & -2.9555 & 0.0062 \\ 
    Gaze\&Pinch - HeadGaze & -2.9895 & 0.0070 \\ 
    \bottomrule
\end{tabular}
\end{table}

\begin{table}[ht]
\centering
\caption{NASA-rTLX: Physical Demand (Mean and Standard Deviation for different interaction methods)}
\label{tab:MeanSdPhysicalDemand} 
\begin{tabular}{lcc}
    \toprule
    Interaction Method & Mean & Standard Deviation \\
    \midrule
    DirectTouch & 68.54 & 25.85 \\
    Gaze\&Pinch & 36.67 & 30.97 \\
    Handray & 55.31 & 26.24 \\
    HeadGaze & 55.52 & 25.08 \\
    \bottomrule
\end{tabular}
\end{table}

\subsection{Filtered Selection Count} \label{c:filteredSelectionCount}
To extend the filtering approach described in \autoref{c:dataPrep}, we analyze the number of filtered selections to assess the influence of vehicle motion and compare the susceptibility across interaction methods.

\begin{figure}[b]
    \centering
    \includegraphics[width=0.5\textwidth]{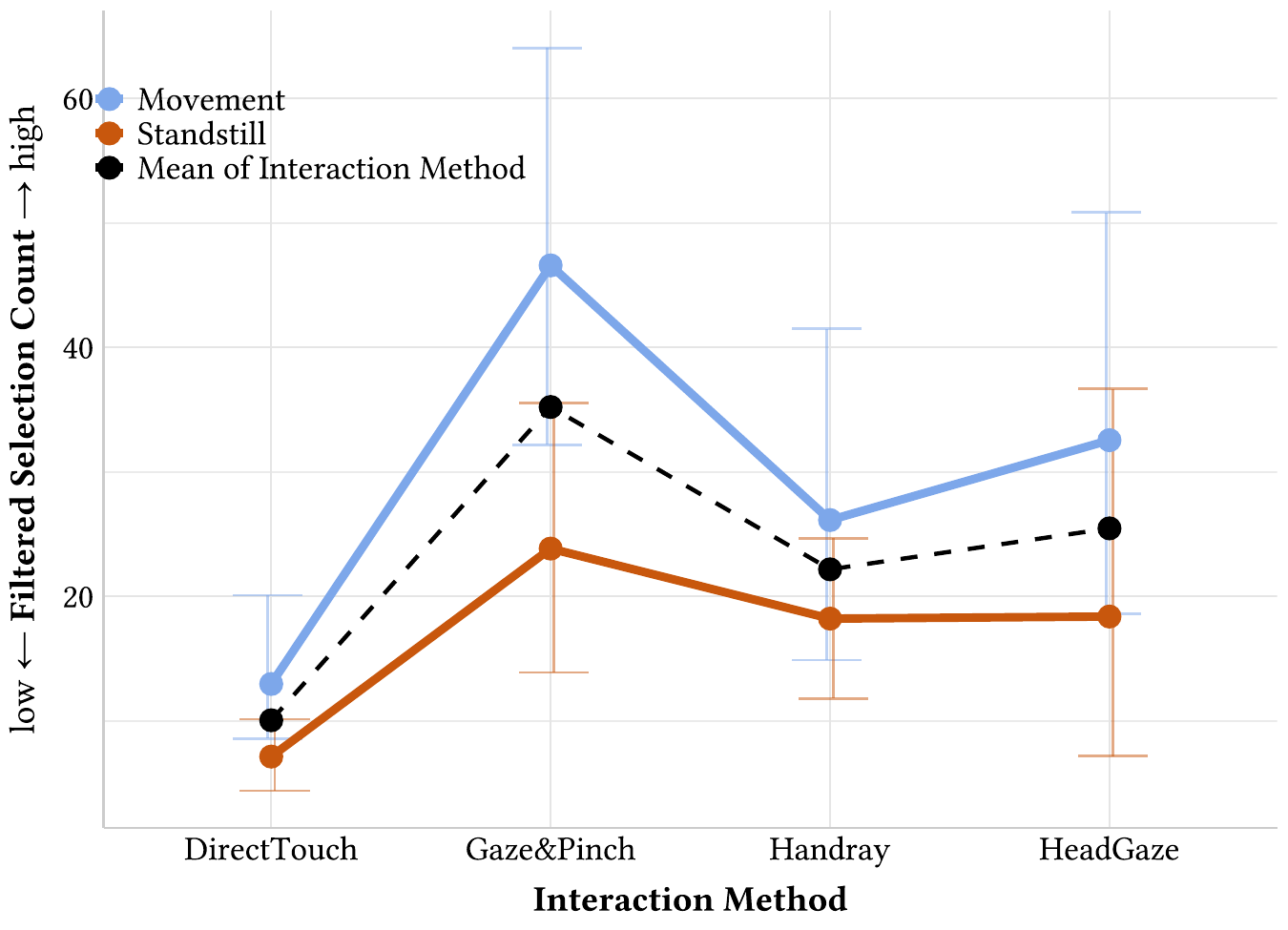}
    \caption{Significant Interaction effect on Filtered Selection Count}
    \label{fig:selection_count_ie}
    \Description{The figure depicts a line graph visualizing the interaction effect on Filtered Selection Count. The Y-Axis represents the count of selections, the X-Axis contains the interaction methods. Values for movement are higher for all interaction methods. The largest count during movement is visible for Gaze\&Pinch, the lowest for DirectTouch. The highest count during standstill is visible for Gaze\&Pinch, the lowest for DirectTouch.}
\end{figure}

The ART found a significant main effect of \condition (\F{3}{69}{11.40}, \pminor{0.001}), and of \movementCondition (\F{1}{23}{64.85}, \pminor{0.001}) on filtered selections. The amount of filtered selections was significantly
higher with movement (\m{29.55}, \sd{36.99}) than without (\m{16.89}, \sd{27.15}). Post-hoc analysis using Dunn's test revealed significant differences (see \autoref{tab:filtered_selection_count_dunn}).
The ART found a significant interaction effect of \condition $\times$ \movementCondition on filtered selections (\F{3}{69}{8.89}, \pminor{0.001}; see \autoref{fig:selection_count_ie}). The largest amount of filtered selections during movement is visible for Gaze\&Pinch, the lowest for DirectTouch. During standstill, the highest count is visible for Gaze\&Pinch, the lowest for DirectTouch.

\subsection{Fitts' Law} 
\label{c:FittsLawArt}

For the analysis of metrics related to the Fitts' Law Task, the data for throughput, movement time, and selection offset was filtered for outliers using Tukey's Inter Quartile Range (IQR). This defined outliers as data points which were either 1.5 times below the first or above the third quartile\footnote{\url{https://easystats.github.io/performance/reference/check_outliers.html}; Accessed 04.12.2024}. We exclude error rate from this approach, as it is derived from already pre-filtered selection data as explained in \autoref{c:dataPrep}. Furthermore, as explained in \autoref{c:FittsTask}, obtained data of the first repetition after a break were excluded from the statistical analysis.

\begin{table}[ht]
\centering
\caption{Post-hoc comparisons for independent variable \condition and dependent variable filtered selection count. Positive Z-values mean that the first-named level is significantly higher than the second-named. For negative Z-values, the opposite is true.} 
\label{tab:filtered_selection_count_dunn}
\begin{tabular}{lrr}
  \toprule
Comparison & Z & p-adjusted \\ 
  \midrule
DirectTouch - Gaze\&Pinch & -5.0636 & 0.0000 \\ 
  DirectTouch - Handray & -2.7748 & 0.0138 \\ 
  DirectTouch - HeadGaze & -2.5038 & 0.0184 \\ 
  Gaze\&Pinch - Handray & 2.2888 & 0.0221 \\ 
  Gaze\&Pinch - HeadGaze & 2.5598 & 0.0209 \\ 
   \bottomrule
\end{tabular}
\end{table}

\begin{table}[ht]
    \centering
    \caption{Fitts' Law: Throughput (Mean and Standard Deviation for different interaction methods)}
    \label{tab:fittsLawThroughputMean}
    \begin{tabular}{lcc}
        \toprule
        Interaction Method &Mean & Standard Deviation \\
        \midrule
        DirectTouch & 2.19 & 0.40 \\
        Gaze\&Pinch & 2.01 & 0.66 \\
        Handray & 2.02 & 0.52 \\
        HeadGaze & 1.93 & 0.47 \\
        \bottomrule
    \end{tabular}
\end{table}

\subsubsection{Throughput} 
\begin{figure}
    \centering
    \includegraphics[width=0.5\textwidth]{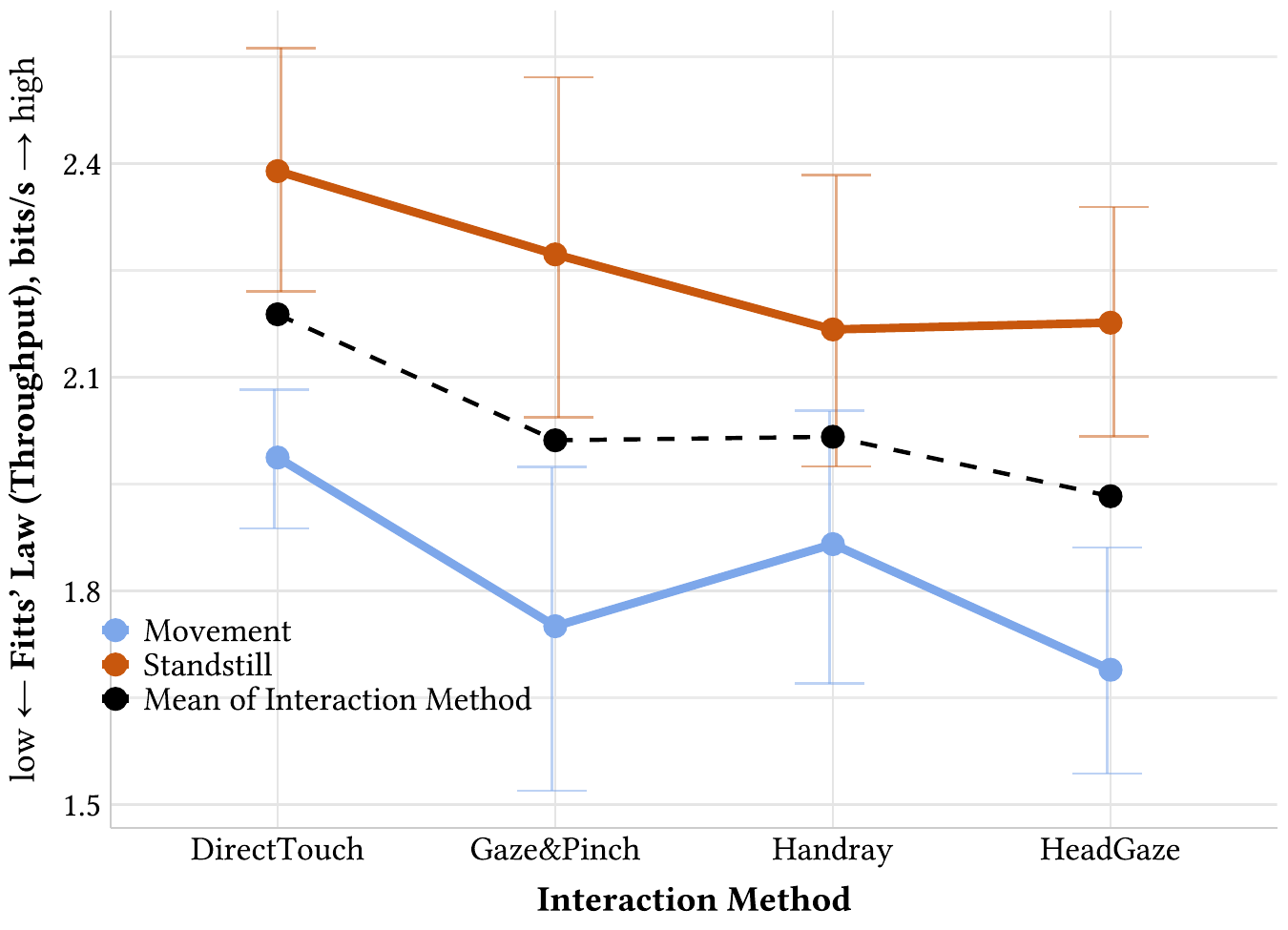}
    \caption{Significant Main effects on Fitts' Law: Throughput}
    \label{fig:fitts_law_throughput}
    \Description{The figure depicts a line graph visualizing the main effects on Fitts' Law throughput. The Y-Axis represents the throughput in bits-per-second, the X-Axis contains the interaction methods. Across all interaction methods, values are higher for standstill than for movement. The highest throughput during movement is exhibited for DirectTouch while HeadGaze features the lowest value. During standstill, DirectTouch shows the highest throughput, while Handray features the lowest value.}
\end{figure}

The ART found a significant main effect of \condition on Fitts' Law throughput (\F{3}{69}{4.23}, \p{0.008}). A post-hoc test found that throughput was significantly higher for DirectTouch (\m{2.19}, \sd{0.40}) than for HeadGaze (\m{1.93}, \sd{0.47}, $p_{\text{adj}}$ = 0.0256, see \autoref{tab:fittsLawThroughputMean}).
The ART found a significant main effect of \movementCondition on Fitts' Law throughput (\F{1}{23}{122.10}, \pminor{0.001}). Throughput was significantly lower with movement (\m{1.82}, \sd{0.45}) than during standstill (\m{2.25}, \sd{0.51}; see \autoref{fig:fitts_law_throughput}).

\subsubsection{Error Rate} \label{c:errorRate}
\begin{figure}
    \centering
    \includegraphics[width=0.5\textwidth]{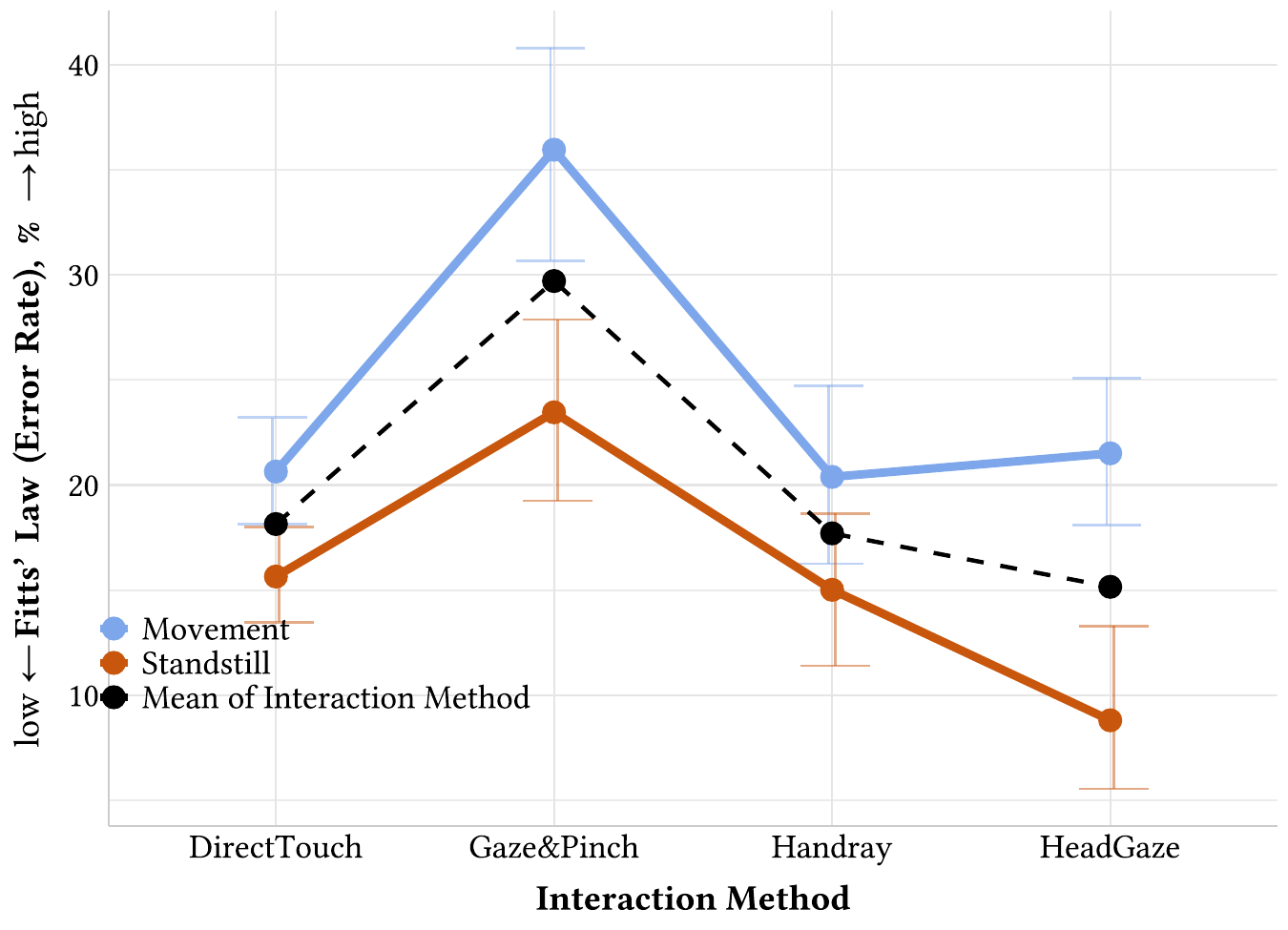}
    \caption{Significant Interaction effect on Fitts' Law: Error Rate}
    \label{fig:fitts_law_error}
    \Description{The figure depicts a line graph visualizing the interaction effect on Fitts' Law: Error Rate. The Y-Axis represents the error rate, the X-Axis contains the interaction methods. Values for movement are higher for all interaction methods. The highest error rate during movement is visible for Gaze\&Pinch, the lowest for Handray. The highest error rate during standstill is visible for Gaze\&Pinch, the lowest for HeadGaze.}
\end{figure}

We calculated the error rate by dividing the number of incorrect selections by the total number of selections made and multiplied this score by 100 to obtain the percentage score. 
The ART found a significant main effect of \condition (\F{3}{69}{17.21}, \pminor{0.001}) and of \movementCondition (\F{1}{23}{111.46}, \pminor{0.001}) on Fitts' Law error rate. The ART found a significant interaction effect of \condition $\times$ \movementCondition on Fitts' Law error rate (\F{3}{69}{15.72}, \pminor{0.001}).
The error rate is higher during movement for every interaction method (see \autoref{fig:fitts_law_error}). Compared to other interaction methods, Gaze\&Pinch features the highest error rate in both movement conditions. HeadGaze has the lowest error rate across interaction methods in Standstill, while Handray has the lowest value in movement (see \autoref{tab:errorRateMethods}).

\begin{table}[ht]
    \centering
    \caption{Fitts' Law: Error Rate (Mean and Standard Deviation for different interaction methods)}
    \label{tab:errorRateMethods}
    \begin{tabular}{lcc}
        \toprule
        Interaction Method & Mean & Standard Deviation \\
        \midrule
        DirectTouch Movement & 20.6 & 6.45 \\
        DirectTouch Standstill & 15.6 & 5.91 \\
        \midrule
        Gaze\&Pinch Movement & 36.0 & 13.4 \\
        Gaze\&Pinch Standstill & 23.5 & 10.6 \\
        \midrule
        Handray Movement & 20.4 & 11.0 \\
        Handray Standstill & 15.0 & 9.23 \\
        \midrule
        HeadGaze Movement & 21.5 & 9.07 \\
        HeadGaze Standstill & 8.80 & 9.90 \\
        \bottomrule
    \end{tabular}
\end{table}

\subsubsection{Movement Time} \label{c:FittsLawArtMT}
For analysis, we calculated the mean movement time per participant for each combination of factors. Movement time refers to the time required to perform one repetition consisting of selecting seven targets.

The ART found a significant main effect of \condition on Fitts' mean movement time (s) (\F{3}{69}{3.43}, \p{0.022}), with post-hoc analysis using Dunn's test not revealing significant differences. The ART found a significant main effect of \movementCondition on Fitts' mean movement time (s) (\F{1}{23}{79.64}, \pminor{0.001}). Movement time was significantly higher with movement (\m{9.38}, \sd{1.93}) than without (\m{7.97}, \sd{1.92}; see \autoref{fig:fitts_law_mt} and \autoref{tab:fittsMovementTimeMean}).

\begin{table}[ht]
    \caption{Fitts' Law: Movement Time in sec (Mean and Standard Deviation for different interaction methods)} 
    \label{tab:fittsMovementTimeMean}
    \centering
    \begin{tabular}{lcc}
            \toprule
            Interaction Method & Mean & Standard Deviation \\ \midrule
            DirectTouch & 8.28 & 1.36 \\
            Gaze\&Pinch & 8.34 & 2.39 \\
            Handray & 8.99 & 2.28 \\
            HeadGaze & 9.08 & 1.93 \\ 
            \bottomrule
        \end{tabular}
\end{table}

\subsubsection{Selection Offset} \label{c:selectionOffsetArt}

\begin{figure}[ht]
    \centering
    \includegraphics[width=0.5\textwidth]{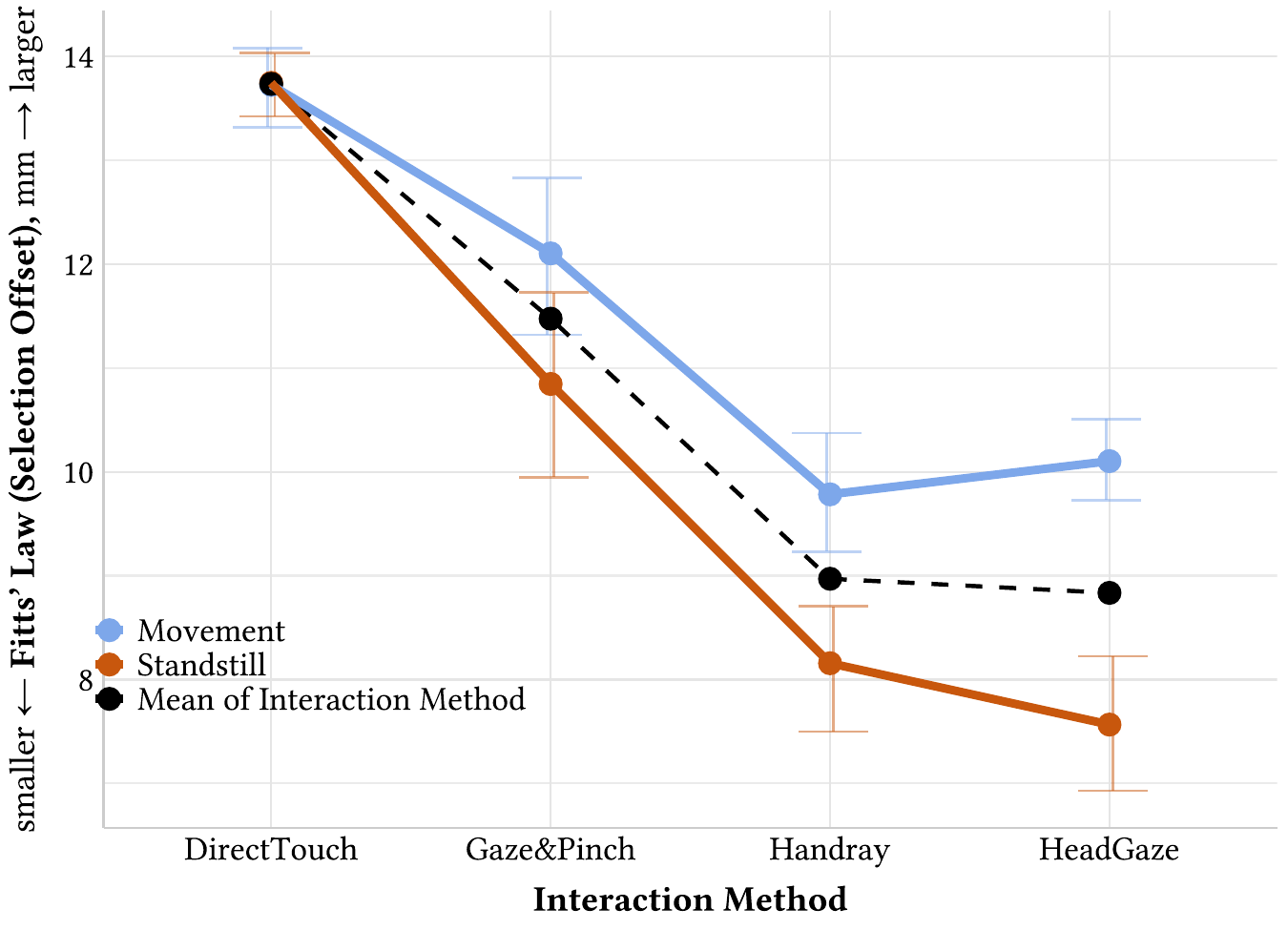}
    \caption{Significant Interaction Effects on Selection Offset}
    \label{fig:ie_gap}
    \Description{The figure depicts a line graph visualizing the interaction effect on selection offset. The Y-Axis represents the selection offset in mm, the X-Axis contains the interaction methods. Values for movement are higher for all interaction methods, except for DirectTouch where they do not differ significantly. The highest selection offset during movement is visible for DirectTouch, the lowest for Handray. The largest selection offset during standstill is visible for DirectTouch, the lowest for HeadGaze.}
\end{figure}

Selection offset describes the distance between the selection point and the center of the currently active target in millimeters. The ART found a significant main effect of \condition (\F{3}{69}{69.63}, \pminor{0.001}), and of \movementCondition (\F{1}{23}{104.40}, \pminor{0.001}) on selection offset. The ART found a significant interaction effect of \condition $\times$ \movementCondition on selection offset (\F{3}{69}{24.61}, \pminor{0.001}; see \autoref{fig:ie_gap}). The distance between selection and target center is larger during movement for all interaction methods, except for DirectTouch which contains comparable values. The difference between movement and standstill are largest for HeadGaze, followed by Handray and Gaze\&Pinch.

\subsection{Impact of Vehicle Motion} 

In this section, we analyze the task metrics selection offset and selection time of each combination of Interaction Method $\times$ Road Type $\times$ Curve Type. This approach allows us to identify which road characteristics affect which interaction method. The first repetition after a break was excluded from analysis (see \autoref{c:FittsTask}). Due to missing sensor data, the recordings of the interaction method DirectTouch of Participant 11 were removed from subsequent analysis.

\subsubsection{Approach}
We fitted a linear mixed model (LMM) (estimated using REML and nloptwrap optimizer) to predict selection offset and selection time per \movementCondition. 
The model always included the participant as random effect (formula: $\sim$1 | participant). The model's corresponded to \condition = DirectTouch, and, for movement to Road Type = SmoothRoad and Curve Type = Straight road. 
For Standstill, we only evaluated the effect of \condition as the course-related independent variables would not alter interaction during Standstill. 
The results obtained in the subsequent analysis are, therefore, always to be viewed relative to DirectTouch.
Standardized parameters were obtained by fitting the model on a standardized version of the dataset. 95\% Confidence Intervals (CIs) and p-values were computed using a Wald t-distribution approximation.
As in \autoref{c:FittsLawArt}, the data for selection offset and selection time were individually filtered for outliers using IQR. Furthermore, the data used contains correct and incorrect selections. In contrast to \autoref{c:FittsLawArtMT} selection time refers to the duration from the start of a target trial to its successful selection in seconds, allowing for increased granularity in the subsequent analysis.

\subsubsection{Selection Offset: Standstill} 

The intercept is represented by DirectTouch and features an average selection offset of 13.83mm. The results show that the interaction methods Handray, HeadGaze and Gaze\&Pinch significantly reduce the selection distance to the target center compared to DirectTouch (see \autoref{fig:gap_standstill}, and \autoref{tab:lmmEstimatesSelectionOffsetStandstill}). HeadGaze performed the best, followed by Handray and Gaze\&Pinch. Furthermore, pairwise comparisons found significant differences as described in \autoref{tab:lmmPairwiseGapStandstill}.

\begin{figure}[ht]
    \centering
    \includegraphics[width=0.5\textwidth]{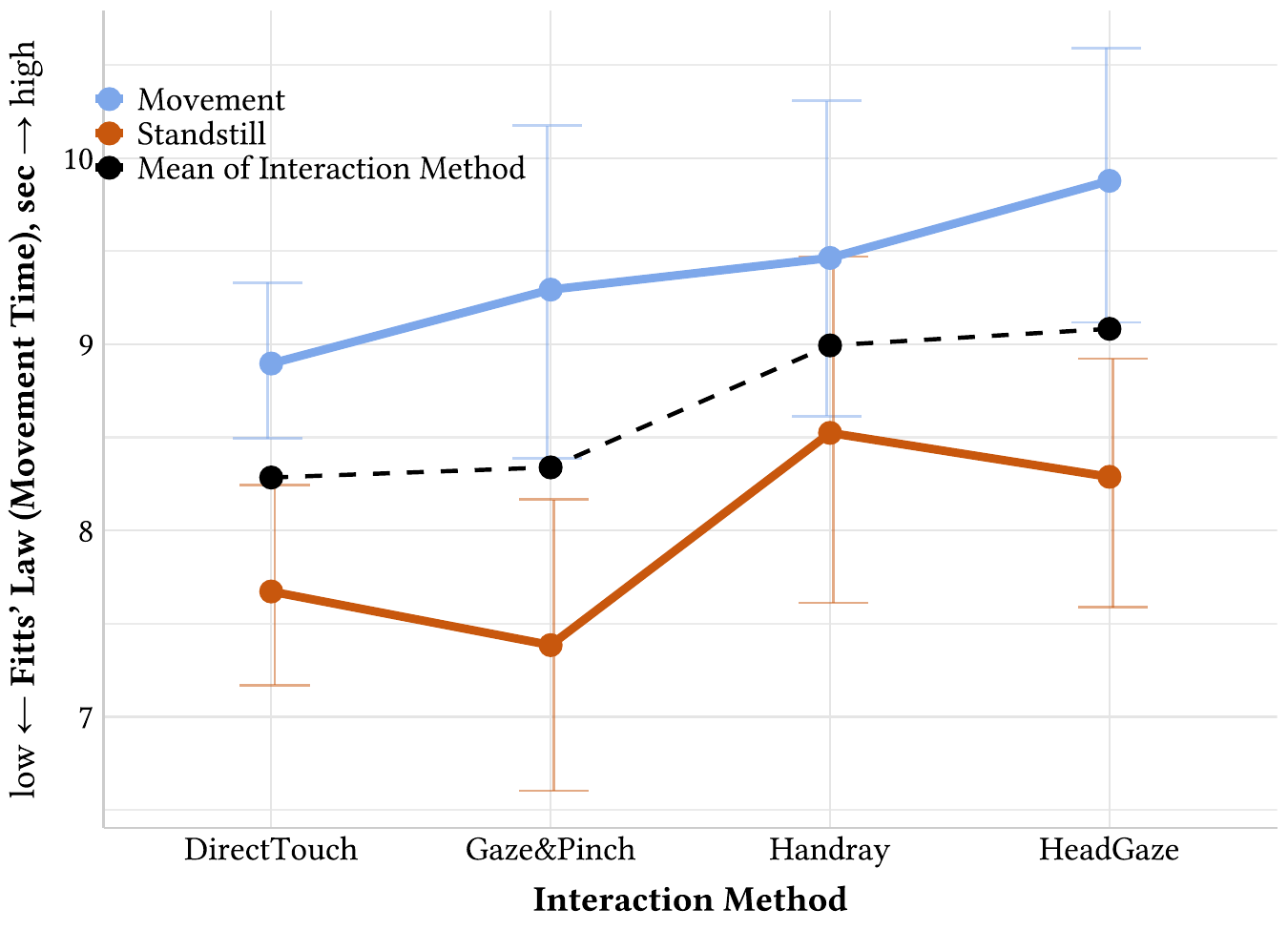}
    \caption{Significant Main effect on Fitts' Law: Movement Time}
    \label{fig:fitts_law_mt}
    \Description{The figure depicts a line graph visualizing the interaction effect on Fitts' Law movement time. The Y-Axis represents the movement time in seconds, the X-Axis contains the interaction methods. Values for movement are higher for all interaction methods. The highest movement time during movement is visible for HeadGaze, the lowest for DirectTouch. The highest movement time during standstill is visible for Handray, the lowest for Gaze\&Pinch.}
\end{figure}

\subsubsection{Selection Offset: Movement}

We evaluate the effects on the selection offset size, representing the distance between the selection point and the target center, the effects of different conditions and their interactions with road and curve types are examined.
The intercept is represented by DirectTouch and features an average selection offset of 13.43mm. During movement, and compared to DirectTouch, the interaction methods Gaze\&Pinch, Handray, and HeadGaze significantly reduce the distance of selections to the target center (see \autoref{fig:gap_movement}, and \autoref{tab:lmmEstimatesSelectionOffsetMovement}). In comparison, these methods provide increased accuracy, with the lowest selection offset present for HeadGaze. Additionally, there is a three-way interaction effect for Handray and HeadGaze during Short-Left Curves with acceleration on BumpyRoad, showing significantly increased accuracy compared to DirectTouch.
Decreased accuracy is found for several others. The road type MixedRoad and the curve type Long-Left-Curve with Acceleration increase the selection offset. BumpyRoad also reduced accuracy for HeadGaze, Handray, and Gaze\&Pinch.
The lowest accuracy is observed with the three-way interaction effect of Gaze\&Pinch during a Short-Left Curve with accelerations performed on a MixedRoad. Here the selection offset is increased by 2.64mm on average compared to DirectTouch.

\subsubsection{Selection Time: Movement}

\begin{figure*}[ht!]
\centering
    \begin{subfigure}[c]{0.47\linewidth}
        \includegraphics[width=\linewidth]{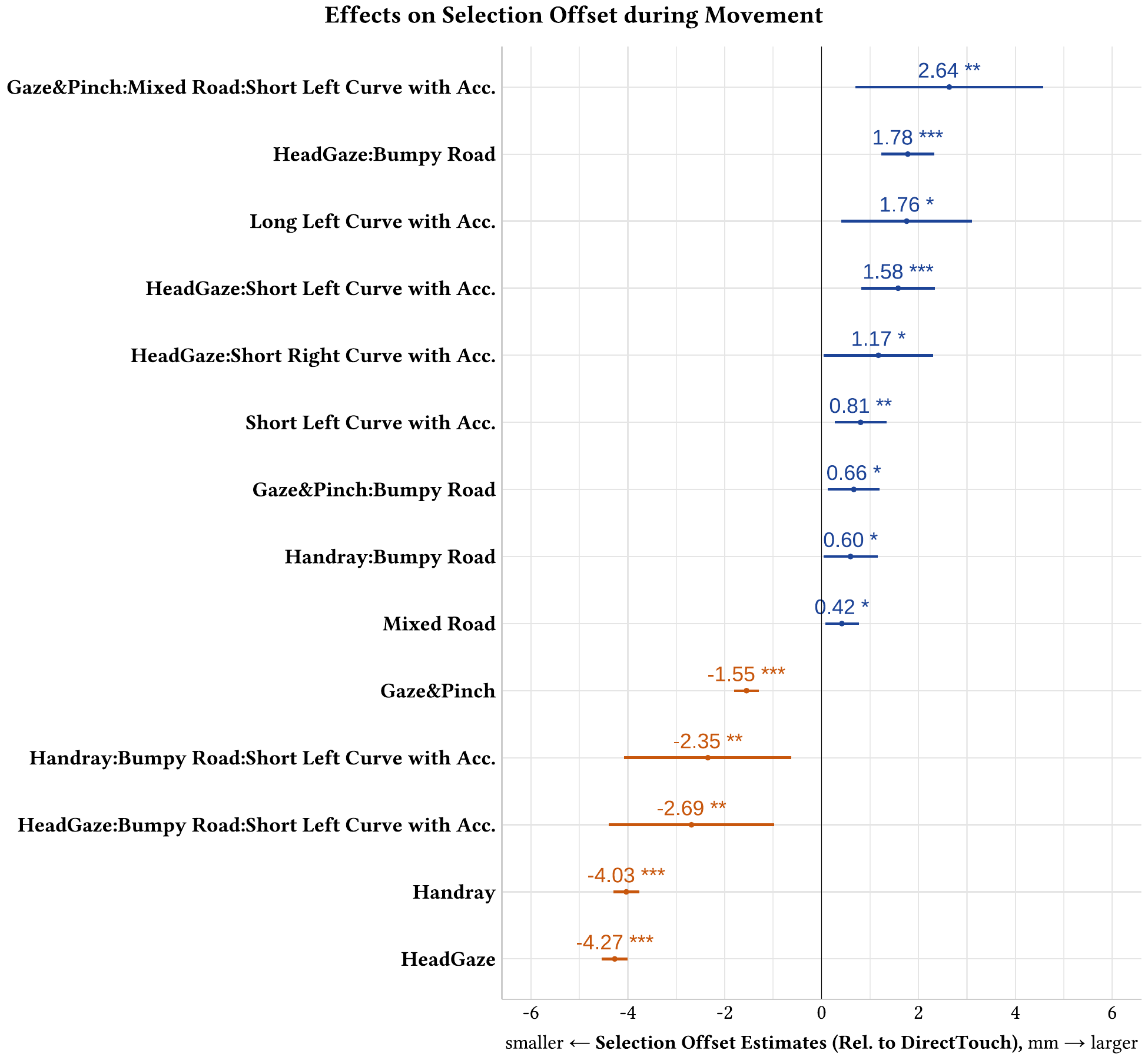}
             \caption{LMM for the average selection offset between the selection location and the center of the current target in millimeters. The estimates describe the distance difference of road and curve types compared to DirectTouch on a straight SmoothRoad (selection offset = 13.43mm). Gaze\&Pinch during a Short-Left Curve with Acceleration on a MixedRoad has the lowest precision with selection offset being 2.64mm larger compared to DirectTouch.}~\label{fig:gap_movement}
        \Description{The figure visualizes significant effects on selection offset during Movement, with multiple combinations of interaction methods, road types and curve types being listed on the y-axis. The x-axis represents Estimates in mm, showing the measured effects, with negative values increasing from the left.}
    \end{subfigure}
    \hfill
    \begin{subfigure}[c]{0.47\linewidth}
        \includegraphics[width=\linewidth]{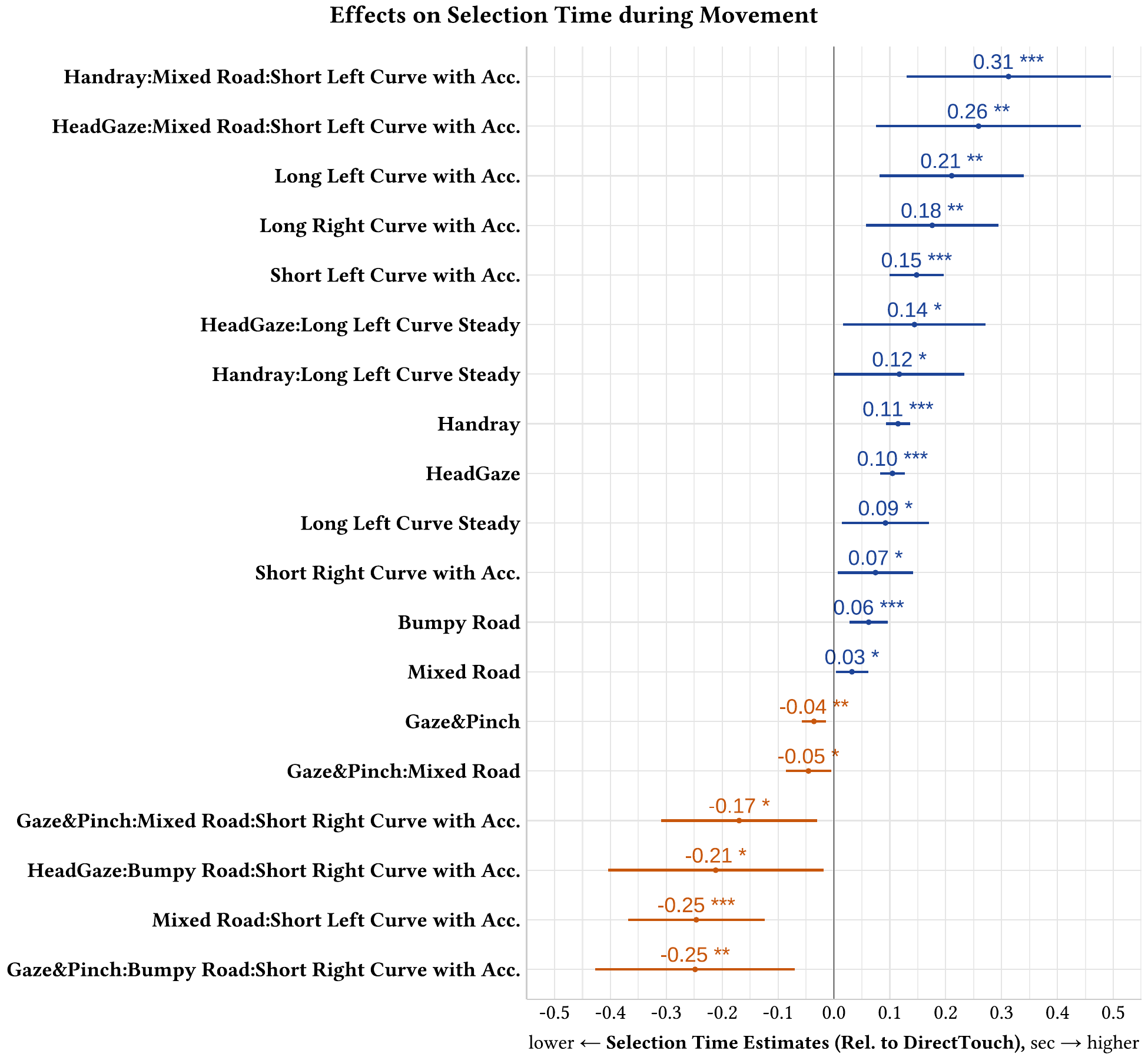}
        \caption{LMM for the average selection time from the start of a target trial to its successful selection in seconds. The estimates describe the time difference of different road and curve types compared to DirectTouch on a straight SmoothRoad (selection time = 1.09s). Selections performed using Handray during a Short-Left Curve with Acceleration on a MixedRoad take the longest with selection times being 0.31s longer compared to DirectTouch.}~\label{fig:time_movement}
        \Description{The figure visualizes significant effects on selection time during Movement, with multiple combinations of interaction methods, road types and curve types being listed on the y-axis. The x-axis represents Estimates, showing the measured effects, with negative values increasing from the left.}
    \end{subfigure}
   \caption{Significant results of LMM for selection offset and selection time during Movement (‘***’ \pminor{ .001}, ‘**’ \pminor{ .01}, ‘*’ \pminor{ .05})}~\label{fig:movement}
   \Description{Overview for significant results of linear mixed models for selection offset and selection time during Movement}
\end{figure*}

The intercept represented by DirectTouch features an average duration of 1.09s.
Using Gaze\&Pinch leads to significantly shorter selection times compared to DirectTouch. This improved performance also applies to the usage of the interaction method during BumpyRoad, and MixedRoad combined with varying curve types. Furthermore, all remaining interaction methods (Handray, HeadGaze) lead to significantly longer durations compared to DirectTouch. The only exception being HeadGaze, which had significantly shorter durations during Short-Right Curves with accelerations on BumpyRoad. Performing selection during Mixed- or BumpyRoad generally leads to a significantly increased time to perform selections. This also applies to multiple curve types (see \autoref{fig:time_movement}, and \autoref{tab:lmmEstimatesSelectionTimeMovement}). 
Three-way interactions found the highest selection duration for the interaction method Handray while being in a Short-Left Curve with acceleration on MixedRoad. Here, the duration increases on average by 0.31s.

\subsubsection{Selection Time: Standstill} 
\begin{figure*}[ht!]
\centering
    \begin{subfigure}[c]{0.47\linewidth}
        \includegraphics[width=\linewidth]{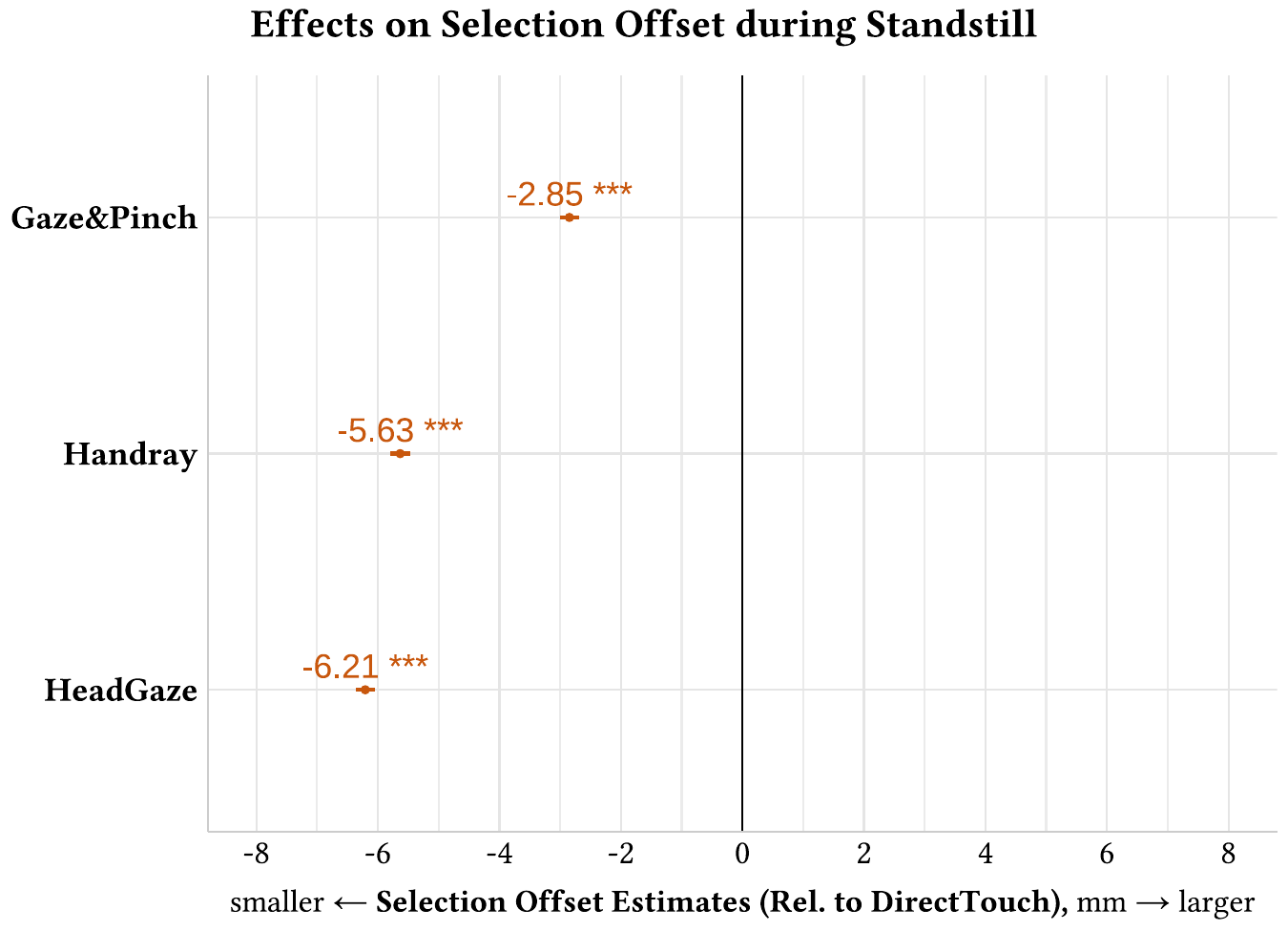}
             \caption{LMM for average selection offset between the selection location and the center of the current target in millimeters. The estimates describe the distance difference between Gaze\&Pinch, Handray, HeadGaze relative to DirectTouch at standstill (selection offset = 13.83mm). Selections were estimated 6.21mm closer to the target center for HeadGaze compared to DirectTouch.}~\label{fig:gap_standstill}
        \Description{The figure visualizes significant effects on selection offset during Standstill, with three interaction methods listed on the y-axis: Gaze\&Pinch, Handray, and HeadGaze. The x-axis represents Estimates in mm, showing the measured effects, with negative values increasing from the left.}
    \end{subfigure}
    \hfill
    \begin{subfigure}[c]{0.47\linewidth}
        \includegraphics[width=\linewidth]{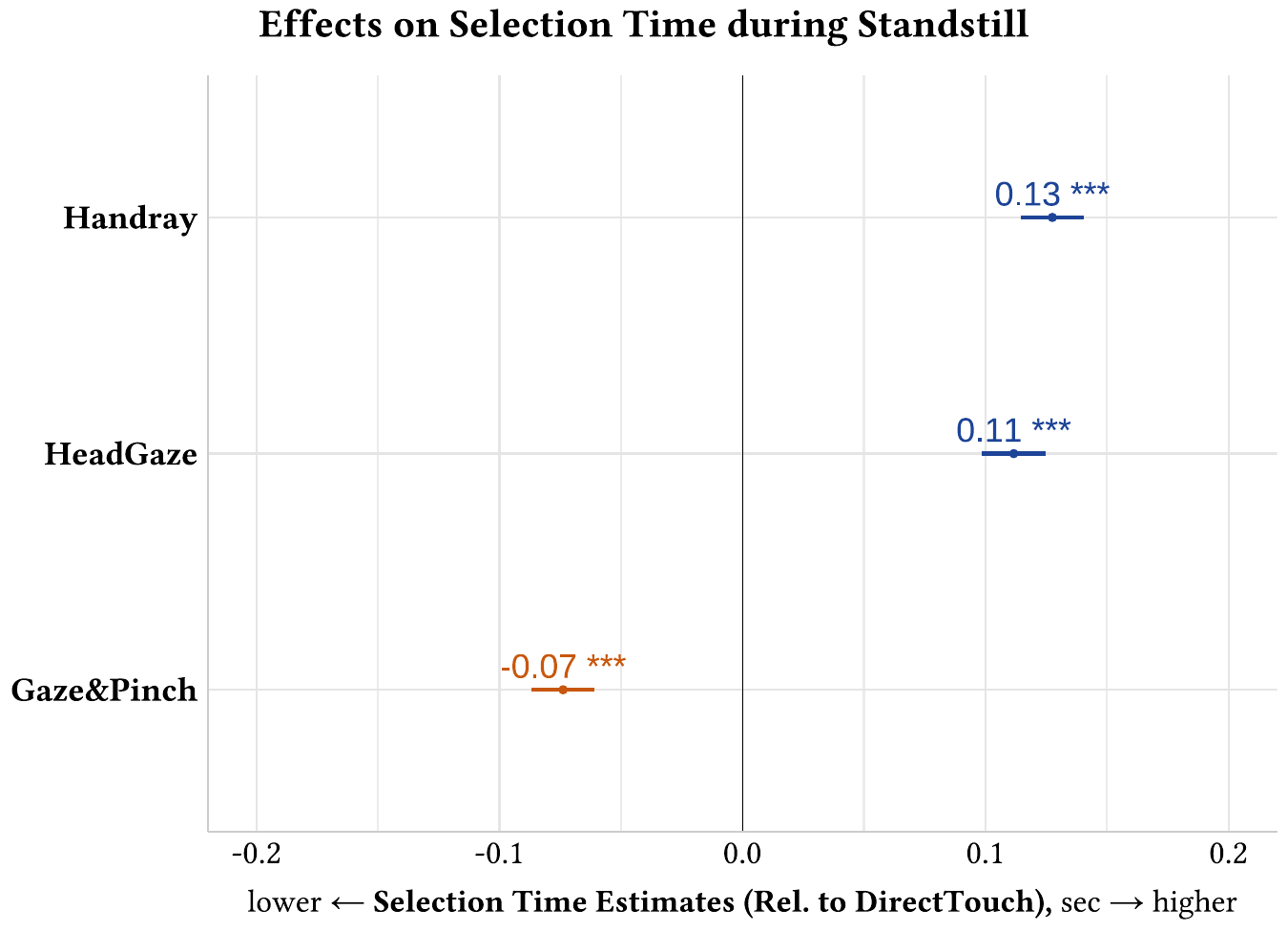}
        \caption{LMM for the average selection time from the start of a target trial to its successful selection in seconds. The estimates describe the time difference between Gaze\&Pinch, Handray, HeadGaze relative to DirectTouch during standstill (selection time = 1.01s). Selections were performed 0.07s faster with Gaze\&Pinch, while Handray increased this value by 0.13s compared to DirectTouch.}~\label{fig:time_standstill}
        \Description{The figure visualizes significant effects on selection time during Standstill, with three interaction methods listed on the y-axis: Gaze\&Pinch, Handray, and HeadGaze. The x-axis represents Estimates, showing the measured effects, with negative values increasing from the left.}
    \end{subfigure}
   \caption{Significant results of LMM for selection offset and selection time during Standstill (‘***’ \pminor{ .001}, ‘**’ \pminor{ .01}, ‘*’ \pminor{ .05})}~\label{fig:standstill}
   \Description{Overview for significant results of linear mixed models for selection offset and selection time during Standstill}
\end{figure*}

The intercept is represented by DirectTouch and features an average duration of 1.01s. Gaze\&Pinch is the only interaction method that performed significantly better by featuring, on average, 0.07s lower selection durations than DirectTouch. Handray and HeadGaze performed significantly worse than DirectTouch, with Handray featuring the highest duration by taking, on average, 0.13s longer (see \autoref{fig:time_standstill}, and \autoref{tab:lmmEstimatesSelectionTimeStandstill}.
Pairwise comparisons found significant differences as described in \autoref{tab:lmmPairwiseSelectionTimeStandstill}.

\begin{table*}[ht!]
\caption{Significant differences in selection offset during standstill. The estimates describe the distance differences between the listed interaction methods in millimeters.}
\label{tab:lmmPairwiseGapStandstill}
\centering
\begin{tabular}{lccccc}
\toprule
\textbf{Contrast} & \textbf{Estimate} & \textbf{SE} & \textbf{df} & \textbf{z-ratio} & \textbf{p-value} \\
\midrule
DirectTouch - Gaze\&Pinch & 2.848 & 0.0791 & 29187 & 36.007 & <.0001 \\
DirectTouch - Handray & 5.633 & 0.0809 & 29183 & 69.602 & <.0001 \\
DirectTouch - HeadGaze & 6.207 & 0.0821 & 29183 & 75.640 & <.0001 \\
Gaze\&Pinch - Handray & 2.785 & 0.0785 & 29176 & 35.465 & <.0001 \\
Gaze\&Pinch - HeadGaze & 3.360 & 0.0797 & 29176 & 42.145 & <.0001 \\
Handray - HeadGaze & 0.574 & 0.0816 & 29175 & 7.037 & <.0001 \\
\bottomrule
\end{tabular}
\end{table*}

\begin{table*}[ht]
\caption{Significant differences in selection time during standstill. The estimates describe the time differences between the listed interaction methods in seconds.}
\label{tab:lmmPairwiseSelectionTimeStandstill}
\centering
\begin{tabular}{lccccc}
\toprule
\textbf{Contrast} & \textbf{Estimate} & \textbf{SE} & \textbf{df} & \textbf{t-ratio} & \textbf{p-value} \\
\midrule
DirectTouch - Gaze\&Pinch & 0.0738 & 0.00667 & 24099 & 11.069 & <.0001 \\
DirectTouch - Handray & -0.1275 & 0.00666 & 24099 & -19.157 & <.0001 \\
DirectTouch - HeadGaze & -0.1116 & 0.00663 & 24099 & -16.847 & <.0001 \\
Gaze\&Pinch - Handray & -0.2013 & 0.00658 & 24097 & -30.614 & <.0001 \\
Gaze\&Pinch - HeadGaze & -0.1854 & 0.00655 & 24097 & -28.328 & <.0001 \\
Handray - HeadGaze & 0.0159 & 0.00653 & 24097 & 2.431 & 0.0715 \\
\bottomrule
\end{tabular}
\end{table*}

\subsubsection{Trajectories}

\begin{figure*}[ht!]
\centering
    %SmoothRoad
    \rotatebox[origin=c]{90}{\makebox[0.5in]{SmoothRoad}}%
    \begin{subfigure}[c]{0.240\linewidth}
        \includegraphics[trim=50 50 50 135, clip,width=\linewidth]{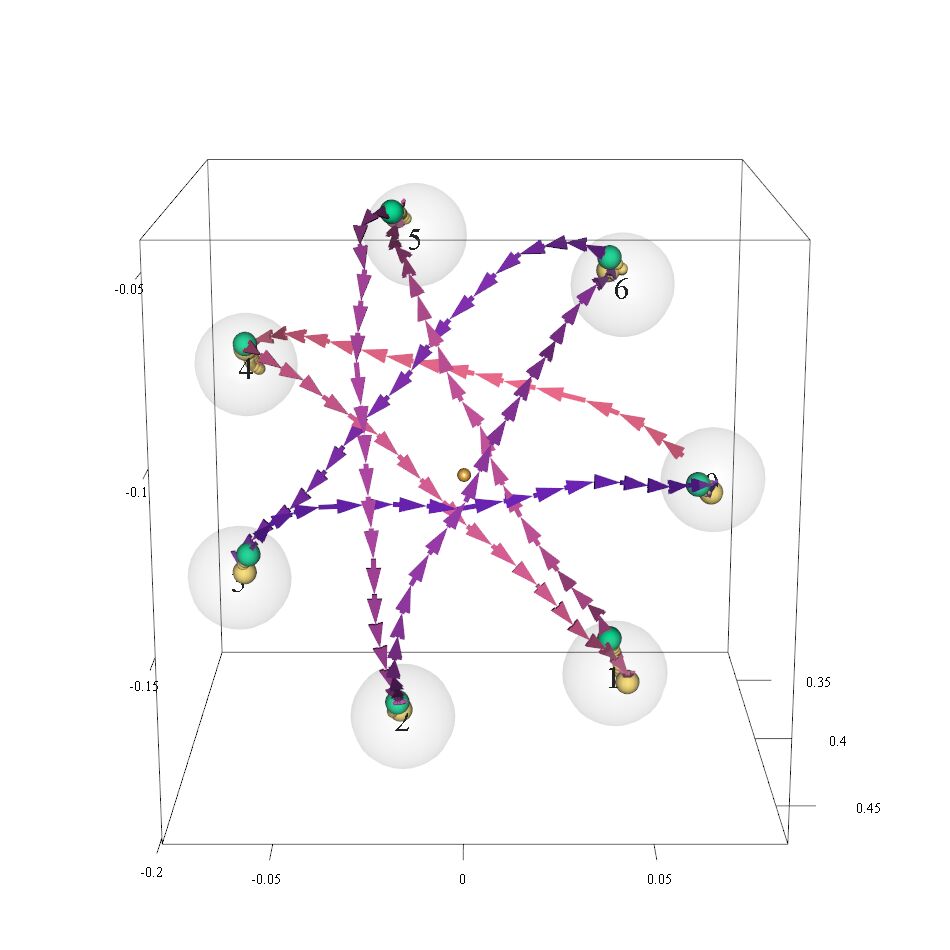}             \caption{DirectTouch (P23)}~\label{fig:dt_smooth_road}
        \Description{The figure visualizes the trajectories between targets for DirectTouch on SmoothRoad. It is visible that the cursor moved with relatively high precision between targets, performing selections near the center.}
    \end{subfigure}
    \begin{subfigure}[c]{0.240\linewidth}
        \includegraphics[trim=50 50 50 135, clip,width=\linewidth]{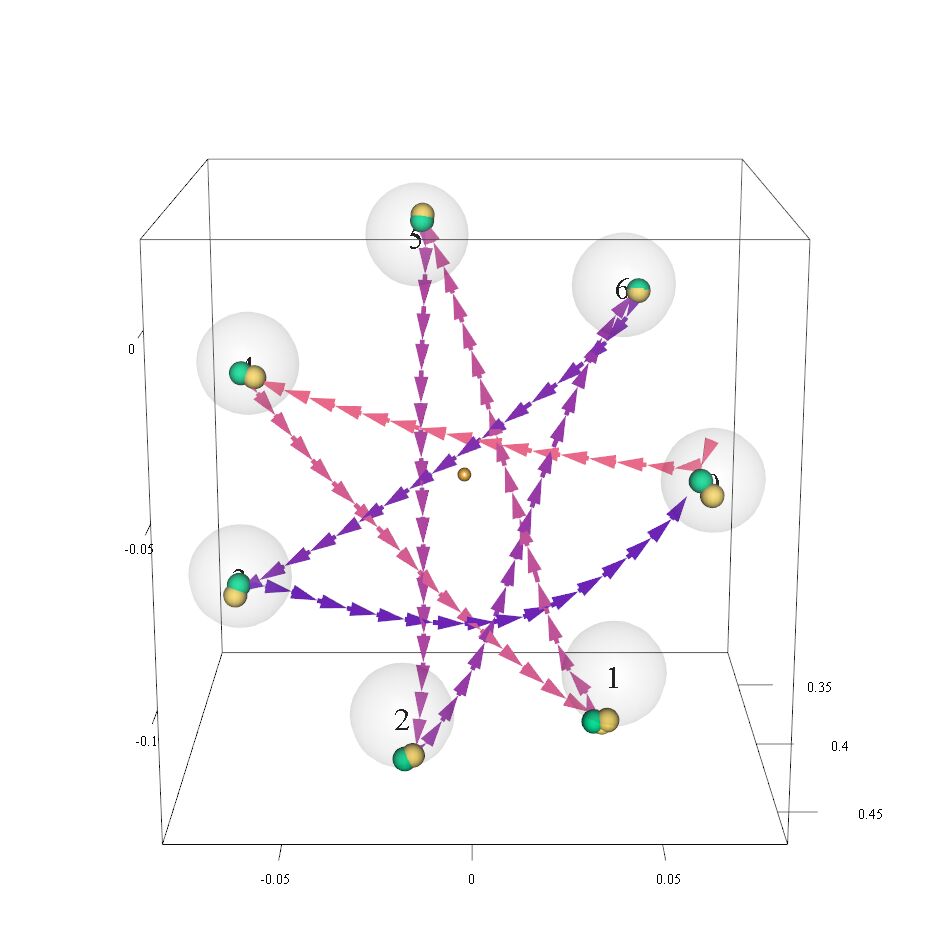}
        \caption{Handray (P25)}~\label{fig:hr_smooth_road}
        \Description{The figure visualizes the trajectories between targets for Handray on SmoothRoad. It is visible that the cursor moved with high precision between targets, performing selections near the center. The exception are the two targets at the bottom, where the cursor nearly had overshooting.}
    \end{subfigure}
    \begin{subfigure}[c]{0.240\linewidth}
        \includegraphics[trim=50 50 50 135, clip,width=\linewidth]{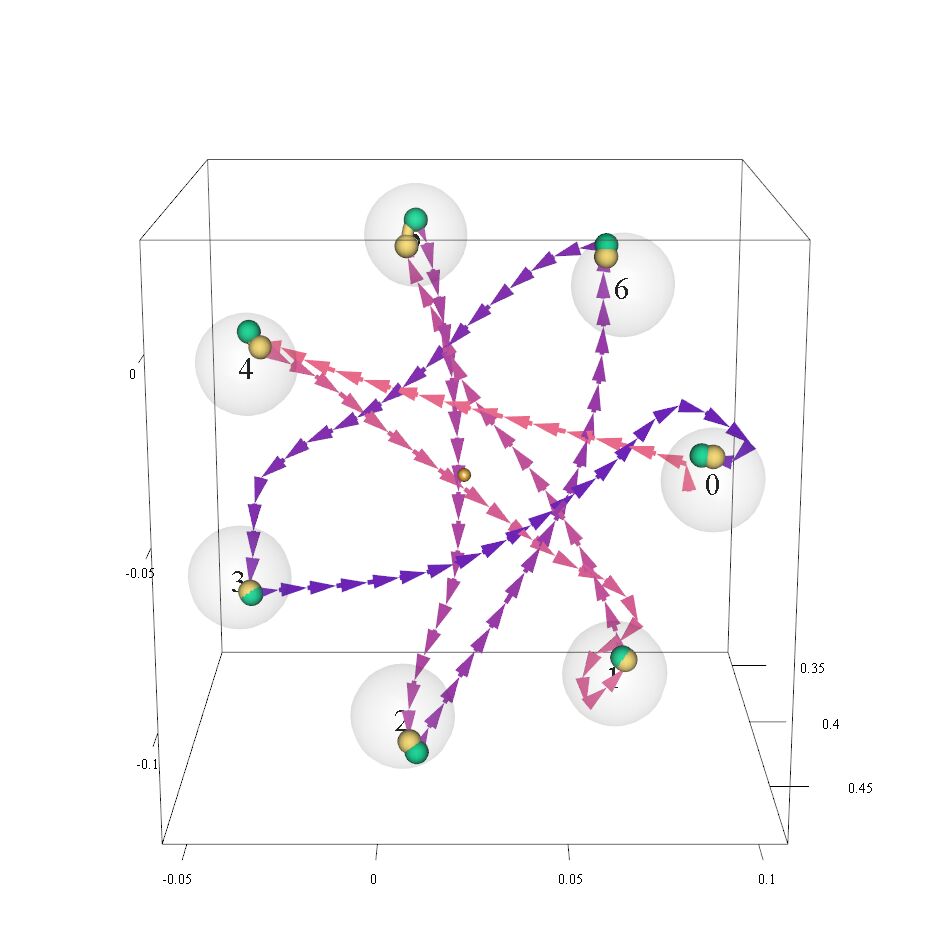}
        \caption{HeadGaze (P20)}~\label{fig:hg_smooth_road}
        \Description{The figure visualizes the trajectories between targets for HeadGaze on SmoothRoad. It is visible that the cursor moved with high precision between targets, but is bend to the right half of the targets. Selections near the outer bounds of targets are visualized, indicating the potential for overshooting with this interaction method.}
    \end{subfigure}
    \begin{subfigure}[c]{0.240\linewidth}
        \includegraphics[trim=50 40 20 135, clip,width=\linewidth]{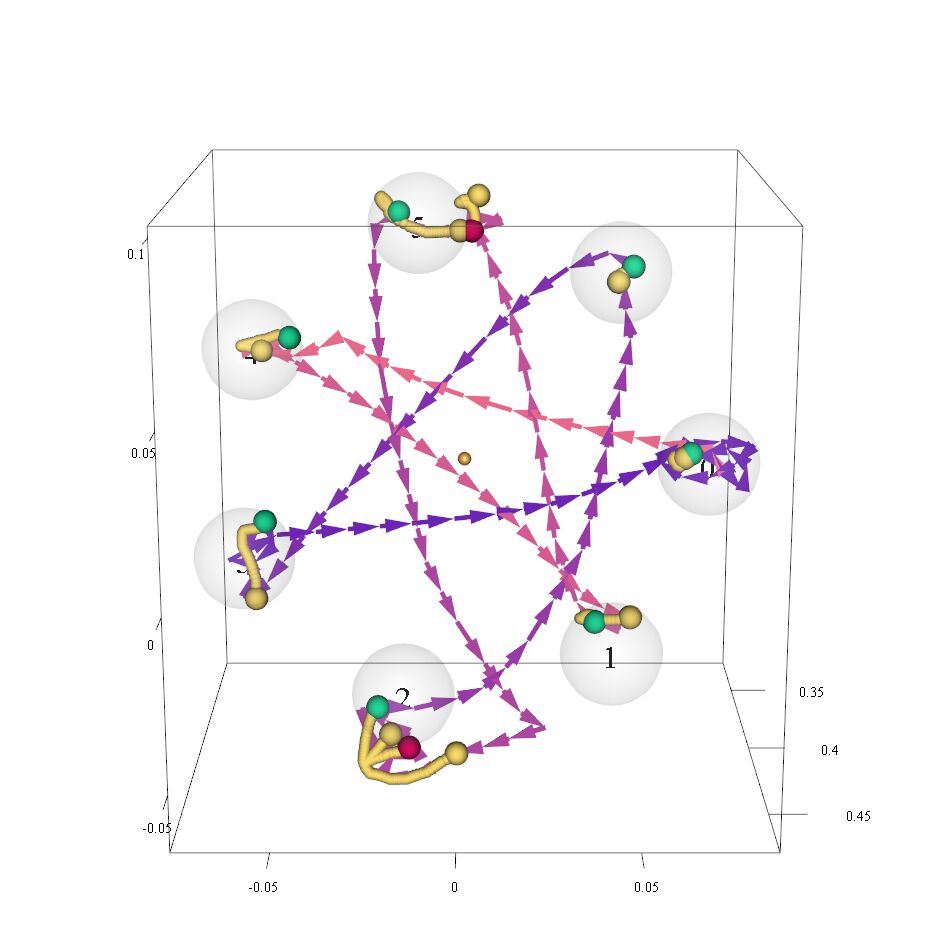}
        \caption{Gaze\&Pinch (P13)}~\label{fig:gp_smooth_road}
        \Description{The figure visualizes the trajectories between targets for Gaze\&Pinch on SmoothRoad. It is visible that the cursor moved with high precision and speed between targets, often performing selections near the center. Corrective movements of the eyes hovering on a target are visible.}
    \end{subfigure}
    %BumpyRoad
    \rotatebox[origin=c]{90}{\makebox[0.5in]{BumpyRoad}}%
    \begin{subfigure}[c]{0.240\linewidth}
        \includegraphics[trim=50 50 50 135, clip,width=\linewidth]{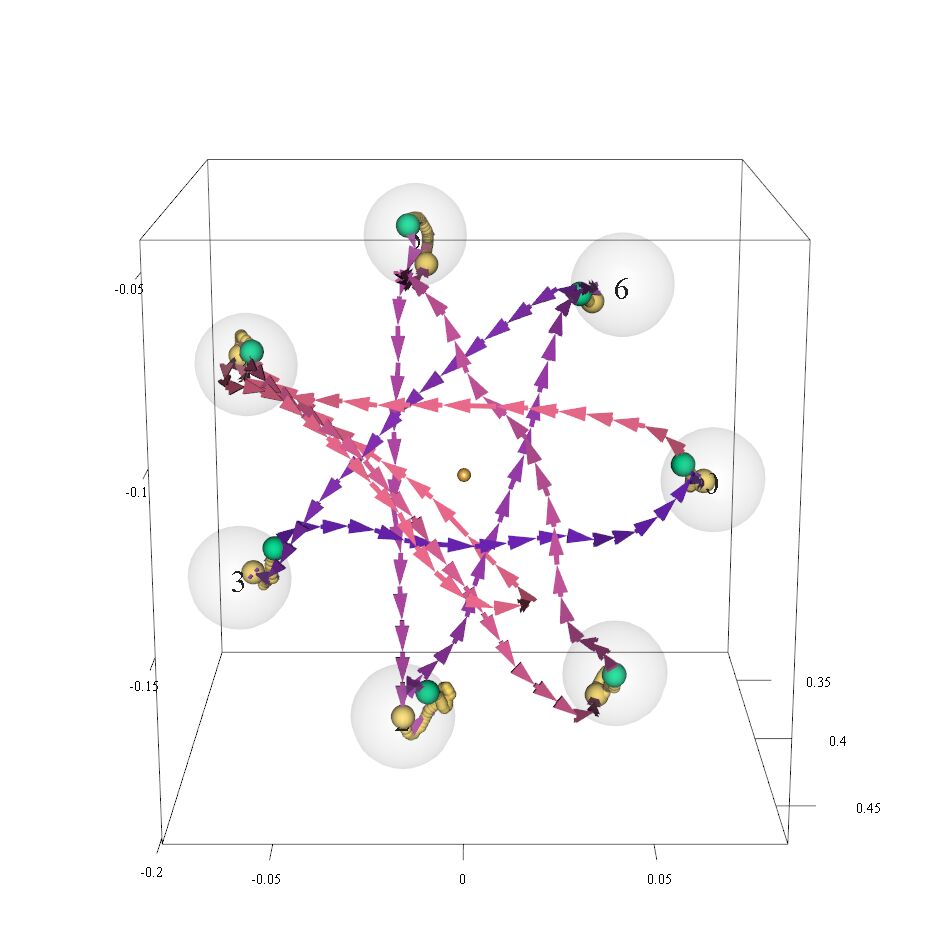}             \caption{DirectTouch (P23)}~\label{fig:dt_bumpy_road}
        \Description{The figure visualizes the trajectories between targets for DirectTouch on BumpyRoad. It is visible that cursor movement is more noisy, with a erroneous selection being visible for the target in the left-top. After the erroneous selection, the cursor first moved away from the target, before returning for a successful selection.}
    \end{subfigure}
    \begin{subfigure}[c]{0.240\linewidth}
        \includegraphics[trim=50 50 50 135, clip,width=\linewidth]{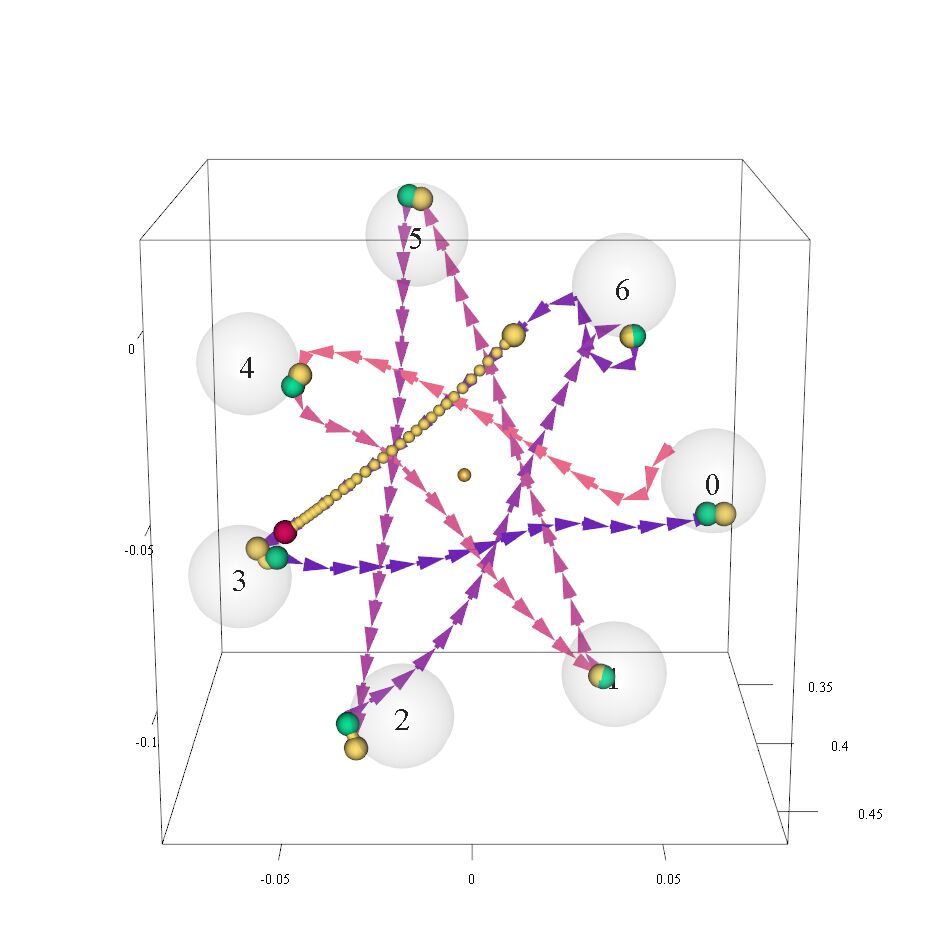}
        \caption{Handray (P25)}~\label{fig:hr_bumpy_road}
        \Description{The figure visualizes the trajectories between targets for Handray on BumpyRoad. Small corrective movements are visible nearby several targets, indicating that road bumps occurred in such situations.}
    \end{subfigure}
    \begin{subfigure}[c]{0.240\linewidth}
        \includegraphics[trim=50 50 50 135, clip,width=\linewidth]{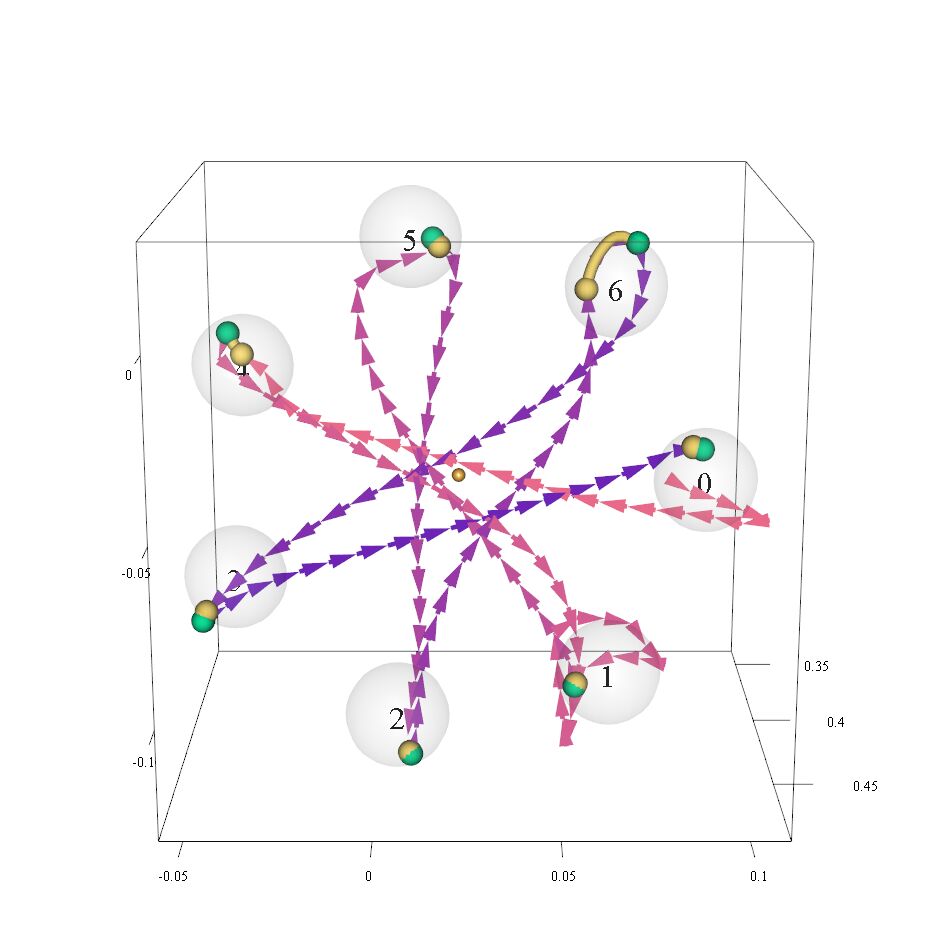}
        \caption{HeadGaze (P20)}~\label{fig:hg_bumpy_road}
        \Description{The figure visualizes the trajectories between targets for HeadGaze on BumpyRoad. It is visible that the cursor moved with high precision between targets, but selections were at the outer bounds of targets indicating potential risk of overshooting. Bumps led to jerky movements of the cursor and required corrective movement afterwards.}
    \end{subfigure}
    \begin{subfigure}[c]{0.240\linewidth}
        \includegraphics[trim=50 40 20 110, clip,width=\linewidth]{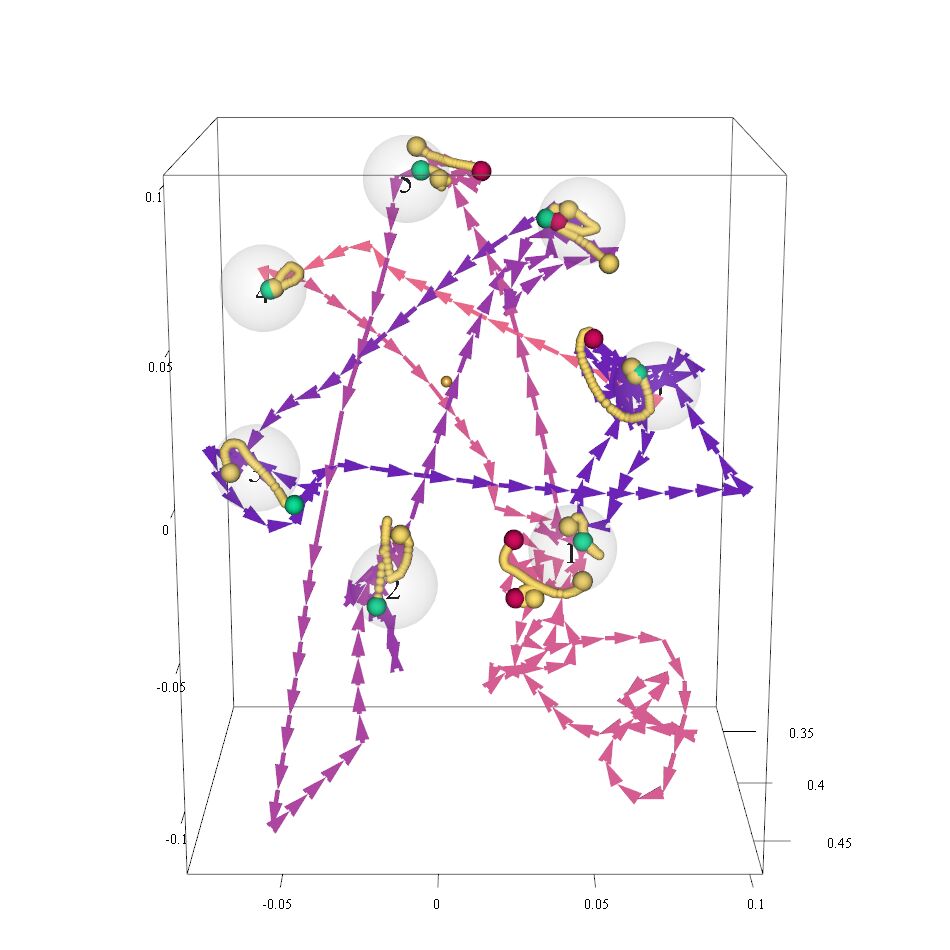}
        \caption{Gaze\&Pinch (P13)}~\label{fig:gp_bumpy_road}
        \Description{The figure visualizes the trajectories between targets for Gaze\&Pinch on BumpyRoad. The cursor movement is highly perturbed, with many cursor positions being outside the bounds of the Fitts' Law Target. Multiple erroneous selections are visible.}
    \end{subfigure}
    %ShortLeftCurve
    \rotatebox[origin=c]{90}{\makebox[0.5in]{Short-Left Curve with Acc.}}%
    \begin{subfigure}[c]{0.240\linewidth}
        \includegraphics[trim=50 50 50 135, clip,width=\linewidth]{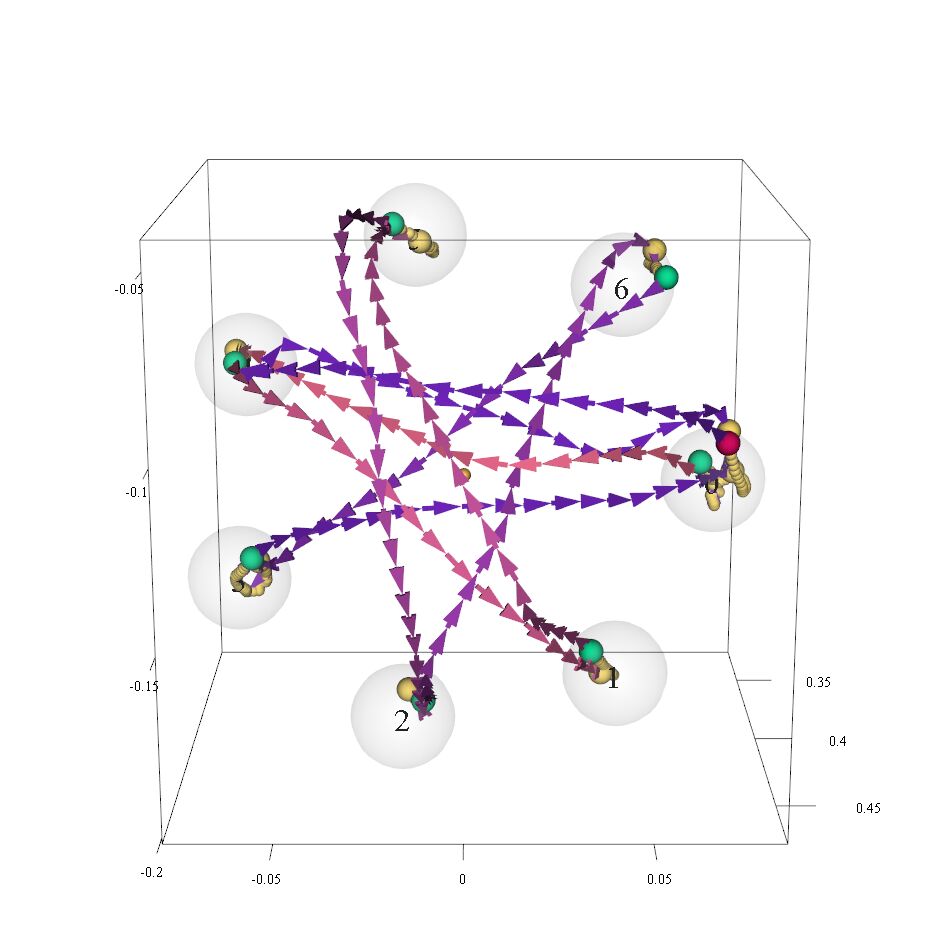}
        \caption{DirectTouch (P23)}~\label{fig:dt_short_left_curve}
        \Description{The figure visualizes the trajectories between targets for DirectTouch on Short-Left Curve with Acceleration. Noisy cursor movement is visible, with the participant having to perform retries for multiple targets.}
    \end{subfigure}
    \begin{subfigure}[c]{0.240\linewidth}
        \includegraphics[trim=50 50 50 135, clip,width=\linewidth]{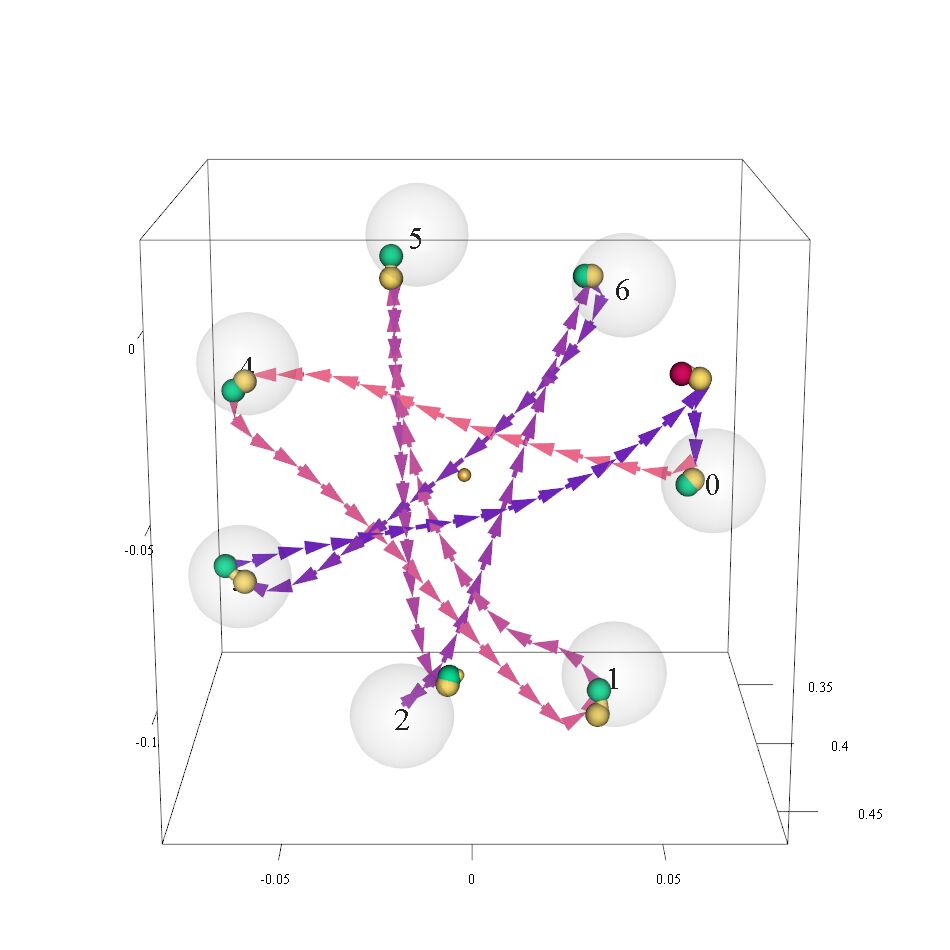}
        \caption{Handray (P25)}~\label{fig:hr_short_left_curve}
        \Description{The figure visualizes the trajectories between targets for Handray during Short-Left Curve with Acceleration. Visible bending is perceivable in the trajectories, with corrective movements and erroneous selections taking place near targets.}
    \end{subfigure}
    \begin{subfigure}[c]{0.240\linewidth}
        \includegraphics[trim=50 50 50 135, clip,width=\linewidth]{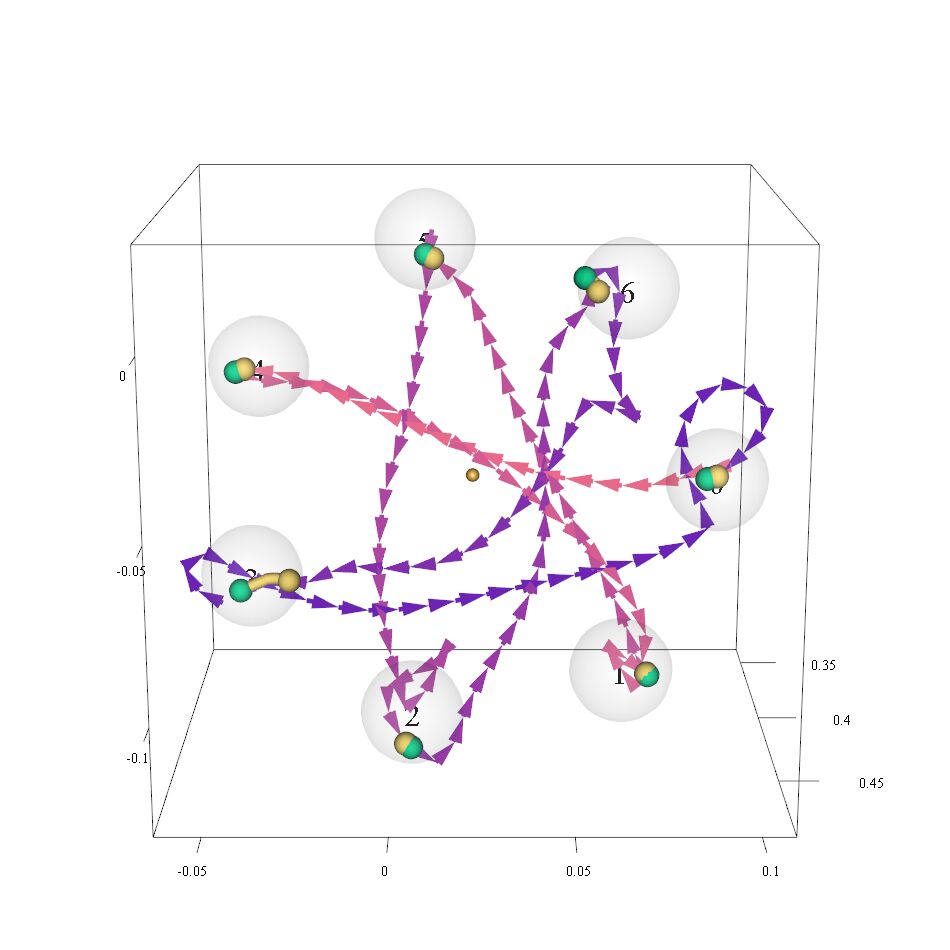}
        \caption{HeadGaze (P20)}~\label{fig:hg_short_left_curve}
        \Description{The figure visualizes the trajectories between targets for HeadGaze during Short-Left Curve with Acceleration. Cursor movement is heavily influenced by vehicle movements, being bend to the right half of the targets.}
    \end{subfigure}
    \begin{subfigure}[c]{0.240\linewidth}
        \includegraphics[trim=50 40 20 135, clip,width=\linewidth]{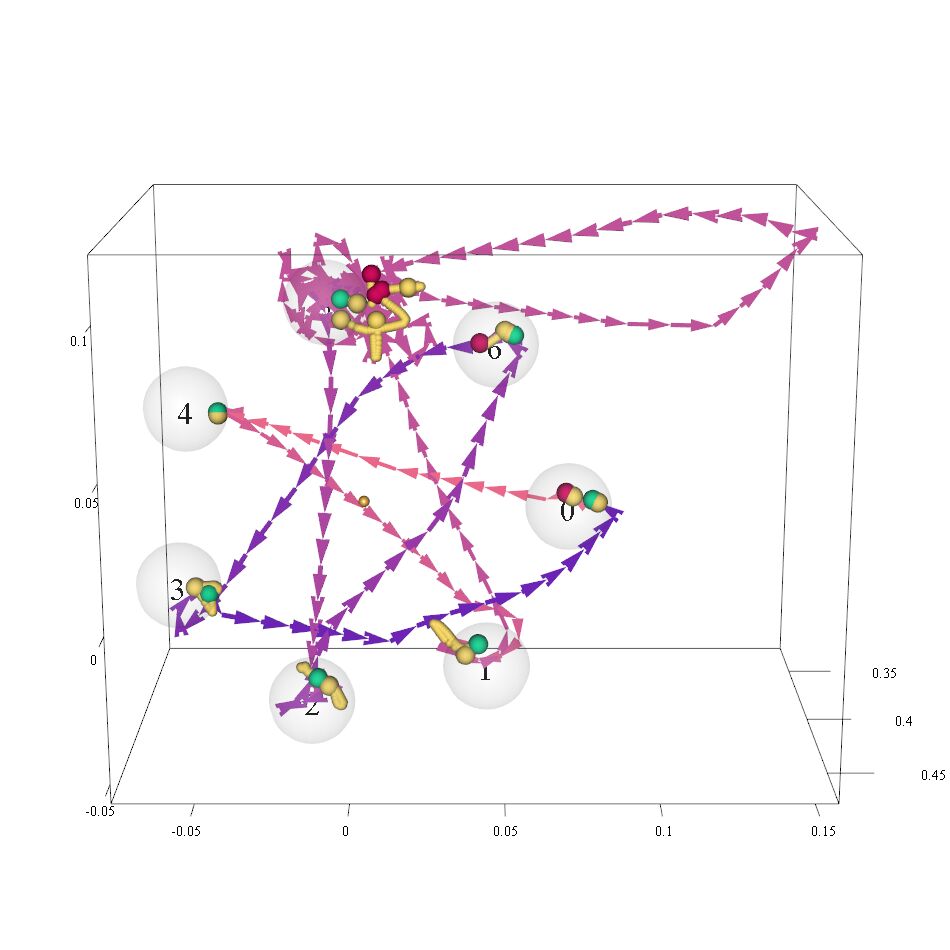}
        \caption{Gaze\&Pinch (P13)}~\label{fig:gp_short_left_curve}
        \Description{The figure visualizes the trajectories between targets for Gaze\&Pinch during Sharp-Left Curve with Acceleration. For most targets it is visible that the cursor moved with high precision, performing selections near the center. The exception is one targets at the top, where multiple erroneous selections were made as we suspected due to issues focusing on the target due to movement.}
    \end{subfigure}
    \caption{Trajectories during user selections. The columns show the four interaction methods. The rows show road conditions. Green spheres represent successful selections, red spheres unsuccessful selections, and yellow spheres cursor movements between pinch-down and release. The sphere for pinch-down is larger, with smaller spheres representing cursor movements. }~\label{fig:trajectoryOverview}
   \Description{Overview for trajectories across the interaction methods Gaze\&Pinch, DirectTouch, HeadGaze, and Handray in combination with SmoothRoad, BumpyRoad and Short-Left Curve with Acceleration. The graphics visualize the impact of different road types and curve types on cursor movements between targets.}
\end{figure*}

To understand how the vehicle movement influenced interactions, we created exemplary 3d visualizations of dynamic situations impacting precision based on the results of \autoref{fig:gap_movement}). 
As expected, the movement condition presented a significant effect on user performance (F(1,23) = 57.44, p<0.001), with users being more accurate in estimating the position of the next target whenever the vehicle was at a standstill.
The examples are based on participants whose number of erroneous selections matched the mean for each interaction method.
The visualizations encompass cursor trajectories between targets within one Fitts' Law repetition and successful (green sphere) and unsuccessful (red sphere) selections (see \autoref{fig:road_conditions}). 
The trajectories let us analyze the quality of the selection approaches and how external forces or environmental conditions could affect normal cursor movements. 
For example, a correct selection movement is characterized by smooth and fast approaches followed by short correction movements towards the target position~(\citet{arizaProximity2018}). 
In contrast, a BumpyRoad or a pronounced curve produces jerky or shifted trajectories.
The trajectories of Gaze\&Pinch were most affected by the vehicle movement, with the SmoothRoad being the least affected (\autoref{fig:gp_smooth_road}). BumpyRoad condition presented the worst problems with numerous inaccuracies translated into the user retrying several times until finally confirming selections inside the target (\autoref{fig:gp_bumpy_road}). SharpLeftCurve presented fewer inaccuracies (e.g., targets 1 and 3) and shiftings (e.g., target 6, \autoref{fig:gp_short_left_curve}). 
In the case of DirectTouch, trajectories recorded in a SmoothRoad depicted normal approaches (\autoref{fig:dt_smooth_road}), similar to BumpyRoad with additional inaccuracies close to target positions (\autoref{fig:dt_bumpy_road}). Finally, the vehicle movement during a SharpLeftCurve affected the trajectories in the shape of occasional incorrect selections followed by retrying.
Regarding HeadGaze, on a SmoothRoad (\autoref{fig:hg_smooth_road}), an increase in jerky directional changes within the trajectory can be identified for BumpyRoad, becoming noticeable in the run-up to the selection of the first target (\autoref{fig:hg_bumpy_road}). The forces exhibited during a SharpLeftCurve with acceleration cause trajectories to bend in the opposite direction of vehicle movement. In \autoref{fig:hg_short_left_curve}, this is visible for the movement between targets 2-6 and 6-3.
On the contrary, Handray featured comparably consistent trajectories for all the road conditions (\autoref{fig:hr_smooth_road}, \autoref{fig:hr_short_left_curve}), showing gaps to the targets, small enough to provide correct selections for BumpyRoad (\autoref{fig:hr_bumpy_road}).

\subsection{Motion Sickness}
%Motion Sickness (MISC)

We analyzed the delta between measurements of the MISC performed before and after each condition. The ART found a significant main effect of \movementCondition on the difference in motion sickness before and after the task (\F{1}{23}{5.75}, \p{0.025}). Motion sickness delta increased significantly more with movement (\m{0.46}, \sd{0.89}) than during standstill (\m{0.03}, \sd{0.57}).

\subsection{Eye-Strain} 
The ART found a significant main effect of \condition on Eye Strain (\F{3}{69}{4.36}, \p{0.007}). A post-hoc test found that Eye-Strain was significantly lower for DirectTouch (\m{2.04}, \sd{1.11}) than for Gaze\&Pinch (\m{2.77}, \sd{1.32}, $p_{\text{adj}}$ = 0.0150).

\subsection{Space Occupation}

\subsubsection{Hand Palm} \label{c:handPalmEuc}%Stats validated 
\begin{figure}[ht]
    \centering
        \includegraphics[width=\linewidth]{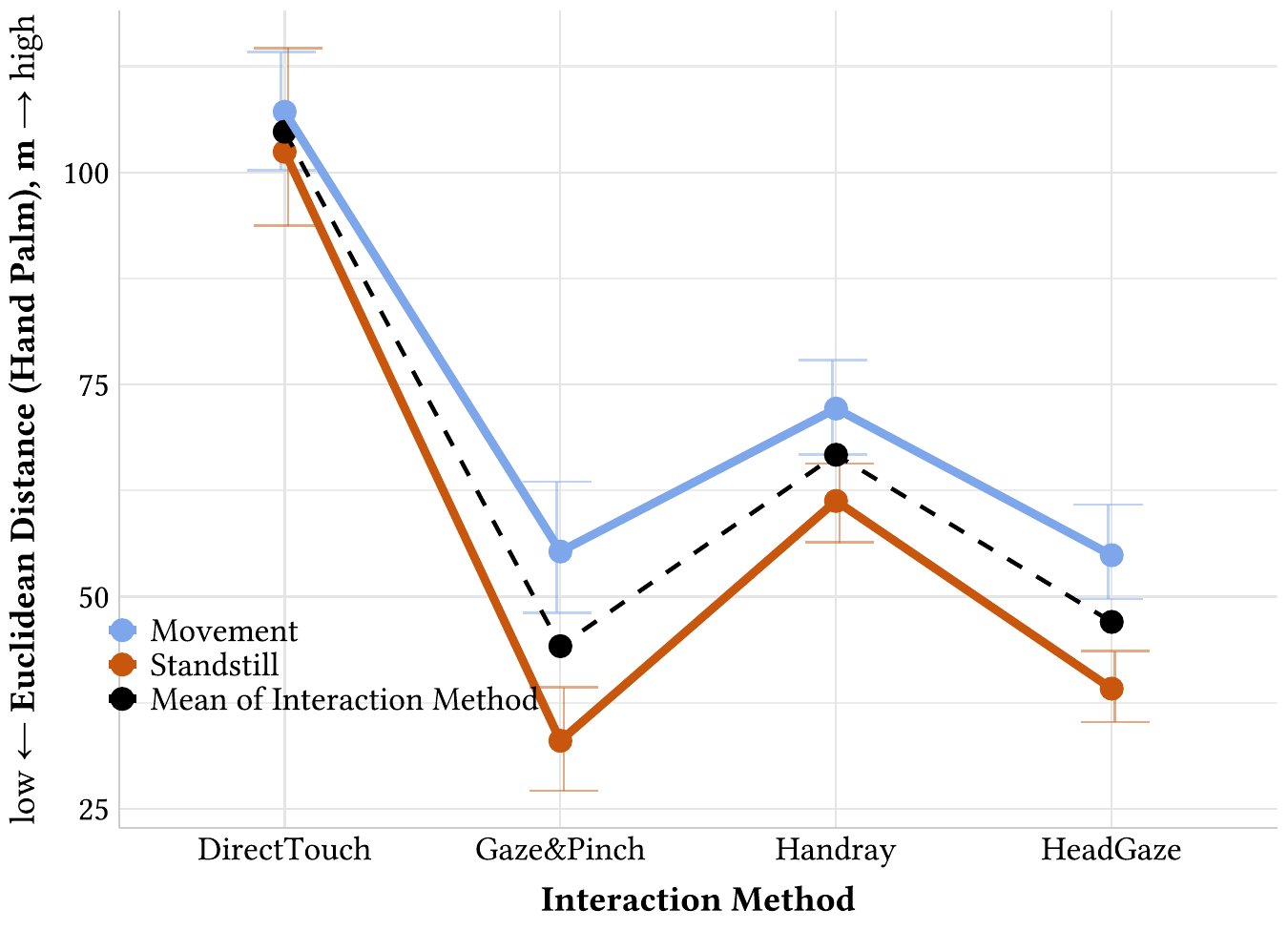}
        \caption{Significant interaction effect on the euclidean distance covered by hand palm movements}
        \label{fig:eucDist_HandPalm}
        \Description{The figure depicts a line graph visualizing the interaction effect on euclidean distance of the hand palm. The Y-Axis represents the distance traversed, the X-Axis contains the interaction methods. The two colors used represent Movement (blue) and Standstill (red) conditions. Values for movement are higher for all interaction methods. The highest distance during movement is visible for DirectTouch, with the two lowest ones being Gaze\&Pinch and HeadGaze. The highest distance during standstill is visible for DirectTouch, the lowest for Gaze\&Pinch.}
\end{figure}

For euclidean distance, the ART found a significant main effect of \condition on hand palm (\F{3}{69}{118.48}, \pminor{0.001}) and of \movementCondition on hand palm (\F{1}{23}{71.24}, \pminor{0.001}). The ART found a significant interaction effect of \condition $\times$ \movementCondition on the euclidean distance of hand palm (\F{3}{69}{5.05}, \p{0.003}; see \autoref{fig:eucDist_HandPalm}). Values of movement condition are similar for DirectTouch but differ more for the other interaction methods. For all interaction methods, euclidean distance is larger during movement than during standstill. DirectTouch is the interaction method which across movement conditions featured the highest values. For standstill, Gaze\&Pinch features the lowest value, while for movement, Gaze\&Pinch and HeadGaze have similarly low values (see \autoref{tab:eucDistHandPalm}).

\subsubsection{Head}
\begin{figure}[ht]
    \centering
    \includegraphics[width=\linewidth]{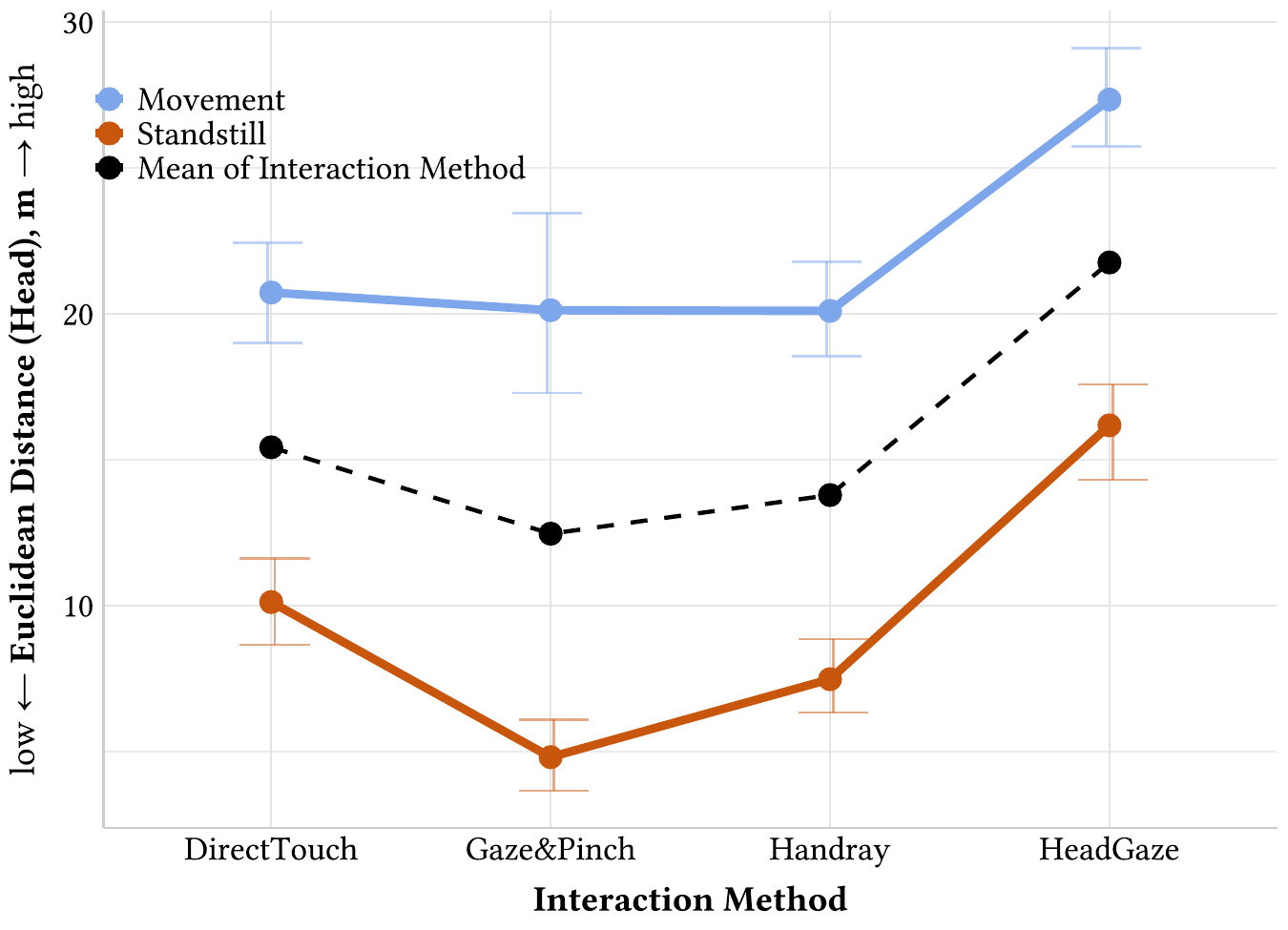}
    \caption{Significant interaction effect on the euclidean distance covered by head movements}
    \label{fig:eucDist_Head}
    \Description{The figure depicts a line graph visualizing the interaction effect on euclidean distance of the head. The Y-Axis represents the distance traversed, the X-Axis contains the interaction methods. The two colors used represent Movement (blue) and Standstill (red) conditions. Values for movement are higher for all interaction methods. The highest distance during movement is visible for HeadGaze, with the two lowest ones being Gaze\&Pinch and Handray. The highest distance during standstill is visible for HeadGaze, the lowest for Gaze\&Pinch.}
\end{figure}

For the euclidean distance, the ART found a significant main effect of \condition (\F{3}{69}{39.58}, \pminor{0.001}), and of \movementCondition (\F{1}{23}{459.08}, \pminor{0.001}) on head . The ART found a significant interaction effect of \condition $\times$ \movementCondition on euclidean distance of the head (\F{3}{69}{4.97}, \p{0.004}; see \autoref{fig:eucDist_Head}). For Head, the values of each interaction method are higher with movement than with standstill. The highest value is represented by HeadGaze during both, movement and standstill. For standstill, the lowest value is assigned to Gaze\&Pinch, while for movement the interaction methods DirectTouch, Gaze\&Pinch, and Handray feature similar values (\autoref{tab:eucDistHead}).

\begin{table}[hb!]
    \centering
    \caption{Euclidean Distance Hand Palm (Mean and Standard Deviation for different interaction methods)}
    \label{tab:eucDistHandPalm}
    \begin{tabular}{lcc}
        \toprule
        Method &Mean & SD \\
        \midrule
        DirectTouch Movement          & 107.0            & 17.3             \\
        DirectTouch Standstill        & 102.0            & 27.9             \\ 
        \midrule
        Gaze\&Pinch Movement            & 55.3             & 19.8             \\
        Gaze\&Pinch Standstill          & 33.0             & 15.3             \\ 
        \midrule
        Handray Movement              & 72.2             & 13.9             \\
        Handray Standstill            & 61.3             & 12.2             \\ 
        \midrule
        HeadGaze Movement             & 54.9             & 14.4             \\
        HeadGaze Standstill           & 39.2             & 10.4             \\ 
        \bottomrule
    \end{tabular}
\end{table}

\begin{table}[hb]
    \centering
    \caption{Euclidean Distance Head (Mean and Standard Deviation for different interaction methods)}
    \label{tab:eucDistHead}
    \begin{tabular}{lcc}
        \toprule
        Method &Mean & SD \\
        \midrule
        DirectTouch Movement          & 20.7             & 4.35             \\
        DirectTouch Standstill        & 10.1             & 3.75             \\ \midrule
        Gaze\&Pinch Movement            & 20.1             & 8.26             \\
        Gaze\&Pinch Standstill          & 4.80             & 3.13             \\ \midrule
        Handray Movement              & 20.1             & 4.07             \\
        Handray Standstill            & 7.47             & 3.25             \\ \midrule
        HeadGaze Movement             & 27.4             & 4.46             \\
        HeadGaze Standstill           & 16.2             & 4.14             \\
        \bottomrule
    \end{tabular}
\end{table}

\subsection{Perceived Trust and Safety}

The ART found a significant main effect of \movementCondition on Trust in Automation (\F{1}{23}{13.57}, \p{0.001}), on understanding (\F{1}{23}{16.31}, \pminor{0.001}), and on perceived safety (\F{1}{23}{8.71}, \p{0.007}). 
Trust (\m{3.98}, \sd{0.80}), understanding (\m{3.65}, \sd{0.86}), and perceived safety (\m{1.60}, \sd{1.28}) were significantly lower with movement than during standstill, where trust (\m{4.35}, \sd{0.78}), understanding (\m{4.15}, \sd{0.68}), and perceived safety (\m{1.91}, \sd{1.06}) were higher.

\subsection{Open User Feedback}
We gathered qualitative feedback from participants to gain deeper insights into their experiences and perceptions.
This encompassed two key areas: (1) the influence of road conditions on the interaction methods, (2) privacy concerns and considerations for public usage, and (3) general preferences between the different methods tested. We summarize the most frequently mentioned concerns and notable observations reported by participants.

\subsubsection{Influence of Road Conditions} \label{c:interviewInfluenceRoadConditions}
Participants highlighted that road conditions significantly impacted their interaction experience, affecting their preferences and effectiveness.

DirectTouch was highly appreciated for its simplicity, with many participants noting how intuitive it was and how it allowed them to stay aware of their surroundings.
Several mentioned the value of understanding the external environment, such as why the vehicle slowed down. However, vehicle movements like bumps and curves caused challenges, leading to errors or misjudgments in interactions. Many participants also experienced physical strain, especially in the arms and shoulders, which was more pronounced during movement but still present at standstill.

Some users found the high level of precision required for effective interaction too demanding, particularly when the vehicle was stationary. While a few participants could compensate for vehicle movements during interactions, with P10 exemplary describing such situations as "(...) whenever there was bumps it was a little hard to get to the right point, but it was not too much and it was easily gone in a few seconds". Others consistently struggled with braking and acceleration, leading to errors. According to P03, breaking and curves led to underestimation, with the participant thinking that he "(...) was making mistake because I was doing the gesture for selection before I was touching the target.", adding furthermore that he "(...) was just going backwards before I was really touching the target". Additionally, technical issues with hand tracking were reported, with tracking loss near the target causing aborted or incorrect selections.
The impact of vehicle movements, combined with physical strain and technical issues, suggests that improvements are needed to enhance the robustness and ergonomic comfort of this interaction method for vehicle-based applications.

The feedback on Gaze\&Pinch highlighted its ease of use and high selection speed, with participants appreciating the method's simplicity and precision. Many felt it allowed for quick, efficient interactions.
However, vehicle motion presented significant challenges, especially during vibrations, which caused participants to lose focus and make incorrect selections.

Bumps were the most disruptive, followed by braking and curves. Many participants struggled to compensate for vehicle movements using only their gaze, feeling that it was much harder than hand-based interactions. This could be due to movement of the glasses leading to decreased tracking quality during sections with high movements. P11 "(...) found it incredibly difficult to select the points. The headset often wobbled. When I looked at a point, but the car wobbled at the same time and so did the glasses, I immediately lost focus on this point again and still pinched, so I had a wrong selection every time." A recurring issue was the coordination between pre-selecting a target with gaze and confirming the selection with the pinch gesture. Several participants found this disconnect between eye and hand actions led to frequent errors, as their gaze would shift too quickly before the pinch gesture was completed.
Even during standstill, some participants reported eye-tracking inaccuracies, requiring them to adjust their gaze multiple times to ensure a target was properly highlighted. Overall, Gaze\&Pinch shows potential for fast and precise interaction, but its effectiveness diminishes significantly with vehicle movement and coordination challenges between gaze and pinch actions.

The feedback for the Handray interaction method highlighted its ease of use and precision, with P20 highlighting its familiarity during standstill due to "(...) the fact that you don't just point, but make two separate movements, I think that reminds me of clicking with the mouse, so it feels very familiar to me".
The biggest challenge during vehicle movement was handling the physical motion, followed by arm strain and issues with cursor smoothing. In the standstill condition, cursor smoothing was mentioned more often, with participants also reporting arm strain and feeling that the method was slow or physically demanding.

Some participants, however, felt that the difficulties were learnable, requiring small corrective movements to fine-tune the cursor's placement.
Bumps were the most problematic vehicle maneuver, followed by curves and braking. In these situations, participants found it harder to compensate for the physical motion. Here, P13 noted that "Especially in curves and when it was bumpy, I was much less accurate. There were several mistakes in a row because I wasn't able to compensate".
While bumps were hard to overcome, braking was seen as somewhat easier to correct for, as P13 stated that "When braking, I have the feeling that I can somehow compensate for this to some extent." Acceleration on the other hand was more difficult to predict or compensate for, as P20 described that "I don't think I have an instinctive feeling for how acceleration affects my movement. And that's why I can't compensate for this with the controls."
Nevertheless, a few participants felt that curves, acceleration, and braking had little influence on their performance, and some mentioned that vehicle movement did not impact the interaction method at all. Overall, while participants found the method intuitive and precise, vehicle movements posed a notable challenge, especially during bumps.

Regarding the HeadGaze method, participants found the method to be precise, especially in the standstill condition, with some describing the interaction as comfortable in standstill, though no one mentioned this during movement.
The biggest challenge during movement was the vehicle motion, which frequently disrupted participants' ability to maintain focus.
Neck strain was a common complaint in both conditions, with the weight of the headset and required head movements being additional issues at a standstill. Cursor smoothing was also a problem, particularly in the standstill condition.
Concerning vehicle maneuvers, bumps and curves were equally problematic, often disrupting the fixation of the cursor.

Regarding curves, P26 described the behavior as one where "(...) the fixation always flew out, corresponding to the curve" indicating that the forces exhibited during such maneuvers could not be compensated for. Adding to the influence of road bumps, P01 observed that "Especially when there was a bit more vertical travel, the cursor would sometimes hop across the entire screen", while humorously describing the method as awkward, likening it to "(...) trying to point at something with a wooden spoon taped to your forehead."
Breaking and acceleration also presented challenges, with P01 suggesting that the system might work better on smoother roads, like highways, but not in urban environments.
However, some participants felt that acceleration, braking, and bumps had little impact on their experience.
Overall, while the HeadGaze method was seen as easy to use and effective during standstill conditions, vehicle movement, neck strain, and physical discomfort posed significant challenges, particularly during more dynamic driving situations like bumps and curves.

\subsubsection{Privacy and Public Usage}

Participants’ perceptions of privacy and public usage of the four interaction methods varied significantly.
Gaze\&Pinch was considered the most comfortable and discreet interaction method when used around other passengers.
It required minimal noticeable movements, with only the pinch gesture being perceptible to others. One participant emphasized that it was "(...) by far the least noticeable (...)" and they felt "(...) totally comfortable (...)" using it in public settings (P10).

DirectTouch was deemed highly noticeable to others due to the large, conspicuous arm movements involved.
This made some participants uncomfortable, as they feared accidentally pointing at or invading the space of other passengers.
One participant expressed concerns about others avoiding them during interactions. Several participants mentioned feeling particularly uncomfortable using this method around unknown passengers.

Handray was considered uncomfortable and very noticeable because of the large arm movements required.
Participants felt this could invade others' personal space, noting that their movements might be distracting or disruptive. However, the movement was seen as natural by some, even if still very visible to those around them.

HeadGaze was viewed as the most noticeable interaction method since it required both head and hand movements, making it highly noticeable to others.
Despite this, one participant indicated they wouldn't feel uncomfortable using it in front of others, particularly during vehicle motion, when the movement might be less apparent due to natural head motion caused by driving. 

\begin{figure*}[h]
    \centering
    \includegraphics[width=0.8\linewidth]{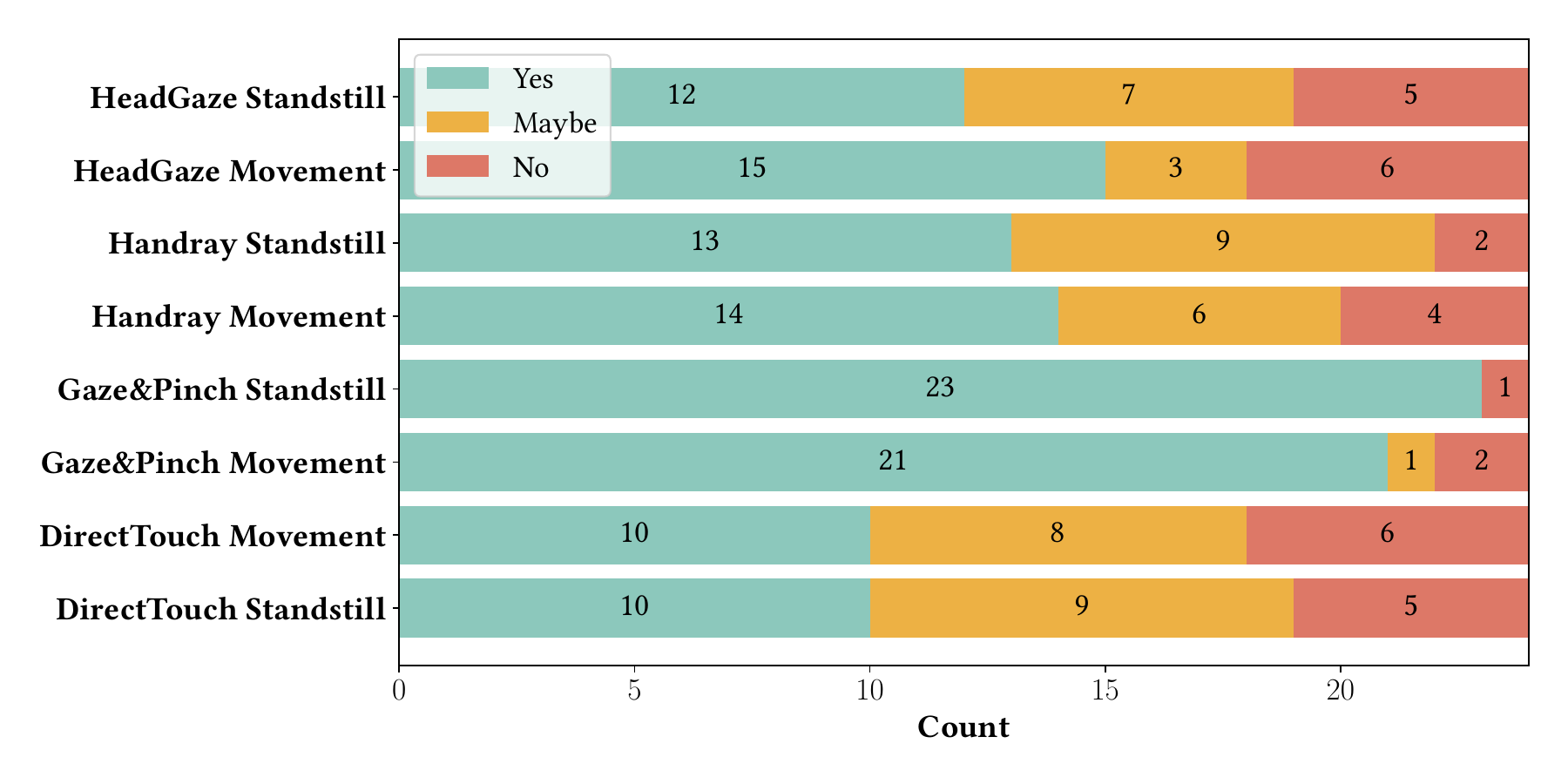}
    \caption{Preference to use a specific interaction method in a shared vehicle, based on participants' perceived comfort}
    \label{fig:comfortPublicUseMethods}
    \Description{This chart shows the user preference of the four interaction methods regarding the usage in contexts with other passengers. Each interaction method is presented in a new row, and is split into the order of preference (Yes, Maybe, No).}
\end{figure*}

\autoref{fig:comfortPublicUseMethods} provides a comprehensive overview of participants' willingness to use specific interaction methods in a shared vehicle, based on their perceived comfort.

\subsubsection{User Preferences}
During standstill, Gaze\&Pinch was perceived as the fastest and most effective method, offering high precision and comfort for the selection task during standstill usage. While this method did not elicit physical demands similar to the other three, its requirement of eye movement for selections resulted in an increased eye-strain compared to its counterparts.

DirectTouch was rated as very strenuous compared to other methods, but allowed participants to achieve a high performance due to its ease of use. It furthermore enabled participants an increased perception of the outside environment. Regarding the implementation of targets, participants mentioned a too high requirement of input precision, as the cursor had to enter and leave the target within its bounds to achieve a successful selection. This should be accommodated for, for example, by only requiring and triggering a selection on touch-down instead of touch-up.
HeadGaze was described as very slow during standstill, while Handray was described as causing arm strain due to its requirement of holding the hand in front of the headset to perform selections.

During vehicle motion, each of the presented interaction methods had its own set of limitations and challenges, influencing their individual ranking. 
The results convey increased user preference for Handray during movement as compared to standstill. Here, the amount of participants rating this interaction method as first preference increased fourfold, while no participants declared it to be their last preference anymore. Handray was described as a method requiring only low mental and physical workload while enabling target selection with high precision. The separation of pointing and selecting with the same extremity was furthermore highlighted as a positive aspect.
This contrasts Gaze\&Pinch, which popularity decreased during movement. While it received the largest count of votes as first preference (\N{10}) during movement, eight participants ranked it as their last preference, indicating that opinions are strongly divided regarding its usage. The method was rated positively due to the low physical workload exhibited. Furthermore, aspects like accuracy and ease of use were mentioned. However, some participants criticized the quality of eye-tracking, finding it imprecise and unstable during vehicle movements, requiring too much concentration.
DirectTouch and HeadGaze on the other hand persist similar user rankings across movement and standstill, featuring a numerical tie regarding the amount of mentions for first and last preference during movement.
While DirectTouch was perceived as intuitive and easy to use by participants, it required high physical effort, which could lead to arm strain. The method, therefore, was perceived as less comfortable than other methods. However, it enabled participants to interact with the system while simultaneously spectating the outside environment and therefore decoupled head movements from task input.

During movement, HeadGaze required a comparatively higher physical effort to compensate for vehicle motion than other methods. The method of steering the cursor by moving the head was perceived as strenuous and cumbersome by participants, with further physical effort being required by keeping the dominant hand in vision of the hand-tracking sensor as to perform the pinch gesture. However, some participants described the interaction method as comfortable but stressed the necessity of compensating vehicular motion to enable more precise input.

\begin{figure}[ht]
    \centering
    \begin{subfigure}[b]{\linewidth}
        \centering
        \includegraphics[width=\linewidth]{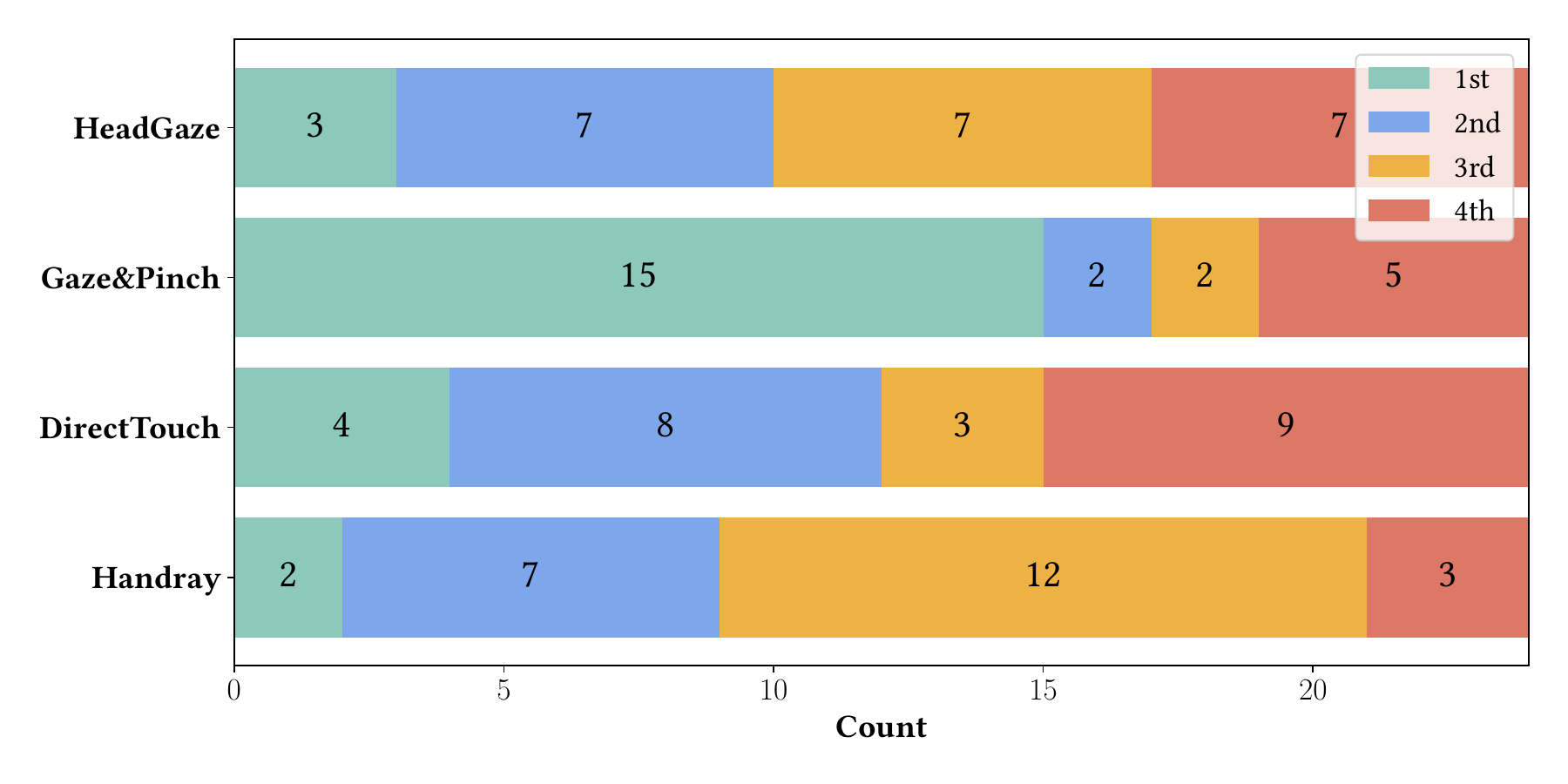}
        \caption{Ranking of interaction methods for Standstill Usage}      \label{fig:subjectiveMethodRankingStandstill}
        \Description{This chart shows the user preference of the four interaction methods for standstill usage. Each interaction method is presented in a new row, and is split into the order of preference (1st, 2nd, 3rd, 4th choice).}
    \end{subfigure}\hfill
    \begin{subfigure}[b]{\linewidth}
        \centering
        \includegraphics[width=\linewidth]{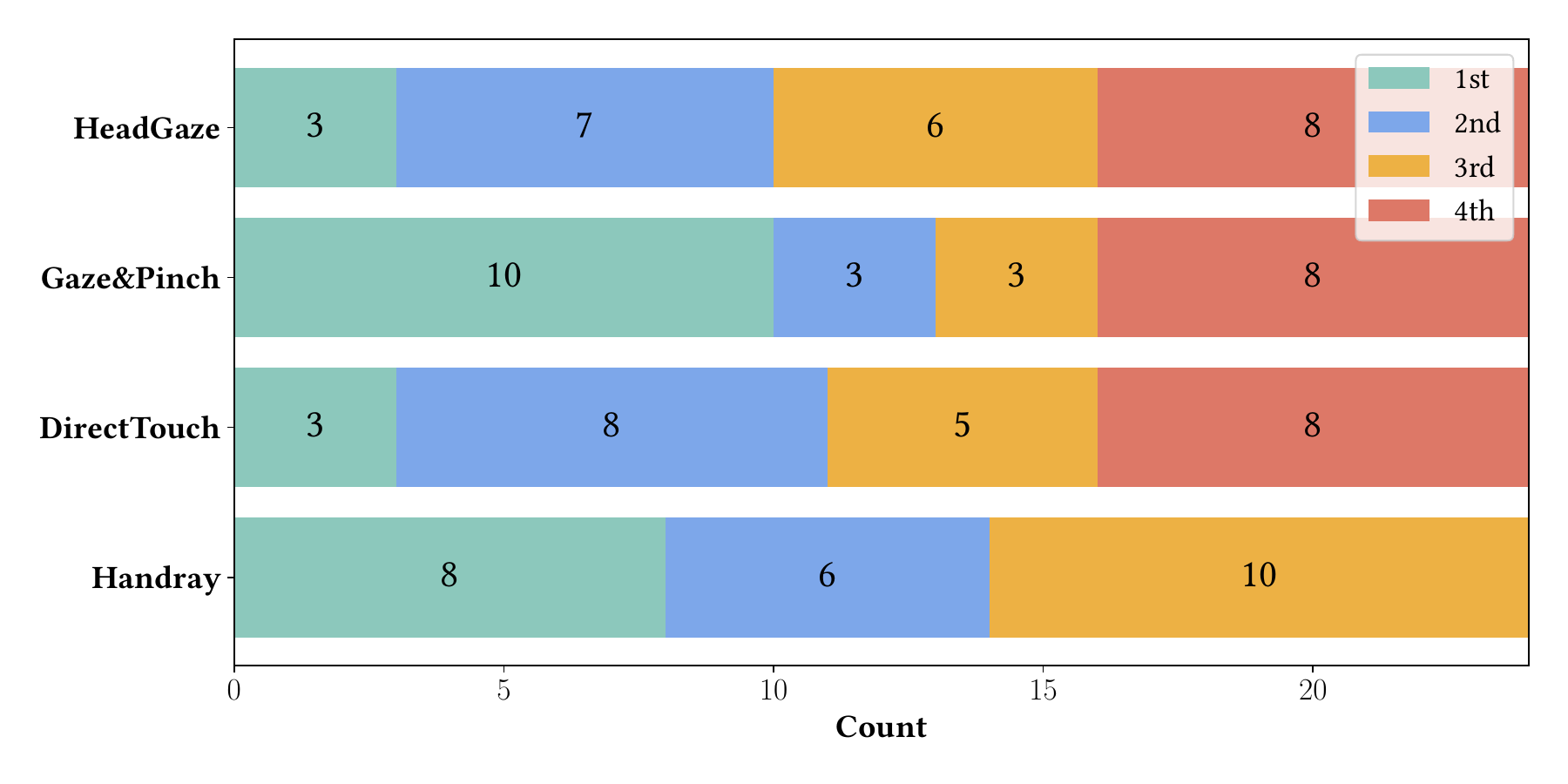}
        \caption{Ranking of interaction methods for Movement Usage}
        \label{fig:subjectiveMethodRankingMovement}
        \Description{This chart shows the user preference of the four interaction methods for movement usage. Each interaction method is presented in a new row, and is split into the order of preference (1st, 2nd, 3rd, 4th choice).}
    \end{subfigure}
    \caption{Ranking of interaction methods}
    \label{fig:subjectiveMethodRankingAll}
    \Description{Overview regarding user preferences of interaction methods. Split between standstill and movement.}
\end{figure}

A complete overview regarding the most and least preferred interaction method across conditions is visualized in \autoref{fig:subjectiveMethodRankingStandstill} for standstill and in \autoref{fig:subjectiveMethodRankingMovement} for movement. For improvements, participants suggested implementing a smoothing mechanism for the pointers of both, HeadGaze and DirectTouch, to improve input precision during vehicular movement.

\section{Limitations}
The headset, including the hand-tracking sensor and optical markers, weighed 1125g, which influenced the performance of interaction methods. 
Furthermore, external factors beyond our control, such as sunlight and varying ambient light conditions, may have influenced the tracking stability of the Ultraleap Leap Motion 2 controller. The eye-tracking system proved sensitive to vehicle vibrations, particularly pronounced on BumpyRoad. These vibrations may have caused relative movement between the HMD and participants' heads, reducing the quality of data collected during vehicle motion. As the shifting of HMDs is not a new observation, we expect this issue to occur for anyone using such eye-tracking systems in a moving context.
This highlights the need for more robust tracking systems to compensate for such external disturbances. As the study was performed under real-life conditions in a traffic-calmed but public environment, external factors such as traffic influenced driving behavior. In rare cases, for example, there were longer braking times at junctions to give way.
Finally, our participant pool was of moderate size, and results might not be generalizable as the group consisted of individuals employed by an automotive company, potentially resulting in an increased affinity towards AR and VR HMDs.

\section{Discussion}
We conducted a field study with 24 participants to investigate the impact of vehicular motion on task performance, perceived workload, usability, perceived safety, and trust in automation using an HMD and a Fitts' Law Task. Selections were performed using Gaze\&Pinch, DirectTouch, Handray, and HeadGaze during standstill and movement. Data was categorized into combinations of Interaction Method $\times$ Road Type $\times$ Curve Type, revealing effects on selection offsets and duration. User preferences and methods were assessed, with suggestions for future improvements.

\subsection{Impact on Usability, Perceived Workload and Task Performance --- \hyperref[rq1]{RQ1}} \label{c:disRQ1}

We found a significant impact of movement, leading to increased perceived workload (see \autoref{c:perceivedWorkloadArt}) and reduced task performance (see \autoref{c:FittsLawArt}), causing lower usability scores (see \autoref{c:usabilityArt}).
%%Usability
According to \citet{bangor2009sus}, our interaction methods featured \textit{good} (> 71.4) usability scores during standstill. However, usability scores decreased significantly during movement, resulting in the descriptive label \textit{ok} (> 50.9), except for Handray, whose usability score remained largely unaffected by vehicle movement. Since Handray featured the lowest error rate (see \autoref{fig:fitts_law_error}) and the best precision values (see \autoref{fig:ie_gap}) during movement, we assume cursor smoothing to be an important factor in maintaining usability scores. Although participants referred to smoothing as cause for perceivable slow interactions (see \autoref{fig:fitts_law_mt} and \autoref{c:interviewInfluenceRoadConditions}), its usage resulted in consistent trajectories (see \autoref{fig:trajectoryOverview}) during otherwise highly perturbed movements, thus allowing precise and intuitive selections (see \autoref{c:interviewInfluenceRoadConditions}).

\citet{schramm2023AssessingAugmentedReality} performed a similar study, but their implementation of HeadGaze and Eye-Gaze required a hardware button to confirm the selection rather than a pinch gesture. Compared to \citet{schramm2023AssessingAugmentedReality}, all interaction methods but Handray received lower usability scores. We assume this to be due to our task requiring more selections in a short time, which led to increased time pressure. In the study by \citet{schramm2023AssessingAugmentedReality} participants had five seconds per target, and breaks between the 70 horizontally arranged targets to be selected. In contrast, our participants had to perform 308 correct selections as quickly and accurately as possible while targets were arranged circularly, requiring diagonal movements. In addition, our course featured mixed and bumpy road segments, increasing complexity.
Similar to usability scores, the impact of vehicle movement on total workload score is more pronounced for HeadGaze and Gaze\&Pinch, while values for DirectTouch and Handray only increase marginally (see \autoref{tab:nasartlxTotalScore}). 

Based on findings by \citet{mayerEffectRoadBumps2018}, \citet{colleySwiVRCarSeatExploringVehicle2021} and \citet{ahmadTouchscreenUsabilityInput2015} 
we expected a stronger impact of external forces on precision and physical demand of the free hand pointing methods DirectTouch and Handray. %as?
%DirectTouch
However, contrary to \citet{colleySwiVRCarSeatExploringVehicle2021}, DirectTouch showed no relation between high physical demand and decreased accuracy during movement (see \autoref{fig:ie_gap}), despite requiring a stretched arm to reach targets. Nonetheless, this requirement elicited significantly higher physical demand than all other interaction methods, thus causing Gorilla Arm fatigue~\cite{hincapie-ramos2014ConsumedEnduranceMetric}. The comparably small impact of movement also becomes apparent in the exemplified trajectories (see \autoref{fig:trajectoryOverview}), containing only occasional incorrect selections during BumpyRoad or Short-Left Curve with Acceleration.
%Handray
Compared to DirectTouch, Handray exhibited significantly less physical demand, required less arm movement (see \autoref{c:handPalmEuc}), and allowed participants to keep the arm closer to their body, increasing stability. This minimized the Gorilla Arm effect in comparison to DirectTouch \cite{mutasim2021PinchClickDwell, hincapie-ramos2014ConsumedEnduranceMetric}. Interestingly, Handray had the second largest decrease in precision during movement. We assume this is due to difficulties in predicting how acceleration affects hand movements and the smoothing behavior necessitating corrective movements for accurate selections (see \autoref{c:interviewInfluenceRoadConditions}).
%HeadGaze
For HeadGaze, participants negatively highlighted the requirement for constant head movement, which caused neck strain. Here, HeadGaze featured the highest value for traversed distance in both movement conditions, with movement featuring an increased value (see \autoref{fig:eucDist_Head}).
%GazePinch
Gaze\&Pinch during movement elicited a higher total workload than during standstill. In line with \citet{colleySwiVRCarSeatExploringVehicle2021}, Gaze\&Pinch requires significantly less physical demand than all other interaction methods while causing increased eye strain. In line with \citet{blattgersteAdvantagesEyegazeHeadgazebased2018}, there was less head movement required for Gaze\&Pinch than for HeadGaze. Furthermore, interesting to note is that Gaze\&Pinch requires the lowest head movement across all conditions during standstill, while also containing the largest increase in head movement from standstill to movement across interaction methods. This could be linked to issues with the eye-tracking system (see \autoref{c:dis_challenges_improvements}), causing participants to more often try and support the eye-gaze selections with additional head movements~\cite{sidenmark2019EyeHeadSynergetic}.
%Througput
Regarding Fitts' Law throughput, our results for Handray and Gaze\&Pinch during standstill are similar to \citet{wagnerFittsLawStudy2023}, but our interaction methods achieved higher throughput. Movement features significantly reduced throughput across interaction methods, with DirectTouch offering the highest values, followed by Handray, Gaze\&Pinch, and HeadGaze (see \autoref{fig:fitts_law_throughput}). Handray featured higher throughput during movement (1.87 bits/s) than the overall equivalent of \citet{wagnerFittsLawStudy2023} during standstill (1.39 bits/s).
Furthermore, movement led to increased erroneous selections for all interaction methods.
In line with \citet{colleySwiVRCarSeatExploringVehicle2021}, Gaze\&Pinch featured the highest error rate across movement conditions.
Although our implementation performed better than the one of \citet{colleySwiVRCarSeatExploringVehicle2021}, it did not outperform HeadGaze and thus opposes findings by \citet{blattgersteAdvantagesEyegazeHeadgazebased2018}. 
Similar to our results, \citet{schramm2023AssessingAugmentedReality} found that HeadGaze resulted in low error rates. However, they identified the highest error rate for their implementation of Handray, while our implementation featured the lowest error rate during movement. An influencing factor for their increased error rate was the arrangement of targets in \citet{schramm2023AssessingAugmentedReality}, as they mention right-handed participants hitting their hand against the vehicle door while trying to perform selections, resulting in failed interactions.

\subsection{Variance Across Vehicle Movements --- \hyperref[rq2]{RQ2}} 
\label{c:disVarianceAcrossVehicleMovements}

Interaction methods generally performed better during standstill than during movement (e.g., see \autoref{fig:fitts_law_throughput} and \autoref{fig:ie_gap}), which was to be expected as external forces were not present \cite{colleySwiVRCarSeatExploringVehicle2021}. 
While \citet{ahmadTouchscreenUsabilityInput2015, ahmadInteractiveDisplaysVehicles2014} investigated the influence of varying road conditions, and \citet{mayerEffectRoadBumps2018} focused on simulated road bumps, we extended these approaches by performing a field study in standstill and in movement on three road conditions (SmoothRoad, MixedRoad, BumpyRoad), additionally introducing three curve types (see \autoref{tab:typesRoadCurveManeuver}) which were driven with and without increasing acceleration.
Our findings align with previous works, as movement significantly impacted factors like selection offset \cite{ahmadTouchscreenUsabilityInput2015, colleySwiVRCarSeatExploringVehicle2021, mayerEffectRoadBumps2018}, and error rate \cite{ahmadInteractiveDisplaysVehicles2014, ahmadTouchscreenUsabilityInput2015,colleySwiVRCarSeatExploringVehicle2021}.
While we did not investigate individual road bumps as performed by \citet{mayerEffectRoadBumps2018}, our findings are in line with \citet{ahmadTouchscreenUsabilityInput2015, mayerEffectRoadBumps2018} as they found selection offset to increase as a result of road bumps.
BumpyRoad caused either multiple retries per selection (Gaze\&Pinch, \autoref{fig:gp_bumpy_road}), inaccuracies nearby targets (DirectTouch, \autoref{fig:dt_bumpy_road}), jerky directional changes in combination with nearly overshooting (HeadGaze, \autoref{fig:hg_bumpy_road}), or could lead to undershooting (Handray, \autoref{fig:hg_bumpy_road}). Effects were most strongly pronounced for HeadGaze, followed by Gaze\&Pinch, and Handray (\autoref{fig:movement}). We assume that BumpyRoad led to decreased eye-tracking quality for Gaze\&Pinch, as participants reported losing focus as the HMD shifted on their heads during vibrations (see \autoref{c:interviewInfluenceRoadConditions}). This occurred even though we instructed participants to tighten the headset firmly but comfortably on their head, indicating two aspects: 1) We assume that the HMD weight was too large to keep it in a stable position during the influence of road bumps, 2) as there is a calibration procedure for a fixed eye-position, employed eye-tracking algorithms might not be optimized for sudden changes in eye-position in relation to the tracking cameras. We further assume that the HMD weight significantly contributed to HeadGaze being the most affected interaction method, making it difficult for participants to compensate the forces exhibited on their neck muscles, leading to neck strain (see \autoref{c:interviewInfluenceRoadConditions}). Even though participants could stabilize their hand with the body while using Handray, they reported the inability to compensate for sudden movements like road bumps (see \autoref{c:interviewInfluenceRoadConditions}), causing decreased accuracy. The precision of DirectTouch was only significantly influenced by MixedRoad, aligning with previous findings that it is only slightly impacted by movement (see \autoref{c:disRQ1}). Adding to previous works~\cite{colleySwiVRCarSeatExploringVehicle2021,ahmadTouchscreenUsabilityInput2015,ahmadInteractiveDisplaysVehicles2014,mayerEffectRoadBumps2018}, we found significant effects for combinations of Short-Curves with Acceleration on BumpyRoad (Handray, HeadGaze), Short-Curves with Accelerations on MixedRoad (Gaze\&Pinch), Short-Curves with Acceleration (DirectTouch, HeadGaze), and Long-Left-Curves with Acceleration (DirectTouch) on precision. Since Short-Curves with Acceleration are the most occurring curve type, we assume that rapid and brief lateral accelerations are responsible for reduced precision. This can be observed in \autoref{fig:trajectoryOverview}, as only the paths between a few targets are affected per curve.

While \citet{ahmadTouchscreenUsabilityInput2015} found increasing variability in task completion times with the amount of noise exhibited by the road profile, we identified this only for curve types. Additionally, all curve types, except for Long-Right-Curve Steady, resulted in significantly increased selection times. Long-Curves with Steady Acceleration primarily affected HeadGaze, Handray, and DirectTouch, with the latter being the only one affected by Long-Curves with Acceleration. Short-Curves with Acceleration impact DirectTouch, and in combination with MixedRoad also HeadGaze and Handray. We expect that selection times are extended by the time required to compensate exhibited lateral accelerations, something participants struggled with (see \autoref{c:interviewInfluenceRoadConditions}). We found rather unexpected results in improved selection times for Gaze\&Pinch and DirectTouch on MixedRoad, and for HeadGaze and Gaze\&Pinch on BumpyRoad, both during Short-Curves with Acceleration. This behavior could have emerged from participants trying to quickly correct previous selection errors. Furthermore, while \citet{mayerEffectRoadBumps2018} found no significant effects of bumps on selection times, we could identify significant influences for DirectTouch and the aforementioned combinations of curve types with Mixed- and BumpyRoad.

Overall, significant findings in \autoref{fig:movement} showed that primarily lateral accelerations influence selection times, while precision was impacted by a more balanced mixture of lateral and vertical accelerations.

\subsection{Improving Interaction Methods --- \hyperref[rq3]{RQ3}} \label{c:dis_challenges_improvements}

While we employed a state-of-the-art sensor (Ultraleap Leap Motion 2) mounted with a 15° downward angle to increase hand visibility, technical issues with hand tracking were reported for all interaction methods, especially for DirectTouch and Gaze\&Pinch. Direct sunlight and infrared light emitted by the utilized Optitrack V120:DUO camera can influence the tracking stability of the Ultraleap Leap Motion 2~\cite{Ultraleap.2024}. This could scarcely result in tracking loss near the targets during DirectTouch, leading to erroneous cursor placement and causing incorrect selections or prohibiting selections.

The system registered unintentional pinch gestures, resulting in multiple selections in a very short interval (see \autoref{c:dataPrep}). This can be due to issues with the \textit{Ultraleap Leap Motion Controller 2} and the threshold defined within the Unity XRI. The gesture requires participants to maintain a pre-defined distance between the thumb and index tip, then bring them together and move them apart to perform the pinch gesture. We suspect that participants could not consistently maintain this distance for the study. As a result, small finger movements could have been sufficient to trigger a selection unconsciously. Similar behavior was also identified by \citet{pfeuffer2017GazePinchInteraction}, where participants brought their hands into a comfortable position, resulting in unintentional selections. Furthermore, the probability of such erroneous selections was increased by vehicle motion (see \autoref{c:filteredSelectionCount} and \autoref{c:errorRate}), presumably as it affects the movement of the body and limbs. Another possible reason for a failed pinch gesture could be that the fingers were not moved far enough from each other after being brought together successfully~\cite{mutasim2021PinchClickDwell}. Furthermore, while this was not mentioned by participants, they had to keep their hands in a position visible to the sensor, a technical limitation that could also have led to unregistered selections during the study. Until such challenges are resolved, a hardware button could provide a viable alternative, offering similar performance while reducing frustration and recognition errors~\cite{mutasim2021PinchClickDwell}.

Regarding Gaze\&Pinch, a recurring theme (see \autoref{c:interviewInfluenceRoadConditions}) was the coordination between pre-selecting a target with eye-gaze and performing the pinch gesture. To mitigate this issue, a short delay or snapping mechanism could be introduced to ensure the target stays active longer and allows the user to react in time. \citet{pfeuffer2024DesignPrinciplesChallenges} suggest using the last fixation and stabilizing the gaze for 200 - 300ms.

While participants reported the cursor smoothing of Handray to be slow and thus challenging to perform selections with (see \autoref{c:interviewInfluenceRoadConditions}), largely unaffected trajectories (see \autoref{fig:trajectoryOverview}) along with highest precision during movement (see \autoref{c:selectionOffsetArt}) indicate that it helped mitigate the influence of accelerations. As Handray was further characterized as easy to use and precise, we suggest keeping the initial concept of the cursor smoothing. However, the degree of smoothing could be dynamically adjusted based on measured accelerations. For example, a low smoothing value could be employed during SmoothRoad, possibly resulting in reduced selection durations and increased throughput. When erratic movements occur (e.g., road bumps), the smoothing value could be increased to stabilize cursor positions. Such behavior could also be adapted for other interaction methods, as suggested by participants regarding HeadGaze and DirectTouch (see \autoref{c:interviewInfluenceRoadConditions}).

\subsection{Practical Implications and Guidelines}

\subsubsection{Which interaction method should I implement?}

Based on our findings, we recommend Handray for selection tasks during movement. While it could lead to arm strain, it featured the lowest error rate and highest precision and was preferred by participants. Handray removes distance constraints regarding positioning digital content as it utilizes a raycast for pointing. However, it might be unsuitable for public transport, as it could involve invading others’ personal space. We also highlight Gaze\&Pinch as a possible alternative in the future because participants rated it as their second preference. This was interesting as it featured the highest error rate but compelled participants through low physical demand and simplicity when it worked well. It is the most accepted interaction method when other passengers are present. For its usage, issues with eye-tracking in highly perturbed situations must first be resolved.

\subsubsection{Movement Considerations for Future Research}

Our findings highlight the importance of including curves in experimental designs due to their impact on selection times and precision. Short-Curves with Acceleration were the most common type, significantly impacting precision and duration values (see \autoref{c:disVarianceAcrossVehicleMovements}).
We also recommend performing interactions during a standstill, or at least on a straight SmoothRoad, to assess the impact of movement precisely.

\subsubsection{Motion Fidelity Considerations}
Interaction methods performed significantly worse during movement than standstill, highlighting the importance of motion. Previous studies investigated interactions during motion using 1-DoF \cite{colleySwiVRCarSeatExploringVehicle2021} and 6-DoF \cite{mayerEffectRoadBumps2018} simulators, or used a real vehicle \cite{ahmadTouchscreenUsabilityInput2015, ahmadInteractiveDisplaysVehicles2014,schramm2023AssessingAugmentedReality}. While we found effects on error rate and selection offset, which were also found for simulator-based studies \cite{colleySwiVRCarSeatExploringVehicle2021, mayerEffectRoadBumps2018}, significant impact of individual road bumps on selection times obtained in a 6-DoF simulator did not align with our results for BumpyRoad \cite{mayerEffectRoadBumps2018}.
Furthermore, no study investigated lateral acceleration in curves. Thus, we can not estimate how suitable motion platforms are.

Nonetheless, previous studies showed that motion simulators can provide comparable results, enabling a lower entry barrier. We argue that introducing any movement is better than including none, as external factors are hard to predict, with results hardly being transferrable from standstill to movement. 

As it was difficult to compare our work to the few previous studies, each with their implementation of interaction methods, we make our configuration for the interaction methods available, hoping to improve reproducibility.

\section{Conclusion and Future Work}

We conducted a field study with 24 participants on a course with varying road types (SmoothRoad, BumpyRoad, MixedRoad) and curve types (Short-Curves with increasing acceleration, Long-Curves with increasing and steady acceleration). We investigated the impact of movement on DirectTouch, Handray, Gaze\&Pinch, and HeadGaze using an HMD and a Fitts' Law Task. During the study, all interactions and movements of participants were recorded along with vehicle movements, and automatically labeled into road and curve types. We used this dataset to precisely analyze which type of movement significantly affected the accuracy and selection time of each interaction method. Interaction methods generally performed worse during movement, compared to standstill. We furthermore identified that each interaction method was affected by movement in a different way. For example, the NASA-rTLX total workload of HeadGaze and Gaze\&Pinch were affected stronger by movement than DirectTouch and Handray. Furthermore task performance, accuracy, and selection time were negatively influenced. Based on our findings we presented practical implications and guidelines, among which we recommend the usage of Handray for selection tasks during movement.

We plan on further evaluating the influence of movement on pointing trajectories, focusing on occurrences of over- and undershooting. As we aim to reduce erroneous selections, we plan to evaluate whether prediction approaches similar to \citet{ahmadInteractiveDisplaysVehicles2014} and \citet{mayerEffectRoadBumps2018} can be used in conjunction with our task and interaction methods.

\section*{Open Science}
Upon acceptance, the source code and the analysis will be released \href{https://github.com/m-sasalovici/BumpyRide-Understanding}{here}.

\begin{acks}
The authors thank Stephan Leenders, Daniel Keßelheim, Okan Sadik Koese, and Lasse Frommelt for their support regarding the technical implementation and the insightful discussions that helped shape this work. We furthermore thank Tuuba Sejfuli and our reviewers for their valuable feedback and comments that helped improve the manuscript. Finally, we sincerely thank all participants who took their time to participate in our study. Your involvement was invaluable to our research.
\end{acks}

\bibliographystyle{ACM-Reference-Format}
\bibliography{bibliography}


\begin{thebibliography}{91}

%%% the Web address.

\ifx \showCODEN    \undefined \def \showCODEN     #1{\unskip}     \fi
\ifx \showISBNx    \undefined \def \showISBNx     #1{\unskip}     \fi
\ifx \showISBNxiii \undefined \def \showISBNxiii  #1{\unskip}     \fi
\ifx \showISSN     \undefined \def \showISSN      #1{\unskip}     \fi
\ifx \showLCCN     \undefined \def \showLCCN      #1{\unskip}     \fi
\ifx \shownote     \undefined \def \shownote      #1{#1}          \fi
\ifx \showarticletitle \undefined \def \showarticletitle #1{#1}   \fi
\ifx \showURL      \undefined \def \showURL       {\relax}        \fi
% invisible to TeX
\providecommand\bibfield[2]{#2}
\providecommand\bibinfo[2]{#2}
\providecommand\natexlab[1]{#1}
\providecommand\showeprint[2][]{arXiv:#2}

\bibitem[AG(2022)]%
        {AUDIAG.2022}
\bibfield{author}{\bibinfo{person}{AUDI AG}.} \bibinfo{year}{2022}\natexlab{}.
\newblock \bibinfo{title}{holoride: Virtual Reality meets the real world}.
\newblock
\urldef\tempurl%
\url{https://www.audi.com/en/innovation/development/holoride-virtual-reality-meets-the-real-world.html}
\showURL{%
\tempurl}
\newblock
\shownote{Accessed: 09.11.2023}.


\bibitem[Ahmad et~al\mbox{.}(2014)]%
        {ahmadInteractiveDisplaysVehicles2014}
\bibfield{author}{\bibinfo{person}{Bashar~I. Ahmad}, \bibinfo{person}{Patrick~M. Langdon}, \bibinfo{person}{Simon~J. Godsill}, \bibinfo{person}{Robert Hardy}, \bibinfo{person}{Eduardo Dias}, {and} \bibinfo{person}{Lee Skrypchuk}.} \bibinfo{year}{2014}\natexlab{}.
\newblock \showarticletitle{Interactive {{Displays}} in {{Vehicles}}: {{Improving Usability}} with a {{Pointing Gesture Tracker}} and {{Bayesian Intent Predictors}}}. In \bibinfo{booktitle}{\emph{Proceedings of the 6th {{International Conference}} on {{Automotive User Interfaces}} and {{Interactive Vehicular Applications}}}} ({Seattle WA USA}, 2014-09-17). \bibinfo{publisher}{ACM}, \bibinfo{address}{New York, NY, USA}, \bibinfo{pages}{1--8}.
\newblock
\showISBNx{978-1-4503-3212-5}
\href{https://doi.org/10.1145/2667317.2667413}{doi:\nolinkurl{10.1145/2667317.2667413}}


\bibitem[Ahmad et~al\mbox{.}(2015)]%
        {ahmadTouchscreenUsabilityInput2015}
\bibfield{author}{\bibinfo{person}{Bashar~I. Ahmad}, \bibinfo{person}{Patrick~M. Langdon}, \bibinfo{person}{Simon~J. Godsill}, \bibinfo{person}{Robert Hardy}, \bibinfo{person}{Lee Skrypchuk}, {and} \bibinfo{person}{Richard Donkor}.} \bibinfo{year}{2015}\natexlab{}.
\newblock \showarticletitle{Touchscreen Usability and Input Performance in Vehicles under Different Road Conditions: An Evaluative Study}. In \bibinfo{booktitle}{\emph{Proceedings of the 7th {{International Conference}} on {{Automotive User Interfaces}} and {{Interactive Vehicular Applications}}}} ({Nottingham United Kingdom}, 2015-09). \bibinfo{publisher}{{ACM}}, \bibinfo{address}{New York, NY, USA}, \bibinfo{pages}{47--54}.
\newblock
\showISBNx{978-1-4503-3736-6}
\href{https://doi.org/10.1145/2799250.2799284}{doi:\nolinkurl{10.1145/2799250.2799284}}


\bibitem[Ariza et~al\mbox{.}(2018)]%
        {arizaProximity2018}
\bibfield{author}{\bibinfo{person}{Oscar Ariza}, \bibinfo{person}{Gerd Bruder}, \bibinfo{person}{Nicholas Katzakis}, {and} \bibinfo{person}{Frank Steinicke}.} \bibinfo{year}{2018}\natexlab{}.
\newblock \showarticletitle{Analysis of Proximity-Based Multimodal Feedback for 3D Selection in Immersive Virtual Environments}. In \bibinfo{booktitle}{\emph{2018 IEEE Conference on Virtual Reality and 3D User Interfaces (VR)}}. \bibinfo{pages}{327--334}.
\newblock
\href{https://doi.org/10.1109/VR.2018.8446317}{doi:\nolinkurl{10.1109/VR.2018.8446317}}


\bibitem[Aslan et~al\mbox{.}(2015)]%
        {aslanLeapTouchProximity2015}
\bibfield{author}{\bibinfo{person}{Ilhan Aslan}, \bibinfo{person}{Alina Krischkowsky}, \bibinfo{person}{Alexander Meschtscherjakov}, \bibinfo{person}{Martin Wuchse}, {and} \bibinfo{person}{Manfred Tscheligi}.} \bibinfo{year}{2015}\natexlab{}.
\newblock \showarticletitle{A leap for touch: proximity sensitive touch targets in cars}. In \bibinfo{booktitle}{\emph{Proceedings of the 7th International Conference on Automotive User Interfaces and Interactive Vehicular Applications}} (Nottingham, United Kingdom) \emph{(\bibinfo{series}{AutomotiveUI '15})}. \bibinfo{publisher}{Association for Computing Machinery}, \bibinfo{address}{New York, NY, USA}, \bibinfo{pages}{39–46}.
\newblock
\showISBNx{9781450337366}
\href{https://doi.org/10.1145/2799250.2799273}{doi:\nolinkurl{10.1145/2799250.2799273}}


\bibitem[Ataya et~al\mbox{.}(2021)]%
        {ataya_how_2021}
\bibfield{author}{\bibinfo{person}{Aya Ataya}, \bibinfo{person}{Won Kim}, \bibinfo{person}{Ahmed Elsharkawy}, {and} \bibinfo{person}{SeungJun Kim}.} \bibinfo{year}{2021}\natexlab{}.
\newblock \showarticletitle{How to {Interact} with a {Fully} {Autonomous} {Vehicle}: {Naturalistic} {Ways} for {Drivers} to {Intervene} in the {Vehicle} {System} {While} {Performing} {Non}-{Driving} {Related} {Tasks}}.
\newblock \bibinfo{journal}{\emph{Sensors}} \bibinfo{volume}{21}, \bibinfo{number}{6} (\bibinfo{date}{Jan.} \bibinfo{year}{2021}), \bibinfo{pages}{2206}.
\newblock
\href{https://doi.org/10.3390/s21062206}{doi:\nolinkurl{10.3390/s21062206}}
\newblock
\shownote{Number: 6 Publisher: Multidisciplinary Digital Publishing Institute}.


\bibitem[B.~Adhanom et~al\mbox{.}(2020)]%
        {b.adhanom2020GazeMetricsOpenSourceTool}
\bibfield{author}{\bibinfo{person}{Isayas B.~Adhanom}, \bibinfo{person}{Samantha~C. Lee}, \bibinfo{person}{Eelke Folmer}, {and} \bibinfo{person}{Paul MacNeilage}.} \bibinfo{year}{2020}\natexlab{}.
\newblock \showarticletitle{{{GazeMetrics}}: {{An Open-Source Tool}} for {{Measuring}} the {{Data Quality}} of {{HMD-based Eye Trackers}}}. In \bibinfo{booktitle}{\emph{{{ACM Symposium}} on {{Eye Tracking Research}} and {{Applications}}}}. \bibinfo{publisher}{ACM}, \bibinfo{address}{Stuttgart Germany}, \bibinfo{pages}{1--5}.
\newblock
\showISBNx{978-1-4503-7134-6}
\href{https://doi.org/10.1145/3379156.3391374}{doi:\nolinkurl{10.1145/3379156.3391374}}


\bibitem[Bach(2022)]%
        {Bach.2022}
\bibfield{author}{\bibinfo{person}{Deborah Bach}.} \bibinfo{year}{2022}\natexlab{}.
\newblock \bibinfo{title}{With their HoloLens 2 project, Microsoft and Volkswagen collaborate to put augmented reality glasses in motion}.
\newblock
\urldef\tempurl%
\url{https://news.microsoft.com/source/features/digital-transformation/with-their-hololens-2-project-microsoft-and-volkswagen-collaborate-to-put-augmented-reality-glasses-in-motion/}
\showURL{%
\tempurl}
\newblock
\shownote{Accessed: 09.11.2023}.


\bibitem[Baltodano et~al\mbox{.}(2015)]%
        {baltodano_rrads_2015}
\bibfield{author}{\bibinfo{person}{Sonia Baltodano}, \bibinfo{person}{Srinath Sibi}, \bibinfo{person}{Nikolas Martelaro}, \bibinfo{person}{Nikhil Gowda}, {and} \bibinfo{person}{Wendy Ju}.} \bibinfo{year}{2015}\natexlab{}.
\newblock \showarticletitle{The {RRADS} platform: a real road autonomous driving simulator}. In \bibinfo{booktitle}{\emph{Proceedings of the 7th {International} {Conference} on {Automotive} {User} {Interfaces} and {Interactive} {Vehicular} {Applications}}} \emph{(\bibinfo{series}{{AutomotiveUI} '15})}. \bibinfo{publisher}{Association for Computing Machinery}, \bibinfo{address}{New York, NY, USA}, \bibinfo{pages}{281--288}.
\newblock
\showISBNx{978-1-4503-3736-6}
\href{https://doi.org/10.1145/2799250.2799288}{doi:\nolinkurl{10.1145/2799250.2799288}}


\bibitem[Bangor et~al\mbox{.}(2009)]%
        {bangor2009sus}
\bibfield{author}{\bibinfo{person}{Aaron Bangor}, \bibinfo{person}{Philip Kortum}, {and} \bibinfo{person}{James Miller}.} \bibinfo{year}{2009}\natexlab{}.
\newblock \showarticletitle{Determining What Individual SUS Scores Mean: Adding an Adjective Rating Scale}.
\newblock \bibinfo{journal}{\emph{J. Usability Studies}} \bibinfo{volume}{4}, \bibinfo{number}{3} (\bibinfo{date}{may} \bibinfo{year}{2009}), \bibinfo{pages}{114–123}.
\newblock


\bibitem[Batmaz and Stuerzlinger(2022)]%
        {batmaz2022EffectiveThroughputAnalysis}
\bibfield{author}{\bibinfo{person}{Anil~Ufuk Batmaz} {and} \bibinfo{person}{Wolfgang Stuerzlinger}.} \bibinfo{year}{2022}\natexlab{}.
\newblock \showarticletitle{Effective {{Throughput Analysis}} of {{Different Task Execution Strategies}} for {{Mid-Air Fitts}}' {{Tasks}} in {{Virtual Reality}}}.
\newblock \bibinfo{journal}{\emph{IEEE Transactions on Visualization and Computer Graphics}} \bibinfo{volume}{28}, \bibinfo{number}{11} (\bibinfo{date}{Nov.} \bibinfo{year}{2022}), \bibinfo{pages}{3939--3947}.
\newblock
\showISSN{1077-2626, 1941-0506, 2160-9306}
\href{https://doi.org/10.1109/TVCG.2022.3203105}{doi:\nolinkurl{10.1109/TVCG.2022.3203105}}


\bibitem[Bengler et~al\mbox{.}(2020)]%
        {bengler_hmi_2020}
\bibfield{author}{\bibinfo{person}{Klaus Bengler}, \bibinfo{person}{Michael Rettenmaier}, \bibinfo{person}{Nicole Fritz}, {and} \bibinfo{person}{Alexander Feierle}.} \bibinfo{year}{2020}\natexlab{}.
\newblock \showarticletitle{From {HMI} to {HMIs}: {Towards} an {HMI} {Framework} for {Automated} {Driving}}.
\newblock \bibinfo{journal}{\emph{Information}} \bibinfo{volume}{11}, \bibinfo{number}{2} (\bibinfo{date}{Feb.} \bibinfo{year}{2020}), \bibinfo{pages}{61}.
\newblock
\href{https://doi.org/10.3390/info11020061}{doi:\nolinkurl{10.3390/info11020061}}
\newblock
\shownote{Number: 2 Publisher: Multidisciplinary Digital Publishing Institute}.


\bibitem[Bergstr{\"o}m et~al\mbox{.}(2021)]%
        {bergstrom2021HowEvaluateObject}
\bibfield{author}{\bibinfo{person}{Joanna Bergstr{\"o}m}, \bibinfo{person}{Tor-Salve Dalsgaard}, \bibinfo{person}{Jason Alexander}, {and} \bibinfo{person}{Kasper Hornb{\ae}k}.} \bibinfo{year}{2021}\natexlab{}.
\newblock \showarticletitle{How to {{Evaluate Object Selection}} and {{Manipulation}} in {{VR}}? {{Guidelines}} from 20 {{Years}} of {{Studies}}}. In \bibinfo{booktitle}{\emph{Proceedings of the 2021 {{CHI Conference}} on {{Human Factors}} in {{Computing Systems}}}}. \bibinfo{publisher}{ACM}, \bibinfo{address}{Yokohama Japan}, \bibinfo{pages}{1--20}.
\newblock
\showISBNx{978-1-4503-8096-6}
\href{https://doi.org/10.1145/3411764.3445193}{doi:\nolinkurl{10.1145/3411764.3445193}}


\bibitem[Blattgerste et~al\mbox{.}(2018)]%
        {blattgersteAdvantagesEyegazeHeadgazebased2018}
\bibfield{author}{\bibinfo{person}{Jonas Blattgerste}, \bibinfo{person}{Patrick Renner}, {and} \bibinfo{person}{Thies Pfeiffer}.} \bibinfo{year}{2018}\natexlab{}.
\newblock \showarticletitle{Advantages of eye-gaze over head-gaze-based selection in virtual and augmented reality under varying field of views}. In \bibinfo{booktitle}{\emph{Proceedings of the Workshop on Communication by Gaze Interaction}} (Warsaw, Poland) \emph{(\bibinfo{series}{COGAIN '18})}. \bibinfo{publisher}{Association for Computing Machinery}, \bibinfo{address}{New York, NY, USA}, Article \bibinfo{articleno}{1}, \bibinfo{numpages}{9}~pages.
\newblock
\showISBNx{9781450357906}
\href{https://doi.org/10.1145/3206343.3206349}{doi:\nolinkurl{10.1145/3206343.3206349}}


\bibitem[Bos et~al\mbox{.}(2005)]%
        {Bos.2005}
\bibfield{author}{\bibinfo{person}{Jelte~E. Bos}, \bibinfo{person}{Scott~N. MacKinnon}, {and} \bibinfo{person}{Anthony Patterson}.} \bibinfo{year}{2005}\natexlab{}.
\newblock \showarticletitle{Motion sickness symptoms in a ship motion simulator: effects of inside, outside, and no view}.
\newblock \bibinfo{journal}{\emph{Aviation, Space, and Environmental Medicine}} \bibinfo{volume}{76}, \bibinfo{number}{12} (\bibinfo{date}{dec} \bibinfo{year}{2005}), \bibinfo{pages}{1111--1118}.
\newblock
\showISSN{0095-6562}


\bibitem[Brooke(1995)]%
        {brookeSUSQuickDirty1995}
\bibfield{author}{\bibinfo{person}{John Brooke}.} \bibinfo{year}{1995}\natexlab{}.
\newblock \showarticletitle{{{SUS}}: {{A}} Quick and Dirty Usability Scale}.
\newblock \bibinfo{journal}{\emph{Usability Eval. Ind.}}  \bibinfo{volume}{189} (\bibinfo{date}{Nov.} \bibinfo{year}{1995}).
\newblock


\bibitem[Bu et~al\mbox{.}(2024)]%
        {bu2024portobello}
\bibfield{author}{\bibinfo{person}{Fanjun Bu}, \bibinfo{person}{Stacey Li}, \bibinfo{person}{David Goedicke}, \bibinfo{person}{Mark Colley}, \bibinfo{person}{Gyanendra Sharma}, {and} \bibinfo{person}{Wendy Ju}.} \bibinfo{year}{2024}\natexlab{}.
\newblock \showarticletitle{Portobello: Extending Driving Simulation from the Lab to the Road}. In \bibinfo{booktitle}{\emph{Proceedings of the CHI Conference on Human Factors in Computing Systems}} (Honolulu, HI, USA) \emph{(\bibinfo{series}{CHI '24})}. \bibinfo{publisher}{Association for Computing Machinery}, \bibinfo{address}{New York, NY, USA}, Article \bibinfo{articleno}{256}, \bibinfo{numpages}{13}~pages.
\newblock
\showISBNx{9798400703300}
\href{https://doi.org/10.1145/3613904.3642341}{doi:\nolinkurl{10.1145/3613904.3642341}}


\bibitem[Colley(2024)]%
        {colley2024rcode}
\bibfield{author}{\bibinfo{person}{Mark Colley}.} \bibinfo{year}{2024}\natexlab{}.
\newblock \bibinfo{title}{rCode: Enhanced R Functions for Statistical Analysis and Reporting}.
\newblock \bibinfo{howpublished}{\url{https://github.com/M-Colley/rCode}}.
\newblock
\newblock
\shownote{A collection of custom R functions for streamlining statistical analysis, visualizations, and APA-compliant reporting.}.


\bibitem[Colley et~al\mbox{.}(2021a)]%
        {colley2021orias}
\bibfield{author}{\bibinfo{person}{Mark Colley}, \bibinfo{person}{Ali Askari}, \bibinfo{person}{Marcel Walch}, \bibinfo{person}{Marcel Woide}, {and} \bibinfo{person}{Enrico Rukzio}.} \bibinfo{year}{2021}\natexlab{a}.
\newblock \showarticletitle{ORIAS: On-The-Fly Object Identification and Action Selection for Highly Automated Vehicles}. In \bibinfo{booktitle}{\emph{13th International Conference on Automotive User Interfaces and Interactive Vehicular Applications}} (Leeds, United Kingdom) \emph{(\bibinfo{series}{AutomotiveUI '21})}. \bibinfo{publisher}{Association for Computing Machinery}, \bibinfo{address}{New York, NY, USA}, \bibinfo{pages}{79–89}.
\newblock
\showISBNx{9781450380638}
\href{https://doi.org/10.1145/3409118.3475134}{doi:\nolinkurl{10.1145/3409118.3475134}}


\bibitem[Colley et~al\mbox{.}(2023a)]%
        {colley2023effectsurgency}
\bibfield{author}{\bibinfo{person}{Mark Colley}, \bibinfo{person}{Cristina Evangelista}, \bibinfo{person}{Tito~Daza Rubiano}, {and} \bibinfo{person}{Enrico Rukzio}.} \bibinfo{year}{2023}\natexlab{a}.
\newblock \showarticletitle{Effects of Urgency and Cognitive Load on Interaction in Highly Automated Vehicles}.
\newblock \bibinfo{journal}{\emph{Proc. ACM Hum.-Comput. Interact.}} \bibinfo{volume}{7}, \bibinfo{number}{MHCI}, Article \bibinfo{articleno}{207} (\bibinfo{date}{sep} \bibinfo{year}{2023}), \bibinfo{numpages}{20}~pages.
\newblock
\href{https://doi.org/10.1145/3604254}{doi:\nolinkurl{10.1145/3604254}}


\bibitem[Colley et~al\mbox{.}(2023b)]%
        {colley2023much}
\bibfield{author}{\bibinfo{person}{Mark Colley}, \bibinfo{person}{Pascal Jansen}, \bibinfo{person}{Jennifer Matthiesen}, \bibinfo{person}{Hanne Hoberg}, \bibinfo{person}{Carmen Reger}, {and} \bibinfo{person}{Isabel Thiermann}.} \bibinfo{year}{2023}\natexlab{b}.
\newblock \showarticletitle{How Much Home Office is Ideal? A Multi-Perspective Algorithm}. In \bibinfo{booktitle}{\emph{Proceedings of the 2nd Annual Meeting of the Symposium on Human-Computer Interaction for Work}}. \bibinfo{publisher}{ACM}, \bibinfo{address}{New York, NY, USA}, \bibinfo{pages}{1--1}.
\newblock


\bibitem[Colley et~al\mbox{.}(2022a)]%
        {colleySwiVRCarSeatExploringVehicle2021}
\bibfield{author}{\bibinfo{person}{Mark Colley}, \bibinfo{person}{Pascal Jansen}, \bibinfo{person}{Enrico Rukzio}, {and} \bibinfo{person}{Jan Gugenheimer}.} \bibinfo{year}{2022}\natexlab{a}.
\newblock \showarticletitle{SwiVR-Car-Seat: Exploring Vehicle Motion Effects on Interaction Quality in Virtual Reality Automated Driving Using a Motorized Swivel Seat}.
\newblock \bibinfo{journal}{\emph{Proc. ACM Interact. Mob. Wearable Ubiquitous Technol.}} \bibinfo{volume}{5}, \bibinfo{number}{4}, Article \bibinfo{articleno}{150} (\bibinfo{date}{Dec.} \bibinfo{year}{2022}), \bibinfo{numpages}{26}~pages.
\newblock
\href{https://doi.org/10.1145/3494968}{doi:\nolinkurl{10.1145/3494968}}


\bibitem[Colley et~al\mbox{.}(2022b)]%
        {colley2022systematic100m}
\bibfield{author}{\bibinfo{person}{Mark Colley}, \bibinfo{person}{Bastian Wankm\"{u}ller}, {and} \bibinfo{person}{Enrico Rukzio}.} \bibinfo{year}{2022}\natexlab{b}.
\newblock \showarticletitle{A Systematic Evaluation of Solutions for the Final 100m Challenge of Highly Automated Vehicles}.
\newblock \bibinfo{journal}{\emph{Proc. ACM Hum.-Comput. Interact.}} \bibinfo{volume}{6}, \bibinfo{number}{MHCI}, Article \bibinfo{articleno}{178} (\bibinfo{date}{sep} \bibinfo{year}{2022}), \bibinfo{numpages}{19}~pages.
\newblock
\href{https://doi.org/10.1145/3546713}{doi:\nolinkurl{10.1145/3546713}}


\bibitem[Colley et~al\mbox{.}(2021b)]%
        {colley2021resync}
\bibfield{author}{\bibinfo{person}{Mark Colley}, \bibinfo{person}{Dennis Wolf}, \bibinfo{person}{Sabrina B{\"o}hm}, \bibinfo{person}{Tobias Lahmann}, \bibinfo{person}{Luca Porta}, {and} \bibinfo{person}{Enrico Rukzio}.} \bibinfo{year}{2021}\natexlab{b}.
\newblock \showarticletitle{Resync: Towards Transferring Somnolent Passengers to Consciousness}. In \bibinfo{booktitle}{\emph{Adjunct Publication of the 23rd International Conference on Mobile Human-Computer Interaction}}. \bibinfo{publisher}{ACM}, \bibinfo{address}{New York, NY, USA}, \bibinfo{pages}{1--6}.
\newblock


\bibitem[Dam and Jeon(2021)]%
        {Dam.2021}
\bibfield{author}{\bibinfo{person}{Abhraneil Dam} {and} \bibinfo{person}{Myounghoon Jeon}.} \bibinfo{year}{2021}\natexlab{}.
\newblock \showarticletitle{A Review of Motion Sickness in Automated Vehicles}. In \bibinfo{booktitle}{\emph{13th International Conference on Automotive User Interfaces and Interactive Vehicular Applications}} (Leeds, United Kingdom) \emph{(\bibinfo{series}{AutomotiveUI '21})}. \bibinfo{publisher}{Association for Computing Machinery}, \bibinfo{address}{New York, NY, USA}, \bibinfo{pages}{39–48}.
\newblock
\showISBNx{9781450380638}
\href{https://doi.org/10.1145/3409118.3475146}{doi:\nolinkurl{10.1145/3409118.3475146}}


\bibitem[Detjen et~al\mbox{.}(2019)]%
        {detjen_user-defined_2019}
\bibfield{author}{\bibinfo{person}{Henrik Detjen}, \bibinfo{person}{Sarah Faltaous}, \bibinfo{person}{Stefan Geisler}, {and} \bibinfo{person}{Stefan Schneegass}.} \bibinfo{year}{2019}\natexlab{}.
\newblock \showarticletitle{User-{Defined} {Voice} and {Mid}-{Air} {Gesture} {Commands} for {Maneuver}-based {Interventions} in {Automated} {Vehicles}}. In \bibinfo{booktitle}{\emph{Proceedings of {Mensch} und {Computer} 2019}} \emph{(\bibinfo{series}{{MuC}'19})}. \bibinfo{publisher}{Association for Computing Machinery}, \bibinfo{address}{New York, NY, USA}, \bibinfo{pages}{341--348}.
\newblock
\showISBNx{978-1-4503-7198-8}
\href{https://doi.org/10.1145/3340764.3340798}{doi:\nolinkurl{10.1145/3340764.3340798}}


\bibitem[Detjen et~al\mbox{.}(2020)]%
        {detjen2020wizard}
\bibfield{author}{\bibinfo{person}{Henrik Detjen}, \bibinfo{person}{Bastian Pfleging}, {and} \bibinfo{person}{Stefan Schneegass}.} \bibinfo{year}{2020}\natexlab{}.
\newblock \showarticletitle{A Wizard of Oz Field Study to Understand Non-Driving-Related Activities, Trust, and Acceptance of Automated Vehicles}. In \bibinfo{booktitle}{\emph{12th International Conference on Automotive User Interfaces and Interactive Vehicular Applications}} (Virtual Event, DC, USA) \emph{(\bibinfo{series}{AutomotiveUI '20})}. \bibinfo{publisher}{Association for Computing Machinery}, \bibinfo{address}{New York, NY, USA}, \bibinfo{pages}{19–29}.
\newblock
\showISBNx{9781450380652}
\href{https://doi.org/10.1145/3409120.3410662}{doi:\nolinkurl{10.1145/3409120.3410662}}


\bibitem[Diels and Bos(2016)]%
        {Diels.2016}
\bibfield{author}{\bibinfo{person}{Cyriel Diels} {and} \bibinfo{person}{Jelte~E. Bos}.} \bibinfo{year}{2016}\natexlab{}.
\newblock \showarticletitle{Self-driving carsickness}.
\newblock \bibinfo{journal}{\emph{Applied Ergonomics}}  \bibinfo{volume}{53} (\bibinfo{year}{2016}), \bibinfo{pages}{374--382}.
\newblock
\showISSN{0003-6870}
\href{https://doi.org/10.1016/j.apergo.2015.09.009}{doi:\nolinkurl{10.1016/j.apergo.2015.09.009}}
\newblock
\shownote{Transport in the 21st Century: The Application of Human Factors to Future User Needs}.


\bibitem[D\"{o}ring et~al\mbox{.}(2011)]%
        {doring2011gestural}
\bibfield{author}{\bibinfo{person}{Tanja D\"{o}ring}, \bibinfo{person}{Dagmar Kern}, \bibinfo{person}{Paul Marshall}, \bibinfo{person}{Max Pfeiffer}, \bibinfo{person}{Johannes Sch\"{o}ning}, \bibinfo{person}{Volker Gruhn}, {and} \bibinfo{person}{Albrecht Schmidt}.} \bibinfo{year}{2011}\natexlab{}.
\newblock \showarticletitle{Gestural Interaction on the Steering Wheel: Reducing the Visual Demand}. In \bibinfo{booktitle}{\emph{Proceedings of the SIGCHI Conference on Human Factors in Computing Systems}} (Vancouver, BC, Canada) \emph{(\bibinfo{series}{CHI '11})}. \bibinfo{publisher}{Association for Computing Machinery}, \bibinfo{address}{New York, NY, USA}, \bibinfo{pages}{483–492}.
\newblock
\showISBNx{9781450302289}
\href{https://doi.org/10.1145/1978942.1979010}{doi:\nolinkurl{10.1145/1978942.1979010}}


\bibitem[Faas et~al\mbox{.}(2020)]%
        {faas2020LongitudinalVideoStudy}
\bibfield{author}{\bibinfo{person}{Stefanie~M. Faas}, \bibinfo{person}{Andrea~C. Kao}, {and} \bibinfo{person}{Martin Baumann}.} \bibinfo{year}{2020}\natexlab{}.
\newblock \showarticletitle{A {{Longitudinal Video Study}} on {{Communicating Status}} and {{Intent}} for {{Self-Driving Vehicle}} {{Pedestrian Interaction}}}. In \bibinfo{booktitle}{\emph{Proceedings of the 2020 {{CHI Conference}} on {{Human Factors}} in {{Computing Systems}}}} ({Honolulu HI USA}, 2020-04-21). \bibinfo{publisher}{ACM}, \bibinfo{address}{New York, NY, USA}, \bibinfo{pages}{1--14}.
\newblock
\showISBNx{978-1-4503-6708-0}
\href{https://doi.org/10.1145/3313831.3376484}{doi:\nolinkurl{10.1145/3313831.3376484}}


\bibitem[Fujimura et~al\mbox{.}(2013)]%
        {fujimura_driver_2013}
\bibfield{author}{\bibinfo{person}{Kikuo Fujimura}, \bibinfo{person}{Lijie Xu}, \bibinfo{person}{Cuong Tran}, \bibinfo{person}{Rishabh Bhandari}, {and} \bibinfo{person}{Victor Ng-Thow-Hing}.} \bibinfo{year}{2013}\natexlab{}.
\newblock \showarticletitle{Driver queries using wheel-constrained finger pointing and 3-{D} head-up display visual feedback}. In \bibinfo{booktitle}{\emph{Proceedings of the 5th {International} {Conference} on {Automotive} {User} {Interfaces} and {Interactive} {Vehicular} {Applications}}} \emph{(\bibinfo{series}{{AutomotiveUI} '13})}. \bibinfo{publisher}{Association for Computing Machinery}, \bibinfo{address}{New York, NY, USA}, \bibinfo{pages}{56--62}.
\newblock
\showISBNx{978-1-4503-2478-6}
\href{https://doi.org/10.1145/2516540.2516551}{doi:\nolinkurl{10.1145/2516540.2516551}}


\bibitem[Ghiur\~{a}u et~al\mbox{.}(2020)]%
        {Ghiurau.2020}
\bibfield{author}{\bibinfo{person}{Florin-Timotei Ghiur\~{a}u}, \bibinfo{person}{Mehmet~Ayd\i{}n Bayta\c{s}}, {and} \bibinfo{person}{Casper Wickman}.} \bibinfo{year}{2020}\natexlab{}.
\newblock \showarticletitle{ARCAR: On-Road Driving in Mixed Reality by Volvo Cars}. In \bibinfo{booktitle}{\emph{Adjunct Proceedings of the 33rd Annual ACM Symposium on User Interface Software and Technology}} (Virtual Event, USA) \emph{(\bibinfo{series}{UIST '20 Adjunct})}. \bibinfo{publisher}{Association for Computing Machinery}, \bibinfo{address}{New York, NY, USA}, \bibinfo{pages}{62–64}.
\newblock
\showISBNx{9781450375153}
\href{https://doi.org/10.1145/3379350.3416186}{doi:\nolinkurl{10.1145/3379350.3416186}}


\bibitem[GmbH(2022)]%
        {BMWMGmbH.2022}
\bibfield{author}{\bibinfo{person}{BMW~M GmbH}.} \bibinfo{year}{2022}\natexlab{}.
\newblock \bibinfo{title}{M MIXED REALITY. Drive the change -- change the drive.}
\newblock
\urldef\tempurl%
\url{https://www.bmw-m.com/en/topics/magazine-article-pool/m-mixed-reality.html}
\showURL{%
\tempurl}
\newblock
\shownote{Accessed: 09.11.2023}.


\bibitem[Goedicke et~al\mbox{.}(2022)]%
        {goedicke2022xroom}
\bibfield{author}{\bibinfo{person}{David Goedicke}, \bibinfo{person}{Alexandra~W.D. Bremers}, \bibinfo{person}{Sam Lee}, \bibinfo{person}{Fanjun Bu}, \bibinfo{person}{Hiroshi Yasuda}, {and} \bibinfo{person}{Wendy Ju}.} \bibinfo{year}{2022}\natexlab{}.
\newblock \showarticletitle{XR-OOM: MiXed Reality driving simulation with real cars for research and design}. In \bibinfo{booktitle}{\emph{Proceedings of the 2022 CHI Conference on Human Factors in Computing Systems}} (New Orleans, LA, USA) \emph{(\bibinfo{series}{CHI '22})}. \bibinfo{publisher}{Association for Computing Machinery}, \bibinfo{address}{New York, NY, USA}, Article \bibinfo{articleno}{107}, \bibinfo{numpages}{13}~pages.
\newblock
\showISBNx{9781450391573}
\href{https://doi.org/10.1145/3491102.3517704}{doi:\nolinkurl{10.1145/3491102.3517704}}


\bibitem[Goedicke et~al\mbox{.}(2018)]%
        {goedicke_vr-oom_2018}
\bibfield{author}{\bibinfo{person}{David Goedicke}, \bibinfo{person}{Jamy Li}, \bibinfo{person}{Vanessa Evers}, {and} \bibinfo{person}{Wendy Ju}.} \bibinfo{year}{2018}\natexlab{}.
\newblock \showarticletitle{{VR}-{OOM}: {Virtual} {Reality} {On}-{rOad} driving {siMulation}}. In \bibinfo{booktitle}{\emph{Proceedings of the 2018 {CHI} {Conference} on {Human} {Factors} in {Computing} {Systems}}} \emph{(\bibinfo{series}{{CHI} '18})}. \bibinfo{publisher}{Association for Computing Machinery}, \bibinfo{address}{New York, NY, USA}, \bibinfo{pages}{1--11}.
\newblock
\showISBNx{978-1-4503-5620-6}
\href{https://doi.org/10.1145/3173574.3173739}{doi:\nolinkurl{10.1145/3173574.3173739}}


\bibitem[Gomaa et~al\mbox{.}(2020)]%
        {gomaa_studying_2020}
\bibfield{author}{\bibinfo{person}{Amr Gomaa}, \bibinfo{person}{Guillermo Reyes}, \bibinfo{person}{Alexandra Alles}, \bibinfo{person}{Lydia Rupp}, {and} \bibinfo{person}{Michael Feld}.} \bibinfo{year}{2020}\natexlab{}.
\newblock \showarticletitle{Studying {Person}-{Specific} {Pointing} and {Gaze} {Behavior} for {Multimodal} {Referencing} of {Outside} {Objects} from a {Moving} {Vehicle}}. In \bibinfo{booktitle}{\emph{Proceedings of the 2020 {International} {Conference} on {Multimodal} {Interaction}}}. \bibinfo{publisher}{ACM}, \bibinfo{address}{Virtual Event Netherlands}, \bibinfo{pages}{501--509}.
\newblock
\showISBNx{978-1-4503-7581-8}
\href{https://doi.org/10.1145/3382507.3418817}{doi:\nolinkurl{10.1145/3382507.3418817}}


\bibitem[Goode et~al\mbox{.}(2012)]%
        {goode_impact_nodate}
\bibfield{author}{\bibinfo{person}{Natassia Goode}, \bibinfo{person}{Michael~G Lenné}, {and} \bibinfo{person}{Paul Salmon}.} \bibinfo{year}{2012}\natexlab{}.
\newblock \showarticletitle{The impact of on-road motion on {BMS} touch screen device operation}.
\newblock \bibinfo{journal}{\emph{Ergonomics}} \bibinfo{volume}{55}, \bibinfo{number}{9} (\bibinfo{year}{2012}), \bibinfo{pages}{12}.
\newblock
\href{https://doi.org/10.1080/00140139.2012.685496}{doi:\nolinkurl{10.1080/00140139.2012.685496}}


\bibitem[Haeling et~al\mbox{.}(2018)]%
        {Haeling.2018}
\bibfield{author}{\bibinfo{person}{Jonas Haeling}, \bibinfo{person}{Christian Winkler}, \bibinfo{person}{Stephan Leenders}, \bibinfo{person}{Daniel Keßelheim}, \bibinfo{person}{Axel Hildebrand}, {and} \bibinfo{person}{Marc Necker}.} \bibinfo{year}{2018}\natexlab{}.
\newblock \showarticletitle{In-Car 6-DoF Mixed Reality for Rear-Seat and Co-Driver Entertainment}. In \bibinfo{booktitle}{\emph{2018 IEEE Conference on Virtual Reality and 3D User Interfaces (VR)}}. \bibinfo{publisher}{IEEE}, \bibinfo{address}{New York, NY, USA}, \bibinfo{pages}{757--758}.
\newblock
\href{https://doi.org/10.1109/VR.2018.8446461}{doi:\nolinkurl{10.1109/VR.2018.8446461}}


\bibitem[Hart(2006)]%
        {Hart.2006}
\bibfield{author}{\bibinfo{person}{Sandra~G. Hart}.} \bibinfo{year}{2006}\natexlab{}.
\newblock \showarticletitle{Nasa-Task Load Index (NASA-TLX); 20 Years Later}.
\newblock \bibinfo{journal}{\emph{Proceedings of the Human Factors and Ergonomics Society Annual Meeting}} \bibinfo{volume}{50}, \bibinfo{number}{9} (\bibinfo{year}{2006}), \bibinfo{pages}{904--908}.
\newblock
\href{https://doi.org/10.1177/154193120605000909}{doi:\nolinkurl{10.1177/154193120605000909}}


\bibitem[Hertel et~al\mbox{.}(2021)]%
        {hertel2021TaxonomyInteractionTechniques}
\bibfield{author}{\bibinfo{person}{Julia Hertel}, \bibinfo{person}{Sukran Karaosmanoglu}, \bibinfo{person}{Susanne Schmidt}, \bibinfo{person}{Julia Braker}, \bibinfo{person}{Martin Semmann}, {and} \bibinfo{person}{Frank Steinicke}.} \bibinfo{year}{2021}\natexlab{}.
\newblock \showarticletitle{A {{Taxonomy}} of {{Interaction Techniques}} for {{Immersive Augmented Reality}} Based on an {{Iterative Literature Review}}}. In \bibinfo{booktitle}{\emph{2021 {{IEEE International Symposium}} on {{Mixed}} and {{Augmented Reality}} ({{ISMAR}})}}. \bibinfo{publisher}{IEEE}, \bibinfo{address}{Bari, Italy}, \bibinfo{pages}{431--440}.
\newblock
\showISBNx{978-1-66540-158-6}
\href{https://doi.org/10.1109/ISMAR52148.2021.00060}{doi:\nolinkurl{10.1109/ISMAR52148.2021.00060}}


\bibitem[{Hincapi{\'e}-Ramos} et~al\mbox{.}(2014)]%
        {hincapie-ramos2014ConsumedEnduranceMetric}
\bibfield{author}{\bibinfo{person}{Juan~David {Hincapi{\'e}-Ramos}}, \bibinfo{person}{Xiang Guo}, \bibinfo{person}{Paymahn Moghadasian}, {and} \bibinfo{person}{Pourang Irani}.} \bibinfo{year}{2014}\natexlab{}.
\newblock \showarticletitle{Consumed Endurance: A Metric to Quantify Arm Fatigue of Mid-Air Interactions}. In \bibinfo{booktitle}{\emph{Proceedings of the {{SIGCHI Conference}} on {{Human Factors}} in {{Computing Systems}}}}. \bibinfo{publisher}{ACM}, \bibinfo{address}{Toronto Ontario Canada}, \bibinfo{pages}{1063--1072}.
\newblock
\showISBNx{978-1-4503-2473-1}
\href{https://doi.org/10.1145/2556288.2557130}{doi:\nolinkurl{10.1145/2556288.2557130}}


\bibitem[Hock et~al\mbox{.}(2022)]%
        {hock2022introducing}
\bibfield{author}{\bibinfo{person}{Philipp Hock}, \bibinfo{person}{Mark Colley}, \bibinfo{person}{Ali Askari}, \bibinfo{person}{Tobias Wagner}, \bibinfo{person}{Martin Baumann}, {and} \bibinfo{person}{Enrico Rukzio}.} \bibinfo{year}{2022}\natexlab{}.
\newblock \showarticletitle{Introducing VAMPIRE – Using Kinaesthetic Feedback in Virtual Reality for Automated Driving Experiments}. In \bibinfo{booktitle}{\emph{Proceedings of the 14th International Conference on Automotive User Interfaces and Interactive Vehicular Applications}} (Seoul, Republic of Korea) \emph{(\bibinfo{series}{AutomotiveUI '22})}. \bibinfo{publisher}{Association for Computing Machinery}, \bibinfo{address}{New York, NY, USA}, \bibinfo{pages}{204–214}.
\newblock
\showISBNx{9781450394154}
\href{https://doi.org/10.1145/3543174.3545252}{doi:\nolinkurl{10.1145/3543174.3545252}}


\bibitem[Inc.(2023)]%
        {AppleInc.2023}
\bibfield{author}{\bibinfo{person}{Apple Inc.}} \bibinfo{year}{2023}\natexlab{}.
\newblock \bibinfo{title}{Introducing Apple Vision Pro: Apple’s first spatial computer}.
\newblock
\urldef\tempurl%
\url{https://www.apple.com/newsroom/2023/06/introducing-apple-vision-pro/}
\showURL{%
\tempurl}
\newblock
\shownote{Accessed: 09.11.2023}.


\bibitem[Jansen et~al\mbox{.}(2022)]%
        {jansenDesignSpaceHuman2022}
\bibfield{author}{\bibinfo{person}{Pascal Jansen}, \bibinfo{person}{Mark Colley}, {and} \bibinfo{person}{Enrico Rukzio}.} \bibinfo{year}{2022}\natexlab{}.
\newblock \showarticletitle{A Design Space for Human Sensor and Actuator Focused In-Vehicle Interaction Based on a Systematic Literature Review}.
\newblock \bibinfo{journal}{\emph{Proc. ACM Interact. Mob. Wearable Ubiquitous Technol.}} \bibinfo{volume}{6}, \bibinfo{number}{2}, Article \bibinfo{articleno}{56} (\bibinfo{date}{July} \bibinfo{year}{2022}), \bibinfo{numpages}{51}~pages.
\newblock
\href{https://doi.org/10.1145/3534617}{doi:\nolinkurl{10.1145/3534617}}


\bibitem[{Jones, Monica Lynn Haumann} et~al\mbox{.}(2018)]%
        {Jones.2018}
\bibfield{author}{\bibinfo{person}{{Jones, Monica Lynn Haumann}}, \bibinfo{person}{{Sienko, Kathleen}}, \bibinfo{person}{{Ebert-Hamilton, Sheila}}, \bibinfo{person}{{Kinnaird, Catherine}}, \bibinfo{person}{{Miller, Carl}}, \bibinfo{person}{{Lin, Brian}}, \bibinfo{person}{{Park, Byoung-Keon}}, \bibinfo{person}{{Sullivan, John}}, \bibinfo{person}{{Reed, Matthew}}, {and} \bibinfo{person}{{Sayer, James}}.} \bibinfo{year}{2018}\natexlab{}.
\newblock \showarticletitle{Development of a Vehicle-Based Experimental Platform for Quantifying Passenger Motion Sickness during Test Track Operations}. In \bibinfo{booktitle}{\emph{WCX World Congress Experience}}. \bibinfo{publisher}{SAE International}.
\newblock
\href{https://doi.org/10.4271/2018-01-0028}{doi:\nolinkurl{10.4271/2018-01-0028}}


\bibitem[Kari and Holz(2023)]%
        {kariHandyCastPhonebasedBimanual2023}
\bibfield{author}{\bibinfo{person}{Mohamed Kari} {and} \bibinfo{person}{Christian Holz}.} \bibinfo{year}{2023}\natexlab{}.
\newblock \showarticletitle{HandyCast: Phone-based Bimanual Input for Virtual Reality in Mobile and Space-Constrained Settings via Pose-and-Touch Transfer}. In \bibinfo{booktitle}{\emph{Proceedings of the 2023 CHI Conference on Human Factors in Computing Systems}} (Hamburg, Germany) \emph{(\bibinfo{series}{CHI '23})}. \bibinfo{publisher}{Association for Computing Machinery}, \bibinfo{address}{New York, NY, USA}, Article \bibinfo{articleno}{528}, \bibinfo{numpages}{15}~pages.
\newblock
\showISBNx{9781450394215}
\href{https://doi.org/10.1145/3544548.3580677}{doi:\nolinkurl{10.1145/3544548.3580677}}


\bibitem[Kauer et~al\mbox{.}(2010)]%
        {kauer_how_2010}
\bibfield{author}{\bibinfo{person}{M. Kauer}, \bibinfo{person}{M. Schreiber}, {and} \bibinfo{person}{R. Bruder}.} \bibinfo{year}{2010}\natexlab{}.
\newblock \showarticletitle{How to conduct a car? {A} design example for maneuver based driver-vehicle interaction}. In \bibinfo{booktitle}{\emph{2010 {IEEE} {Intelligent} {Vehicles} {Symposium}}}. \bibinfo{publisher}{IEEE}, \bibinfo{address}{New York, NY, USA}, \bibinfo{pages}{1214--1221}.
\newblock
\href{https://doi.org/10.1109/IVS.2010.5548099}{doi:\nolinkurl{10.1109/IVS.2010.5548099}}
\newblock
\shownote{ISSN: 1931-0587}.


\bibitem[Kim and Song(2014)]%
        {kim_evaluation_2014}
\bibfield{author}{\bibinfo{person}{Huhn Kim} {and} \bibinfo{person}{Haewon Song}.} \bibinfo{year}{2014}\natexlab{}.
\newblock \showarticletitle{Evaluation of the safety and usability of touch gestures in operating in-vehicle information systems with visual occlusion}.
\newblock \bibinfo{journal}{\emph{Applied Ergonomics}} \bibinfo{volume}{45}, \bibinfo{number}{3} (\bibinfo{year}{2014}), \bibinfo{pages}{789--798}.
\newblock
\showISSN{0003-6870}
\href{https://doi.org/10.1016/j.apergo.2013.10.013}{doi:\nolinkurl{10.1016/j.apergo.2013.10.013}}


\bibitem[Kim et~al\mbox{.}(2020)]%
        {kim_cascaded_2020}
\bibfield{author}{\bibinfo{person}{Myeongseop Kim}, \bibinfo{person}{Eunjin Seong}, \bibinfo{person}{Younkyung Jwa}, \bibinfo{person}{Jieun Lee}, {and} \bibinfo{person}{Seungjun Kim}.} \bibinfo{year}{2020}\natexlab{}.
\newblock \showarticletitle{A {Cascaded} {Multimodal} {Natural} {User} {Interface} to {Reduce} {Driver} {Distraction}}.
\newblock \bibinfo{journal}{\emph{IEEE Access}}  \bibinfo{volume}{8} (\bibinfo{year}{2020}), \bibinfo{pages}{112969--112984}.
\newblock
\showISSN{2169-3536}
\href{https://doi.org/10.1109/ACCESS.2020.3002775}{doi:\nolinkurl{10.1109/ACCESS.2020.3002775}}
\newblock
\shownote{Conference Name: IEEE Access}.


\bibitem[Kim and Xiong(2022)]%
        {kim2022PseudohapticButtonImproving}
\bibfield{author}{\bibinfo{person}{Woojoo Kim} {and} \bibinfo{person}{Shuping Xiong}.} \bibinfo{year}{2022}\natexlab{}.
\newblock \showarticletitle{Pseudo-Haptic Button for Improving User Experience of Mid-Air Interaction in {{VR}}}.
\newblock \bibinfo{journal}{\emph{International Journal of Human-Computer Studies}}  \bibinfo{volume}{168} (\bibinfo{date}{Dec.} \bibinfo{year}{2022}), \bibinfo{pages}{102907}.
\newblock
\showISSN{10715819}
\href{https://doi.org/10.1016/j.ijhcs.2022.102907}{doi:\nolinkurl{10.1016/j.ijhcs.2022.102907}}


\bibitem[K{\"o}rber(2019)]%
        {korber2019TheoreticalConsiderationsDevelopment}
\bibfield{author}{\bibinfo{person}{Moritz K{\"o}rber}.} \bibinfo{year}{2019}\natexlab{}.
\newblock \showarticletitle{Theoretical Considerations and Development of a Questionnaire to Measure Trust in Automation}.
\newblock In \bibinfo{booktitle}{\emph{Proceedings of the 20th Congress of the International Ergonomics Association (IEA 2018)}}, \bibfield{editor}{\bibinfo{person}{Sebastiano Bagnara}, \bibinfo{person}{Riccardo Tartaglia}, \bibinfo{person}{Sara Albolino}, \bibinfo{person}{Thomas Alexander}, {and} \bibinfo{person}{Yushi Fujita}} (Eds.). \bibinfo{publisher}{Springer International Publishing}, \bibinfo{address}{Cham}, \bibinfo{pages}{13--30}.
\newblock
\showISBNx{978-3-319-96074-6}


\bibitem[Koyama et~al\mbox{.}(2014)]%
        {koyama2014multi}
\bibfield{author}{\bibinfo{person}{Shunsuke Koyama}, \bibinfo{person}{Yuta Sugiura}, \bibinfo{person}{Masa Ogata}, \bibinfo{person}{Anusha Withana}, \bibinfo{person}{Yuji Uema}, \bibinfo{person}{Makoto Honda}, \bibinfo{person}{Sayaka Yoshizu}, \bibinfo{person}{Chihiro Sannomiya}, \bibinfo{person}{Kazunari Nawa}, {and} \bibinfo{person}{Masahiko Inami}.} \bibinfo{year}{2014}\natexlab{}.
\newblock \showarticletitle{Multi-Touch Steering Wheel for in-Car Tertiary Applications Using Infrared Sensors}. In \bibinfo{booktitle}{\emph{Proceedings of the 5th Augmented Human International Conference}} (Kobe, Japan) \emph{(\bibinfo{series}{AH '14})}. \bibinfo{publisher}{Association for Computing Machinery}, \bibinfo{address}{New York, NY, USA}, Article \bibinfo{articleno}{5}, \bibinfo{numpages}{4}~pages.
\newblock
\showISBNx{9781450327619}
\href{https://doi.org/10.1145/2582051.2582056}{doi:\nolinkurl{10.1145/2582051.2582056}}


\bibitem[Kuiper et~al\mbox{.}(2018)]%
        {Kuiper.2018}
\bibfield{author}{\bibinfo{person}{Ouren~X. Kuiper}, \bibinfo{person}{Jelte~E. Bos}, {and} \bibinfo{person}{Cyriel Diels}.} \bibinfo{year}{2018}\natexlab{}.
\newblock \showarticletitle{Looking forward: In-vehicle auxiliary display positioning affects carsickness}.
\newblock \bibinfo{journal}{\emph{Applied Ergonomics}}  \bibinfo{volume}{68} (\bibinfo{year}{2018}), \bibinfo{pages}{169--175}.
\newblock
\showISSN{0003-6870}
\href{https://doi.org/10.1016/j.apergo.2017.11.002}{doi:\nolinkurl{10.1016/j.apergo.2017.11.002}}


\bibitem[Manawadu et~al\mbox{.}(2016)]%
        {manawadu_hand_2016}
\bibfield{author}{\bibinfo{person}{Udara~E. Manawadu}, \bibinfo{person}{Mitsuhiro Kamezaki}, \bibinfo{person}{Masaaki Ishikawa}, \bibinfo{person}{Takahiro Kawano}, {and} \bibinfo{person}{Shigeki Sugano}.} \bibinfo{year}{2016}\natexlab{}.
\newblock \showarticletitle{A hand gesture based driver-vehicle interface to control lateral and longitudinal motions of an autonomous vehicle}. In \bibinfo{booktitle}{\emph{2016 {IEEE} {International} {Conference} on {Systems}, {Man}, and {Cybernetics} ({SMC})}}. \bibinfo{publisher}{IEEE}, \bibinfo{address}{New York, NY, USA}, \bibinfo{pages}{001785--001790}.
\newblock
\href{https://doi.org/10.1109/SMC.2016.7844497}{doi:\nolinkurl{10.1109/SMC.2016.7844497}}


\bibitem[Mathis et~al\mbox{.}(2021)]%
        {Mathis.2021}
\bibfield{author}{\bibinfo{person}{Lesley-Ann Mathis}, \bibinfo{person}{Harald Widlroither}, {and} \bibinfo{person}{Nico Traub}.} \bibinfo{year}{2021}\natexlab{}.
\newblock \showarticletitle{Towards Future Interior Concepts: User Perception and Requirements for the Use Case Working in the Autonomou Car}.
\newblock In \bibinfo{booktitle}{\emph{Advances in Human Aspects of Transportation}}, \bibfield{editor}{\bibinfo{person}{Neville Stanton}} (Ed.). \bibinfo{series}{Lecture Notes in Networks and Systems}, Vol.~\bibinfo{volume}{270}. \bibinfo{publisher}{Springer International Publishing}, \bibinfo{address}{Cham}, \bibinfo{pages}{315--322}.
\newblock
\showISBNx{978-3-030-80012-3}
\href{https://doi.org/10.1007/978-3-030-80012-3_37}{doi:\nolinkurl{10.1007/978-3-030-80012-3_37}}


\bibitem[Maxwell et~al\mbox{.}(2017)]%
        {maxwell2017DesigningExperimentsAnalyzing}
\bibfield{author}{\bibinfo{person}{Scott~E. Maxwell}, \bibinfo{person}{Harold~D. Delaney}, {and} \bibinfo{person}{Ken Kelley}.} \bibinfo{year}{2017}\natexlab{}.
\newblock \bibinfo{booktitle}{\emph{Designing {{Experiments}} and {{Analyzing Data}}: {{A Model Comparison Perspective}}} (\bibinfo{edition}{3} ed.)}.
\newblock \bibinfo{publisher}{Routledge}, \bibinfo{address}{Third edition / Scott E. Maxwell, Harold D. Delaney, and Ken Kelley. {\textbar} New York,}.
\newblock
\showISBNx{978-1-315-64295-6}
\href{https://doi.org/10.4324/9781315642956}{doi:\nolinkurl{10.4324/9781315642956}}


\bibitem[Mayer et~al\mbox{.}(2018)]%
        {mayerEffectRoadBumps2018}
\bibfield{author}{\bibinfo{person}{Sven Mayer}, \bibinfo{person}{Huy~Viet Le}, \bibinfo{person}{Alessandro Nesti}, \bibinfo{person}{Niels Henze}, \bibinfo{person}{Heinrich~H. Bülthoff}, {and} \bibinfo{person}{Lewis~L. Chuang}.} \bibinfo{year}{2018}\natexlab{}.
\newblock \showarticletitle{The {{Effect}} of {{Road Bumps}} on {{Touch Interaction}} in {{Cars}}}. In \bibinfo{booktitle}{\emph{Proceedings of the 10th {{International Conference}} on {{Automotive User Interfaces}} and {{Interactive Vehicular Applications}}}} (2018-09-23). \bibinfo{publisher}{ACM}, \bibinfo{address}{New York, NY, USA}, \bibinfo{pages}{85--93}.
\newblock
\showISBNx{978-1-4503-5946-7}
\href{https://doi.org/10.1145/3239060.3239071}{doi:\nolinkurl{10.1145/3239060.3239071}}


\bibitem[Mcgill et~al\mbox{.}(2020)]%
        {Mcgill.2020b}
\bibfield{author}{\bibinfo{person}{Mark Mcgill}, \bibinfo{person}{Aidan Kehoe}, \bibinfo{person}{Euan Freeman}, {and} \bibinfo{person}{Stephen Brewster}.} \bibinfo{year}{2020}\natexlab{}.
\newblock \showarticletitle{Expanding the Bounds of Seated Virtual Workspaces}.
\newblock \bibinfo{journal}{\emph{ACM Transactions on Computer-Human Interaction}} \bibinfo{volume}{27}, \bibinfo{number}{3} (\bibinfo{year}{2020}), \bibinfo{pages}{1--40}.
\newblock
\showISSN{1073-0516}
\href{https://doi.org/10.1145/3380959}{doi:\nolinkurl{10.1145/3380959}}


\bibitem[McGill et~al\mbox{.}(2022)]%
        {mcgill20222passengxr}
\bibfield{author}{\bibinfo{person}{Mark McGill}, \bibinfo{person}{Graham Wilson}, \bibinfo{person}{Daniel Medeiros}, {and} \bibinfo{person}{Stephen~Anthony Brewster}.} \bibinfo{year}{2022}\natexlab{}.
\newblock \showarticletitle{PassengXR: A Low Cost Platform for Any-Car, Multi-User, Motion-Based Passenger XR Experiences}. In \bibinfo{booktitle}{\emph{Proceedings of the 35th Annual ACM Symposium on User Interface Software and Technology}} (Bend, OR, USA) \emph{(\bibinfo{series}{UIST '22})}. \bibinfo{publisher}{Association for Computing Machinery}, \bibinfo{address}{New York, NY, USA}, Article \bibinfo{articleno}{2}, \bibinfo{numpages}{15}~pages.
\newblock
\showISBNx{9781450393201}
\href{https://doi.org/10.1145/3526113.3545657}{doi:\nolinkurl{10.1145/3526113.3545657}}


\bibitem[Meta(2023)]%
        {Meta.2023}
\bibfield{author}{\bibinfo{person}{Meta}.} \bibinfo{year}{2023}\natexlab{}.
\newblock \bibinfo{title}{Meet Meta Quest 3, Our Mixed Reality Headset Starting at \$499.99}.
\newblock
\urldef\tempurl%
\url{https://about.fb.com/news/2023/09/meet-meta-quest-3-mixed-reality-headset/}
\showURL{%
\tempurl}
\newblock
\shownote{Accessed: 09.11.2023}.


\bibitem[Microsoft(2021)]%
        {Microsoft.2021}
\bibfield{author}{\bibinfo{person}{Microsoft}.} \bibinfo{year}{2021}\natexlab{}.
\newblock \bibinfo{title}{Comfort}.
\newblock
\urldef\tempurl%
\url{https://learn.microsoft.com/en-us/windows/mixed-reality/design/comfort}
\showURL{%
\tempurl}
\newblock
\shownote{Accessed: 22.06.2024}.


\bibitem[Mutasim et~al\mbox{.}(2021)]%
        {mutasim2021PinchClickDwell}
\bibfield{author}{\bibinfo{person}{Aunnoy~K Mutasim}, \bibinfo{person}{Anil~Ufuk Batmaz}, {and} \bibinfo{person}{Wolfgang Stuerzlinger}.} \bibinfo{year}{2021}\natexlab{}.
\newblock \showarticletitle{Pinch, {{Click}}, or {{Dwell}}: {{Comparing Different Selection Techniques}} for {{Eye-Gaze-Based Pointing}} in {{Virtual Reality}}}. In \bibinfo{booktitle}{\emph{{{ACM Symposium}} on {{Eye Tracking Research}} and {{Applications}}}}. \bibinfo{publisher}{ACM}, \bibinfo{address}{Virtual Event Germany}, \bibinfo{pages}{1--7}.
\newblock
\showISBNx{978-1-4503-8345-5}
\href{https://doi.org/10.1145/3448018.3457998}{doi:\nolinkurl{10.1145/3448018.3457998}}


\bibitem[Mühlbacher et~al\mbox{.}(2020)]%
        {Muhlbacher.2020}
\bibfield{author}{\bibinfo{person}{Dominik Mühlbacher}, \bibinfo{person}{Markus Tomzig}, \bibinfo{person}{Katharina Reinmüller}, {and} \bibinfo{person}{Lena Rittger}.} \bibinfo{year}{2020}\natexlab{}.
\newblock \showarticletitle{Methodological Considerations Concerning Motion Sickness Investigations during Automated Driving}.
\newblock \bibinfo{journal}{\emph{Information}} \bibinfo{volume}{11}, \bibinfo{number}{5} (\bibinfo{year}{2020}).
\newblock
\showISSN{2078-2489}
\href{https://doi.org/10.3390/info11050265}{doi:\nolinkurl{10.3390/info11050265}}


\bibitem[Neßelrath et~al\mbox{.}(2016)]%
        {neselrath_combining_2016}
\bibfield{author}{\bibinfo{person}{Robert Neßelrath}, \bibinfo{person}{Mohammad~Mehdi Moniri}, {and} \bibinfo{person}{Michael Feld}.} \bibinfo{year}{2016}\natexlab{}.
\newblock \showarticletitle{Combining Speech, Gaze, and Micro-gestures for the Multimodal Control of In-Car Functions}. In \bibinfo{booktitle}{\emph{2016 12th {International} {Conference} on {Intelligent} {Environments} ({IE})}}. \bibinfo{publisher}{IEEE}, \bibinfo{address}{New York, NY, USA}, \bibinfo{pages}{190--193}.
\newblock
\href{https://doi.org/10.1109/IE.2016.42}{doi:\nolinkurl{10.1109/IE.2016.42}}
\newblock
\shownote{ISSN: 2472-7571}.


\bibitem[Ng and Brewster(2016)]%
        {ng2016investigating}
\bibfield{author}{\bibinfo{person}{Alexander Ng} {and} \bibinfo{person}{Stephen~A. Brewster}.} \bibinfo{year}{2016}\natexlab{}.
\newblock \showarticletitle{Investigating Pressure Input and Haptic Feedback for In-Car Touchscreens and Touch Surfaces}. In \bibinfo{booktitle}{\emph{Proceedings of the 8th International Conference on Automotive User Interfaces and Interactive Vehicular Applications}} (Ann Arbor, MI, USA) \emph{(\bibinfo{series}{Automotive'UI 16})}. \bibinfo{publisher}{Association for Computing Machinery}, \bibinfo{address}{New York, NY, USA}, \bibinfo{pages}{121–128}.
\newblock
\showISBNx{9781450345330}
\href{https://doi.org/10.1145/3003715.3005420}{doi:\nolinkurl{10.1145/3003715.3005420}}


\bibitem[Pampel et~al\mbox{.}(2019)]%
        {pampelFittsGoesAutobahn2019}
\bibfield{author}{\bibinfo{person}{Sanna~M. Pampel}, \bibinfo{person}{Gary Burnett}, \bibinfo{person}{Chrisminder Hare}, \bibinfo{person}{Harpreet Singh}, \bibinfo{person}{Arber Shabani}, \bibinfo{person}{Lee Skrypchuk}, {and} \bibinfo{person}{Alex Mouzakitis}.} \bibinfo{year}{2019}\natexlab{}.
\newblock \showarticletitle{Fitts Goes Autobahn: Assessing the Visual Demand of Finger-Touch Pointing Tasks in an On-Road Study}. In \bibinfo{booktitle}{\emph{Proceedings of the 11th International Conference on Automotive User Interfaces and Interactive Vehicular Applications}} (Utrecht, Netherlands) \emph{(\bibinfo{series}{AutomotiveUI '19})}. \bibinfo{publisher}{Association for Computing Machinery}, \bibinfo{address}{New York, NY, USA}, \bibinfo{pages}{254–261}.
\newblock
\showISBNx{9781450368841}
\href{https://doi.org/10.1145/3342197.3344538}{doi:\nolinkurl{10.1145/3342197.3344538}}


\bibitem[Petersen(2019)]%
        {Petersen.2019b}
\bibfield{author}{\bibinfo{person}{Klaus Petersen}.} \bibinfo{year}{2019}\natexlab{}.
\newblock \bibinfo{title}{LPVR Middleware a Full Solution for AR / VR: Introducing LPVR Middleware}.
\newblock
\urldef\tempurl%
\url{https://lp-research.com/middleware-full-solution-ar-vr/}
\showURL{%
\tempurl}


\bibitem[Pfeiffer et~al\mbox{.}(2010)]%
        {pfeiffer2010multi}
\bibfield{author}{\bibinfo{person}{Max Pfeiffer}, \bibinfo{person}{Dagmar Kern}, \bibinfo{person}{Johannes Sch\"{o}ning}, \bibinfo{person}{Tanja D\"{o}ring}, \bibinfo{person}{Antonio Kr\"{u}ger}, {and} \bibinfo{person}{Albrecht Schmidt}.} \bibinfo{year}{2010}\natexlab{}.
\newblock \showarticletitle{A Multi-Touch Enabled Steering Wheel: Exploring the Design Space}.
\newblock In \bibinfo{booktitle}{\emph{CHI '10 Extended Abstracts on Human Factors in Computing Systems}}. \bibinfo{publisher}{Association for Computing Machinery}, \bibinfo{address}{New York, NY, USA}, \bibinfo{pages}{3355–3360}.
\newblock
\showISBNx{9781605589305}
\href{https://doi.org/10.1145/1753846.1753984}{doi:\nolinkurl{10.1145/1753846.1753984}}


\bibitem[Pfeuffer et~al\mbox{.}(2024)]%
        {pfeuffer2024DesignPrinciplesChallenges}
\bibfield{author}{\bibinfo{person}{Ken Pfeuffer}, \bibinfo{person}{Hans Gellersen}, {and} \bibinfo{person}{Mar {Gonzalez-Franco}}.} \bibinfo{year}{2024}\natexlab{}.
\newblock \showarticletitle{Design {{Principles}} and {{Challenges}} for {{Gaze}} + {{Pinch Interaction}} in {{XR}}}.
\newblock \bibinfo{journal}{\emph{IEEE Computer Graphics and Applications}} \bibinfo{volume}{44}, \bibinfo{number}{3} (\bibinfo{date}{May} \bibinfo{year}{2024}), \bibinfo{pages}{74--81}.
\newblock
\showISSN{0272-1716, 1558-1756}
\href{https://doi.org/10.1109/MCG.2024.3382961}{doi:\nolinkurl{10.1109/MCG.2024.3382961}}


\bibitem[Pfeuffer et~al\mbox{.}(2017)]%
        {pfeuffer2017GazePinchInteraction}
\bibfield{author}{\bibinfo{person}{Ken Pfeuffer}, \bibinfo{person}{Benedikt Mayer}, \bibinfo{person}{Diako Mardanbegi}, {and} \bibinfo{person}{Hans Gellersen}.} \bibinfo{year}{2017}\natexlab{}.
\newblock \showarticletitle{Gaze + Pinch Interaction in Virtual Reality}. In \bibinfo{booktitle}{\emph{Proceedings of the 5th {{Symposium}} on {{Spatial User Interaction}}}}. \bibinfo{publisher}{ACM}, \bibinfo{address}{Brighton United Kingdom}, \bibinfo{pages}{99--108}.
\newblock
\showISBNx{978-1-4503-5486-8}
\href{https://doi.org/10.1145/3131277.3132180}{doi:\nolinkurl{10.1145/3131277.3132180}}


\bibitem[Poitschke et~al\mbox{.}(2011)]%
        {poitschke_gaze-based_2011}
\bibfield{author}{\bibinfo{person}{Tony Poitschke}, \bibinfo{person}{Florian Laquai}, \bibinfo{person}{Stilyan Stamboliev}, {and} \bibinfo{person}{Gerhard Rigoll}.} \bibinfo{year}{2011}\natexlab{}.
\newblock \showarticletitle{Gaze-based interaction on multiple displays in an automotive environment}. In \bibinfo{booktitle}{\emph{2011 {IEEE} {International} {Conference} on {Systems}, {Man}, and {Cybernetics}}}. \bibinfo{publisher}{IEEE}, \bibinfo{address}{New York, NY, USA}, \bibinfo{pages}{543--548}.
\newblock
\href{https://doi.org/10.1109/ICSMC.2011.6083740}{doi:\nolinkurl{10.1109/ICSMC.2011.6083740}}
\newblock
\shownote{ISSN: 1062-922X}.


\bibitem[Raffelh{\"u}schen and Schlinkert(2018)]%
        {raffelhuschen2018deutsche}
\bibfield{author}{\bibinfo{person}{B. Raffelh{\"u}schen} {and} \bibinfo{person}{R. Schlinkert}.} \bibinfo{year}{2018}\natexlab{}.
\newblock \bibinfo{booktitle}{\emph{{Deutsche Post Gl{\"u}cksatlas 2018}}}.
\newblock \bibinfo{publisher}{Penguin Verlag}, \bibinfo{address}{Munich, Germany}.
\newblock
\showISBNx{9783641238902}
\urldef\tempurl%
\url{https://books.google.de/books?id=x3ByDwAAQBAJ}
\showURL{%
\tempurl}


\bibitem[Riegler et~al\mbox{.}(2020)]%
        {riegler2020gaze}
\bibfield{author}{\bibinfo{person}{Andreas Riegler}, \bibinfo{person}{Bilal Aksoy}, \bibinfo{person}{Andreas Riener}, {and} \bibinfo{person}{Clemens Holzmann}.} \bibinfo{year}{2020}\natexlab{}.
\newblock \showarticletitle{Gaze-Based Interaction with Windshield Displays for Automated Driving: Impact of Dwell Time and Feedback Design on Task Performance and Subjective Workload}. In \bibinfo{booktitle}{\emph{12th International Conference on Automotive User Interfaces and Interactive Vehicular Applications}} (Virtual Event, DC, USA) \emph{(\bibinfo{series}{AutomotiveUI '20})}. \bibinfo{publisher}{Association for Computing Machinery}, \bibinfo{address}{New York, NY, USA}, \bibinfo{pages}{151–160}.
\newblock
\showISBNx{9781450380652}
\href{https://doi.org/10.1145/3409120.3410654}{doi:\nolinkurl{10.1145/3409120.3410654}}


\bibitem[Roider and Gross(2018)]%
        {roiderSeeYourPoint2018}
\bibfield{author}{\bibinfo{person}{Florian Roider} {and} \bibinfo{person}{Tom Gross}.} \bibinfo{year}{2018}\natexlab{}.
\newblock \showarticletitle{I {{See Your Point}}: {{Integrating Gaze}} to {{Enhance Pointing Gesture Accuracy While Driving}}}. In \bibinfo{booktitle}{\emph{Proceedings of the 10th {{International Conference}} on {{Automotive User Interfaces}} and {{Interactive Vehicular Applications}}}} ({Toronto ON Canada}, 2018-09-23). \bibinfo{publisher}{ACM}, \bibinfo{address}{New York, NY, USA}, \bibinfo{pages}{351--358}.
\newblock
\showISBNx{978-1-4503-5946-7}
\href{https://doi.org/10.1145/3239060.3239084}{doi:\nolinkurl{10.1145/3239060.3239084}}


\bibitem[Roider et~al\mbox{.}(2017)]%
        {roider_effects_2017}
\bibfield{author}{\bibinfo{person}{Florian Roider}, \bibinfo{person}{Sonja Rümelin}, \bibinfo{person}{Bastian Pfleging}, {and} \bibinfo{person}{Tom Gross}.} \bibinfo{year}{2017}\natexlab{}.
\newblock \showarticletitle{The {Effects} of {Situational} {Demands} on {Gaze}, {Speech} and {Gesture} {Input} in the {Vehicle}}. In \bibinfo{booktitle}{\emph{Proceedings of the 9th {International} {Conference} on {Automotive} {User} {Interfaces} and {Interactive} {Vehicular} {Applications}}}. \bibinfo{publisher}{ACM}, \bibinfo{address}{Oldenburg Germany}, \bibinfo{pages}{94--102}.
\newblock
\showISBNx{978-1-4503-5150-8}
\href{https://doi.org/10.1145/3122986.3122999}{doi:\nolinkurl{10.1145/3122986.3122999}}


\bibitem[Rümelin et~al\mbox{.}(2013)]%
        {rumelin_free-hand_2013}
\bibfield{author}{\bibinfo{person}{Sonja Rümelin}, \bibinfo{person}{Chadly Marouane}, {and} \bibinfo{person}{Andreas Butz}.} \bibinfo{year}{2013}\natexlab{}.
\newblock \showarticletitle{Free-hand pointing for identification and interaction with distant objects}. In \bibinfo{booktitle}{\emph{Proceedings of the 5th {International} {Conference} on {Automotive} {User} {Interfaces} and {Interactive} {Vehicular} {Applications}}} \emph{(\bibinfo{series}{{AutomotiveUI} '13})}. \bibinfo{publisher}{Association for Computing Machinery}, \bibinfo{address}{New York, NY, USA}, \bibinfo{pages}{40--47}.
\newblock
\showISBNx{978-1-4503-2478-6}
\href{https://doi.org/10.1145/2516540.2516556}{doi:\nolinkurl{10.1145/2516540.2516556}}


\bibitem[Salmon et~al\mbox{.}(2011)]%
        {salmon_effects_2011}
\bibfield{author}{\bibinfo{person}{Paul~M. Salmon}, \bibinfo{person}{Michael~G. Lenné}, \bibinfo{person}{Tom Triggs}, \bibinfo{person}{Natassia Goode}, \bibinfo{person}{Miranda Cornelissen}, {and} \bibinfo{person}{Victor Demczuk}.} \bibinfo{year}{2011}\natexlab{}.
\newblock \showarticletitle{The effects of motion on in-vehicle touch screen system operation: {A} battle management system case study}.
\newblock \bibinfo{journal}{\emph{Transportation Research Part F: Traffic Psychology and Behaviour}} \bibinfo{volume}{14}, \bibinfo{number}{6} (\bibinfo{date}{Nov.} \bibinfo{year}{2011}), \bibinfo{pages}{494--503}.
\newblock
\showISSN{1369-8478}
\href{https://doi.org/10.1016/j.trf.2011.08.002}{doi:\nolinkurl{10.1016/j.trf.2011.08.002}}


\bibitem[Sasalovici et~al\mbox{.}(2023)]%
        {sasalovici2023InCarOfficeCan}
\bibfield{author}{\bibinfo{person}{Markus Sasalovici}, \bibinfo{person}{Stephan Leenders}, \bibinfo{person}{Robin~Connor Schramm}, \bibinfo{person}{Jann~Philipp Freiwald}, \bibinfo{person}{Hannes~Frederic Botzet}, \bibinfo{person}{Daniel Ke\ss{}elheim}, \bibinfo{person}{Thomas Krach}, {and} \bibinfo{person}{Christian Winkler}.} \bibinfo{year}{2023}\natexlab{}.
\newblock \showarticletitle{In-Car Office: Can HMD-Based AR Alleviate Passenger Motion Sickness?}. In \bibinfo{booktitle}{\emph{Adjunct Proceedings of the 15th International Conference on Automotive User Interfaces and Interactive Vehicular Applications}} (Ingolstadt, Germany) \emph{(\bibinfo{series}{AutomotiveUI '23 Adjunct})}. \bibinfo{publisher}{Association for Computing Machinery}, \bibinfo{address}{New York, NY, USA}, \bibinfo{pages}{133–138}.
\newblock
\showISBNx{9798400701122}
\href{https://doi.org/10.1145/3581961.3609869}{doi:\nolinkurl{10.1145/3581961.3609869}}


\bibitem[Schramm et~al\mbox{.}(2023)]%
        {schramm2023AssessingAugmentedReality}
\bibfield{author}{\bibinfo{person}{Robin~Connor Schramm}, \bibinfo{person}{Markus Sasalovici}, \bibinfo{person}{Axel Hildebrand}, {and} \bibinfo{person}{Ulrich Schwanecke}.} \bibinfo{year}{2023}\natexlab{}.
\newblock \showarticletitle{Assessing {{Augmented Reality Selection Techniques}} for {{Passengers}} in {{Moving Vehicles}}: {{A Real-World User Study}}}. In \bibinfo{booktitle}{\emph{Proceedings of the 15th {{International Conference}} on {{Automotive User Interfaces}} and {{Interactive Vehicular Applications}}}} ({Ingolstadt Germany}, 2023-09-18). \bibinfo{publisher}{ACM}, \bibinfo{address}{New York, NY, USA}, \bibinfo{pages}{22--31}.
\newblock
\showISBNx{9798400701054}
\href{https://doi.org/10.1145/3580585.3607152}{doi:\nolinkurl{10.1145/3580585.3607152}}


\bibitem[Sezgin et~al\mbox{.}(2009)]%
        {sezgin_multimodal_2009}
\bibfield{author}{\bibinfo{person}{Tevfik~Metin Sezgin}, \bibinfo{person}{Ian Davies}, {and} \bibinfo{person}{Peter Robinson}.} \bibinfo{year}{2009}\natexlab{}.
\newblock \showarticletitle{Multimodal inference for driver-vehicle interaction}. In \bibinfo{booktitle}{\emph{Proceedings of the 2009 international conference on {Multimodal} interfaces}} \emph{(\bibinfo{series}{{ICMI}-{MLMI} '09})}. \bibinfo{publisher}{Association for Computing Machinery}, \bibinfo{address}{New York, NY, USA}, \bibinfo{pages}{193--198}.
\newblock
\showISBNx{978-1-60558-772-1}
\href{https://doi.org/10.1145/1647314.1647348}{doi:\nolinkurl{10.1145/1647314.1647348}}


\bibitem[Sidenmark and Gellersen(2019)]%
        {sidenmark2019EyeHeadSynergetic}
\bibfield{author}{\bibinfo{person}{Ludwig Sidenmark} {and} \bibinfo{person}{Hans Gellersen}.} \bibinfo{year}{2019}\natexlab{}.
\newblock \showarticletitle{Eye\&{{Head}}: {{Synergetic Eye}} and {{Head Movement}} for {{Gaze Pointing}} and {{Selection}}}. In \bibinfo{booktitle}{\emph{Proceedings of the 32nd {{Annual ACM Symposium}} on {{User Interface Software}} and {{Technology}}}}. \bibinfo{publisher}{ACM}, \bibinfo{address}{New Orleans LA USA}, \bibinfo{pages}{1161--1174}.
\newblock
\showISBNx{978-1-4503-6816-2}
\href{https://doi.org/10.1145/3332165.3347921}{doi:\nolinkurl{10.1145/3332165.3347921}}


\bibitem[Speicher et~al\mbox{.}(2019)]%
        {speicher2019PseudohapticControlsMidair}
\bibfield{author}{\bibinfo{person}{Marco Speicher}, \bibinfo{person}{Jan Ehrlich}, \bibinfo{person}{Vito Gentile}, \bibinfo{person}{Donald Degraen}, \bibinfo{person}{Salvatore Sorce}, {and} \bibinfo{person}{Antonio Kr{\"u}ger}.} \bibinfo{year}{2019}\natexlab{}.
\newblock \showarticletitle{Pseudo-Haptic {{Controls}} for {{Mid-air Finger-based Menu Interaction}}}. In \bibinfo{booktitle}{\emph{Extended {{Abstracts}} of the 2019 {{CHI Conference}} on {{Human Factors}} in {{Computing Systems}}}}. \bibinfo{publisher}{ACM}, \bibinfo{address}{Glasgow Scotland Uk}, \bibinfo{pages}{1--6}.
\newblock
\showISBNx{978-1-4503-5971-9}
\href{https://doi.org/10.1145/3290607.3312927}{doi:\nolinkurl{10.1145/3290607.3312927}}


\bibitem[Tseng et~al\mbox{.}(2023)]%
        {tsengFingerMapperMappingFinger2023}
\bibfield{author}{\bibinfo{person}{Wen-Jie Tseng}, \bibinfo{person}{Samuel Huron}, \bibinfo{person}{Eric Lecolinet}, {and} \bibinfo{person}{Jan Gugenheimer}.} \bibinfo{year}{2023}\natexlab{}.
\newblock \showarticletitle{FingerMapper: Mapping Finger Motions onto Virtual Arms to Enable Safe Virtual Reality Interaction in Confined Spaces}. In \bibinfo{booktitle}{\emph{Proceedings of the 2023 CHI Conference on Human Factors in Computing Systems}} (Hamburg, Germany) \emph{(\bibinfo{series}{CHI '23})}. \bibinfo{publisher}{Association for Computing Machinery}, \bibinfo{address}{New York, NY, USA}, Article \bibinfo{articleno}{874}, \bibinfo{numpages}{14}~pages.
\newblock
\showISBNx{9781450394215}
\href{https://doi.org/10.1145/3544548.3580736}{doi:\nolinkurl{10.1145/3544548.3580736}}


\bibitem[Ultraleap(2024)]%
        {Ultraleap.2024}
\bibfield{author}{\bibinfo{person}{Ultraleap}.} \bibinfo{year}{2024}\natexlab{}.
\newblock \bibinfo{title}{Camera Placement}.
\newblock
\urldef\tempurl%
\url{https://docs.ultraleap.com/TouchFree/touchfree-user-manual/camera-placement.html}
\showURL{%
\tempurl}
\newblock
\shownote{Accessed: 05.12.2024}.


\bibitem[{U.S. Census Bureau}(2022)]%
        {uscensus_commutetime_2022}
\bibfield{author}{\bibinfo{person}{{U.S. Census Bureau}}.} \bibinfo{year}{2022}\natexlab{}.
\newblock \bibinfo{title}{Commute Time in the United States}.
\newblock \bibinfo{howpublished}{\url{https://www.census.gov/programs-surveys/acs}}.
\newblock
\urldef\tempurl%
\url{https://www.census.gov/programs-surveys/acs}
\showURL{%
\tempurl}
\newblock
\shownote{2006 to 2022 American Community Survey, 1-year estimates, Table S0801}.


\bibitem[Wagner et~al\mbox{.}(2023)]%
        {wagnerFittsLawStudy2023}
\bibfield{author}{\bibinfo{person}{Uta Wagner}, \bibinfo{person}{Mathias~N. Lystb\ae{}k}, \bibinfo{person}{Pavel Manakhov}, \bibinfo{person}{Jens Emil~Sloth Gr\o{}nb\ae{}k}, \bibinfo{person}{Ken Pfeuffer}, {and} \bibinfo{person}{Hans Gellersen}.} \bibinfo{year}{2023}\natexlab{}.
\newblock \showarticletitle{A Fitts’ Law Study of Gaze-Hand Alignment for Selection in 3D User Interfaces}. In \bibinfo{booktitle}{\emph{Proceedings of the 2023 CHI Conference on Human Factors in Computing Systems}} (Hamburg, Germany) \emph{(\bibinfo{series}{CHI '23})}. \bibinfo{publisher}{Association for Computing Machinery}, \bibinfo{address}{New York, NY, USA}, Article \bibinfo{articleno}{252}, \bibinfo{numpages}{15}~pages.
\newblock
\showISBNx{9781450394215}
\href{https://doi.org/10.1145/3544548.3581423}{doi:\nolinkurl{10.1145/3544548.3581423}}


\bibitem[Walch et~al\mbox{.}(2019)]%
        {walch2019cooperation}
\bibfield{author}{\bibinfo{person}{Marcel Walch}, \bibinfo{person}{Mark Colley}, {and} \bibinfo{person}{Michael Weber}.} \bibinfo{year}{2019}\natexlab{}.
\newblock \showarticletitle{CooperationCaptcha: On-The-Fly Object Labeling for Highly Automated Vehicles}. In \bibinfo{booktitle}{\emph{Extended Abstracts of the 2019 CHI Conference on Human Factors in Computing Systems}} (Glasgow, Scotland Uk) \emph{(\bibinfo{series}{CHI EA '19})}. \bibinfo{publisher}{Association for Computing Machinery}, \bibinfo{address}{New York, NY, USA}, \bibinfo{pages}{1–6}.
\newblock
\showISBNx{9781450359719}
\href{https://doi.org/10.1145/3290607.3313022}{doi:\nolinkurl{10.1145/3290607.3313022}}


\bibitem[Walch et~al\mbox{.}(2017)]%
        {walch_touch_2017}
\bibfield{author}{\bibinfo{person}{Marcel Walch}, \bibinfo{person}{Lorenz Jaksche}, \bibinfo{person}{Philipp Hock}, \bibinfo{person}{Martin Baumann}, {and} \bibinfo{person}{Michael Weber}.} \bibinfo{year}{2017}\natexlab{}.
\newblock \showarticletitle{Touch {Screen} {Maneuver} {Approval} {Mechanisms} for {Highly} {Automated} {Vehicles}: {A} {First} {Evaluation}}. In \bibinfo{booktitle}{\emph{Proceedings of the 9th {International} {Conference} on {Automotive} {User} {Interfaces} and {Interactive} {Vehicular} {Applications} {Adjunct}}} \emph{(\bibinfo{series}{{AutomotiveUI} '17})}. \bibinfo{publisher}{Association for Computing Machinery}, \bibinfo{address}{New York, NY, USA}, \bibinfo{pages}{206--211}.
\newblock
\showISBNx{978-1-4503-5151-5}
\href{https://doi.org/10.1145/3131726.3131756}{doi:\nolinkurl{10.1145/3131726.3131756}}


\bibitem[Walch et~al\mbox{.}(2016)]%
        {walch_towards_2016}
\bibfield{author}{\bibinfo{person}{Marcel Walch}, \bibinfo{person}{Tobias Sieber}, \bibinfo{person}{Philipp Hock}, \bibinfo{person}{Martin Baumann}, {and} \bibinfo{person}{Michael Weber}.} \bibinfo{year}{2016}\natexlab{}.
\newblock \showarticletitle{Towards {Cooperative} {Driving}: {Involving} the {Driver} in an {Autonomous} {Vehicle}'s {Decision} {Making}}. In \bibinfo{booktitle}{\emph{Proceedings of the 8th {International} {Conference} on {Automotive} {User} {Interfaces} and {Interactive} {Vehicular} {Applications}}} \emph{(\bibinfo{series}{Automotive'{UI} 16})}. \bibinfo{publisher}{Association for Computing Machinery}, \bibinfo{address}{New York, NY, USA}, \bibinfo{pages}{261--268}.
\newblock
\showISBNx{978-1-4503-4533-0}
\href{https://doi.org/10.1145/3003715.3005458}{doi:\nolinkurl{10.1145/3003715.3005458}}


\bibitem[Wobbrock et~al\mbox{.}(2011)]%
        {wobbrock2011AlignedRankTransform}
\bibfield{author}{\bibinfo{person}{Jacob~O. Wobbrock}, \bibinfo{person}{Leah Findlater}, \bibinfo{person}{Darren Gergle}, {and} \bibinfo{person}{James~J. Higgins}.} \bibinfo{year}{2011}\natexlab{}.
\newblock \showarticletitle{The Aligned Rank Transform for Nonparametric Factorial Analyses Using Only Anova Procedures}. In \bibinfo{booktitle}{\emph{Proceedings of the {{SIGCHI Conference}} on {{Human Factors}} in {{Computing Systems}}}}. \bibinfo{publisher}{ACM}, \bibinfo{address}{Vancouver BC Canada}, \bibinfo{pages}{143--146}.
\newblock
\showISBNx{978-1-4503-0228-9}
\href{https://doi.org/10.1145/1978942.1978963}{doi:\nolinkurl{10.1145/1978942.1978963}}


\bibitem[Wolf et~al\mbox{.}(2020)]%
        {wolf2020understanding}
\bibfield{author}{\bibinfo{person}{Dennis Wolf}, \bibinfo{person}{Jan Gugenheimer}, \bibinfo{person}{Marco Combosch}, {and} \bibinfo{person}{Enrico Rukzio}.} \bibinfo{year}{2020}\natexlab{}.
\newblock \showarticletitle{Understanding the heisenberg effect of spatial interaction: A selection induced error for spatially tracked input devices}. In \bibinfo{booktitle}{\emph{Proceedings of the 2020 CHI conference on human factors in computing systems}}. \bibinfo{publisher}{ACM}, \bibinfo{address}{New York, NY, USA}, \bibinfo{pages}{1--10}.
\newblock


\end{thebibliography}


\appendix

\section{Semi-Structured Interview} \label{c:appendixInterview}
\subsection{Post Condition Interview}
The following questions were asked every time a participant finished an input method:
\begin{itemize}
  \item Which challenges did you notice during task execution?
  \item What did you notice positively, and what negatively?
  \item Could you imagine using this method in the future, and for which use-case?
  \item What did you perceive of the outside environment?
  \item How noticeable did you feel your interaction with the system was to other passengers?
  \item How comfortable would you feel using this interaction method in a vehicle with other passengers?
  \item \textit{After a condition has been performed in both movement and stillstand:}
  \begin{itemize}
      \item By comparing the usage of the input method between driving and stillstand, which differences did you notice?
    \end{itemize}
  \item \textit{If input method was performed in a moving vehicle:}
  \begin{itemize}
      \item Did you notice any vehicle movements that influenced the interaction with the system?
    \end{itemize}
\end{itemize}

\subsection{Final Interview}
The following questions were asked after participants finished all conditions:
\begin{itemize}
  \item Please rank the utilized input methods once for usage in a moving vehicle, and once for usage in a standing vehicle.
  \item If you could make any changes to any of the provided input methods, what would you modify and why?
  \item While driving, were there any situations where you felt a specific input method was less suitable or inefficient than another, and why?
\end{itemize}

\section{Ultraleap Hyperion Configuration}
\label{sec:hyperion_hinting}

Hints provided to the Hinting API. In Order:
\begin{itemize}
    \item high\_background\_illumination
    \item ultra\_performance\_mode
    \item microgestures
    \item high\_hand\_fidelity
    \item user\_input
\end{itemize}

\clearpage
\onecolumn
\section{LMM Estimates}

%Selection Time Standstill
\begin{table}[h]
    \centering
    \caption{Significant estimates of the LMM for Selection Time during Standstill. The intercept is represented by DirectTouch. The estimates describe the time differences between the listed terms and the intercept in seconds.}
    \label{tab:lmmEstimatesSelectionTimeStandstill}
    \begin{tabular}{lrrrr}
        \toprule
        Term & Estimate & SE & t & p-Value \\
        \midrule
        (Intercept) &  1.008  &  0.038  &  26.197  & $<$ .001 \\
        Gaze\&Pinch & -0.074  &  0.007  & -11.069  & $<$ .001 \\
        Handray     &  0.128  &  0.007  &  19.157  & $<$ .001 \\
        HeadGaze    &  0.112  &  0.007  &  16.847  & $<$ .001 \\
        \bottomrule
    \end{tabular}
\end{table}

%Selection Time movement
\begin{table}[h]
    \centering
    \caption{Significant estimates of the LMM for Selection Time during Movement. The intercept is represented by DirectTouch with road type SmoothRoad, and curve type Straight Road. The estimates describe the time differences between the listed terms and the intercept in seconds.}
    \label{tab:lmmEstimatesSelectionTimeMovement}
    \begin{tabular}{lrrrr}
        \toprule
        Term &      Estimate  & SE      & t        & p-Value \\
        \midrule
        (Intercept) &  1.087  &  0.033  &  33.253  & $<$ .001 \\
        Gaze\&Pinch & -0.036  &  0.011  &  -3.215  &  0.001 \\
        Handray     &  0.115  &  0.011  &  10.258  & $<$ .001 \\
        HeadGaze    &  0.105  &  0.011  &   9.409  & $<$ .001 \\
        Bumpy Road  &  0.062  &  0.017  &   3.587  & $<$ .001 \\
        Mixed Road  &  0.032  &  0.015  &   2.175  &  0.030 \\
        Long Left Curve with Acc. &  0.211  &  0.066  &   3.207  &  0.001 \\
        Long Left Curve Steady    &  0.092  &  0.040  &   2.325  &  0.020 \\
        Long Right Curve with Acc. &  0.176  &  0.061  &   2.902  &  0.004 \\
        Short Left Curve with Acc. &  0.148  &  0.025  &   5.985  & $<$ .001 \\
        Short Right Curve with Acc. &  0.074  &  0.035  &   2.153  &  0.031 \\
        Gaze\&Pinch:Mixed Road  & -0.046  &  0.021  &  -2.199  &  0.028 \\
        Handray:Long Left Curve Steady  &  0.117  &  0.060  &   1.966  &  0.049 \\
        HeadGaze:Long Left Curve Steady &  0.144  &  0.065  &   2.216  &  0.027 \\
        Mixed Road:Short Left Curve with Acc.  & -0.246  &  0.062  &  -3.945  & $<$ .001 \\
        Handray:Mixed Road:Short Left Curve with Acc.  &  0.312  &  0.093  &   3.351  & $<$ .001 \\
        HeadGaze:Mixed Road:Short Left Curve with Acc. &  0.259  &  0.093  &   2.769  &  0.006 \\
        Gaze\&Pinch:Bumpy Road:Short Right Curve with Acc. & -0.248  &  0.091  &  -2.724  &  0.006 \\
        HeadGaze:Bumpy Road:Short Right Curve with Acc.   & -0.212  &  0.098  &  -2.151  &  0.031 \\
        Gaze\&Pinch:Mixed Road:Short Right Curve with Acc. & -0.170  &  0.071  &  -2.375  &  0.018 \\
        \bottomrule
    \end{tabular}
\end{table}

\clearpage
%SelectionOffset Standstill
\begin{table}[h]
    \centering
    \caption{Significant estimates of the LMM for Selection Offset during Standstill. The intercept is represented by DirectTouch. The estimates describe the distance differences between the listed terms and the intercept in millimeters.}
    \label{tab:lmmEstimatesSelectionOffsetStandstill}
    \begin{tabular}{lrrrr}
        \toprule
        Term        & Estimate & SE    & t        & p-Value \\
        \midrule
        (Intercept) &  13.826 &  0.210 &  65.961  & $<$ .001 \\
        Gaze\&Pinch &  -2.848 &  0.079 & -36.008  & $<$ .001 \\
        Handray     &  -5.633 &  0.081 & -69.603  & $<$ .001 \\
        HeadGaze    &  -6.207 &  0.082 & -75.641  & $<$ .001 \\
        \bottomrule
    \end{tabular}
\end{table}

%SelectionOffset Movement
\begin{table}[h]
    \caption{Significant estimates of the LMM for Selection Offset during Movement. The intercept is represented by DirectTouch with road type SmoothRoad, and curve type Straight Road. The estimates describe the distance differences between the listed terms and the intercept in millimeters.}
    \label{tab:lmmEstimatesSelectionOffsetMovement}
    \centering
    \begin{tabular}{lrrrr}
        \toprule
        Term                                             & Estimate & SE    & t        & p-Value \\
        \midrule
        (Intercept)                                      &  13.429 &  0.197 &  68.065  & $<$ .001 \\
        Gaze\&Pinch                                      &  -1.550 &  0.129 & -12.060  & $<$ .001 \\
        Handray                                          &  -4.032 &  0.136 & -29.612  & $<$ .001 \\
        HeadGaze                                         &  -4.271 &  0.137 & -31.289  & $<$ .001 \\
        Mixed Road                                       &   0.419 &  0.177 &   2.368  &  0.018 \\
        Long Left Curve with Acc.                        &   1.756 &  0.687 &   2.554  &  0.011 \\
        Short Left Curve with Acc.                       &   0.806 &  0.274 &   2.936  &  0.003 \\
        Gaze\&Pinch:Bumpy Road                           &   0.665 &  0.273 &   2.436  &  0.015 \\
        Handray:Bumpy Road                               &   0.598 &  0.286 &   2.087  &  0.037 \\
        HeadGaze:Bumpy Road                              &   1.779 &  0.280 &   6.348  & $<$ .001 \\
        HeadGaze:Short Left Curve with Acc.              &   1.581 &  0.390 &   4.056  & $<$ .001 \\
        HeadGaze:Short Right Curve with Acc.             &   1.173 &  0.576 &   2.034  &  0.042 \\
        Handray:Bumpy Road:Short Left Curve with Acc.    &  -2.348 &  0.881 &  -2.665  &  0.008 \\
        HeadGaze:Bumpy Road:Short Left Curve with Acc.   &  -2.686 &  0.871 &  -3.083  &  0.002 \\
        Gaze\&Pinch:Mixed Road:Short Left Curve with Acc. &   2.637 &  0.990 &   2.662  &  0.008 \\
        \bottomrule
    \end{tabular}
\end{table}

\clearpage

\section{GazeMetrics: Vision Aids}
\begin{table*}[ht!]
\caption{GazeMetrics: Accuracy and Precision filtered by utilized vision aids}
\label{tab:gazemetrics_visionaids}
\centering
\begin{tabular}{llcccccccccccc}
\toprule
 &  & \multicolumn{2}{c}{vision aid} & \multicolumn{2}{c}{RmsPrecision} & \multicolumn{2}{c}{AverageAccuracy} & \multicolumn{2}{c}{SdPrecision.X} & \multicolumn{2}{c}{SdPrecision.Y} & \multicolumn{2}{c}{SdPrecision.Z} \\
\midrule
N &  & Glasses &  & 36 &  & 36 &  & 36 &  & 36 &  & 36 &  \\
 &  & Nothing &  & 378 &  & 378 &  & 378 &  & 378 &  & 378 &  \\
Mean &  & Glasses &  & 0.131 &  & 1.47 &  & 0.00510 &  & 0.00808 &  & 0.0305 &  \\
 &  & Nothing &  & 0.0805 &  & 1.30 &  & 0.00718 &  & 0.00659 &  & 0.0404 &  \\
Median &  & Glasses &  & 0.0354 &  & 1.10 &  & 0.00152 &  & 0.00170 &  & 0.0103 &  \\
 &  & Nothing &  & 0.0339 &  & 1.10 &  & 0.00228 &  & 0.00176 &  & 0.00985 &  \\
SD &  & Glasses &  & 0.292 &  & 1.01 &  & 0.0104 &  & 0.0211 &  & 0.0578 &  \\
 &  & Nothing &  & 0.165 &  & 0.889 &  & 0.0214 &  & 0.0169 &  & 0.0991 &  \\
\bottomrule
\end{tabular}
\end{table*}

\end{document}